\documentclass[twoside, journal]{IEEEtran}

\usepackage{epsfig}
\usepackage{graphicx}
\usepackage{amsmath}
\usepackage{amssymb}
\usepackage{float}
\usepackage{url}
\newtheorem{definition}{Definition}
\newtheorem{lemma}{Lemma}
\newtheorem{proposition}{Proposition}
\newtheorem{corollary}{Corollary}
\newtheorem{theorem}{Theorem}
\newtheorem{conjecture}{Conjecture}
\newtheorem{remark}{Remark}
\newtheorem{example}{Example}
\newtheorem{discussion}{Discussion}

\newcommand{\dsum}{\displaystyle\sum}

\newcommand{\naturals}{\ensuremath{\mathbb{N}}}
\newcommand{\reals}{\ensuremath{\mathbb{R}}}

\newcommand{\pr}{\ensuremath{\mathbb{P}}}
\newcommand{\expectation}{\ensuremath{\mathbb{E}}}
\newcommand{\LL}{\ensuremath{\mathbb{L}}}

\begin{document}
\markboth{SURVEY}{I. SASON: On Refined
Versions of the Azuma-Hoeffding Inequality with Applications in
Information Theory}

\title{ \Huge{On Refined Versions of the Azuma-Hoeffding Inequality
with Applications in Information Theory}}

\author{
Igal~Sason\thanks{I. Sason is with the Department of Electrical Engineering,
Technion -- Israel Institute of Technology, Haifa 32000, Israel
(e-mail: sason@ee.technion.ac.il).}}

\maketitle

\begin{abstract}
This paper derives some refined versions of the Azuma-Hoeffding
inequality for discrete-parameter martingales with uniformly
bounded jumps, and it considers some of their potential applications in
information theory and related topics. The first part of this
paper derives these refined inequalities, followed by a
discussion on their relations to some classical results in
probability theory. It also considers a geometric interpretation
of some of these inequalities, providing an insight on the
inter-connections between them. The second part
exemplifies the use of these refined inequalities in the context of
hypothesis testing and information theory, communication, and
coding theory. The paper is concluded with a discussion on some
directions for further research. This work is meant to stimulate
the use of some refined versions of the Azuma-Hoeffding inequality in
information-theoretic aspects.
\end{abstract}


\section{Introduction}
\label{section: introduction} Inequalities providing upper bounds
on probabilities of the type $\pr(|X-\overline{x}| \geq t)$ (or
$\pr(X-\overline{x} \geq t)$ for a random variable (RV) $X$, where
$\overline{x}$ denotes the expectation or median of $X$, have been
among the main tools of probability theory. These inequalities are
known as concentration inequalities, and they have been subject to
interesting developments in probability theory. Very roughly
speaking, the concentration of measure phenomenon can be stated in
the following simple way: ``A random variable that depends in a
smooth way on many independent random variables (but not too much
on any of them) is essentially constant'' \cite{Talagrand96}. The
exact meaning of such a statement clearly needs to be clarified
rigorously, but it will often mean that such a random variable $X$
concentrates around $\overline{x}$ in a way that the probability
of the event $\{|X-\overline{x}| > t\}$ decays exponentially in $t$
(for $t \geq 0$). The foundations in concentration of
measures have been introduced, e.g., in
\cite[Chapter~7]{AlonS_tpm3}, \cite[Chapter~2]{Chung_LU2006},
\cite{survey2006}, \cite{Ledoux}, \cite{Lugosi_Lecture_Notes},
\cite[Chapter~5]{Massart_book}, \cite{McDiarmid_tutorial},
\cite{Talagrand95} and \cite{Talagrand96}. Concentration inequalities
are also at the core of probabilistic analysis of randomized
algorithms (see, e.g., \cite{AlonS_tpm3}, \cite{DubashiP09_book},
\cite{MitzenmacherU_book}, \cite{MotwaniR_book}, \cite{RiU_book}).

The Chernoff bounds provide sharp concentration inequalities when
the considered RV $X$ can be expressed as a sum of $n$ independent
and bounded RVs. However, the situation is clearly more complex
for non-product measures where the concentration property may not
exist. Several techniques have been developed to prove
concentration of measures. Among several methodologies, these
concentration inequalities include isoperimetric inequalities for product measures
(e.g., \cite{Talagrand95} and \cite{Talagrand96}),
logarithmic-Sobolev inequalities (e.g.,
\cite{Gross75}, \cite{ITW04} and \cite[Chapter~5]{Ledoux}),
transportation-cost inequalities (e.g., \cite[Chapter~6]{Ledoux}),
and the Azuma-Hoeffding inequality that is used to derive
concentration inequalities for discrete-parameter martingales
with bounded jumps
(e.g., \cite[Chapter~7]{AlonS_tpm3}, \cite{Azuma}, \cite{McDiarmid_tutorial}).
The focus of this paper is on the last methodology.

The Azuma-Hoeffding inequality is by now a well-known methodology
that has been often used to prove concentration phenomena.
It is due to Hoeffding \cite{Hoeffding} who
proved this inequality for $X = \sum_{i=1}^n X_i$ where
$\{X_i\}$ are independent and bounded RVs, and Azuma \cite{Azuma}
later extended it to bounded-difference martingales. Some relative
entropy and exponential deviation bounds  were derived in \cite{ISIT05}
for an important class of Markov chains, and these bounds are
essentially identical to the Hoeffding inequality in
the special case of i.i.d. RVs. A common method for proving
concentration of a function $f: \reals^n \rightarrow \reals$
of $n$ independent RVs, around the expected value $\expectation[f]$,
where the function $f$ is characterized by
bounded differences whenever the $n$-dimensional vectors differ
in only one coordinate, is called McDiarmid's inequality (see
\cite[Theorem~3.1]{McDiarmid_tutorial}). Some of the applications
of this inequality are exemplified in \cite[Section~3]{McDiarmid_tutorial}.
The derivation of McDiarmid's inequality is based on introducing
a martingale-difference sequence whose jumps are proved to be
bounded almost surely (a.s.), and then the rest of the proof relies
on the Azuma-Hoeffding inequality.

The use of the Azuma-Hoeffding inequality was introduced to the
computer science literature in \cite{ShamirS87} in order to prove
concentration, around the expected value, of the chromatic number
for random graphs. The chromatic number of a graph is defined to
be the minimal number of colors that is required to color all the
vertices of this graph so that no two vertices which are connected
by an edge have the same color, and the ensemble for which
concentration was demonstrated in \cite{ShamirS87} was the ensemble
of random graphs with $n$ vertices such that any ordered pair of
vertices in the graph is connected by an edge with a fixed probability
$p$ for some $p \in (0,1)$. It is noted that the concentration result in
\cite{ShamirS87} was established without knowing the expected value
over this ensemble. The migration of this bounding inequality into
coding theory, especially for exploring some concentration phenomena
that are related to the analysis of codes defined on graphs and
iterative message-passing decoding algorithms, was initiated in
\cite{LubyMSS_IT01}, \cite{RichardsonU2001} and \cite{SipserS96}.
During the last decade, the Azuma-Hoeffding inequality has been
extensively used for proving concentration of measures in
coding theory (see, e.g., \cite{Kavcic_IT2003},
\cite{MeassonMU08}, \cite{Montanari05}, \cite{RiU_book} and
\cite{Varshney_IT11}). In general, all these concentration
inequalities serve to justify theoretically the ensemble
approach of codes defined on graphs. However, much stronger
concentration phenomena are observed in practice. The
Azuma-Hoeffding inequality was also recently used in
\cite{WagnerPK_IT11} for the analysis of probability estimation in
the rare-events regime where it was assumed that an observed string is drawn i.i.d.
from an unknown distribution, but the alphabet size and the source
distribution both scale with the block length (so the empirical
distribution does not converge to the true distribution as the
block length tends to infinity). In another recent work
\cite{Volterra_ITW09}, Azuma's inequality was used to
derive achievable rates and random coding error exponents for
non-linear additive white Gaussian noise channels. This was followed
by another work of the same authors \cite{Volterra_IT_March11} who used some
other concentration inequalities, for discrete-parameter
martingales with bounded jumps, to derive achievable rates and
random coding error exponents for non-linear Volterra channels
(where their bounding technique can be also applied to intersymbol-interference
(ISI) channels, as was noted in \cite{Volterra_IT_March11}).

This work derives some refined versions of the Azuma-Hoeffding
inequality, and it exemplifies some of their possible applications
in information theory and related topics. The paper is structured
as follows: Section~\ref{section: Preliminaries} presents briefly
some background that is essential to the analysis in this work.
The core of the paper is divided into two parts. The first part
includes Sections~\ref{section: Refined Versions of Azuma's
Inequality} and~\ref{section: relation to previously reported
bound}. Section~\ref{section: Refined Versions of Azuma's
Inequality} is focused on the derivation of some refined versions
of Azuma's inequality, and it considers interconnections between
these bounds. Section~\ref{section: relation to previously
reported bound} considers some relations between concentration
inequalities that are introduced in Section~\ref{section: Refined
Versions of Azuma's Inequality} to the method of types, central
limit theorem, law of iterated logarithm, moderate deviations
principle, and some previously-reported concentration inequalities
for discrete-parameter martingales with bounded jumps. The second
part of this work includes Sections~\ref{section: Applications}
and~\ref{Section: Summary and Outlook}. Section~\ref{section:
Applications} is focused on some of the applications of these
concentration inequalities to hypothesis testing and information
theory, communications and coding. This paper is summarized in
Section~\ref{Section: Summary and Outlook}, followed by a
discussion on some topics for further research (mainly in Shannon
theory and coding). Various mathematical details of the analysis
are relegated to the appendices. This work is meant to stimulate
the derivation and use of some refined versions of the
Azuma-Hoeffding inequality in various information-theoretic
aspects.

\section{Preliminaries}
\label{section: Preliminaries} In the following, we present
briefly some background that is essential to the analysis in this
work, followed by some examples that serve to motivate the
continuation of this paper.

\subsection{Doob's Martingales} \label{subsection: Martingales}
This sub-section provides a short background on martingales to set
definitions and notation. For a more thorough study of
martingales, the reader it referred to standard textbooks, e.g.,
\cite{Billingsley}, \cite{Rosenthal} and \cite{Williams}.

\begin{definition}{\bf[Doob's Martingale]} Let $(\Omega, \mathcal{F},
\pr)$ be a probability space. A Doob's martingale sequence is a
sequence $X_0, X_1, \ldots$ of random variables (RVs) and
corresponding sub $\sigma$-algebras $\mathcal{F}_0, \mathcal{F}_1,
\ldots$ that satisfy the following conditions:
\begin{enumerate}
\item $X_i \in \LL^1(\Omega, \mathcal{F}_i, \pr)$
for every $i$, i.e., each $X_i$ is defined on the same sample
space $\Omega$, it is measurable with respect to the
$\sigma$-algebra $\mathcal{F}_i$ (i.e., $X_i$ is
$\mathcal{F}_i$-measurable) and $\expectation [|X_i|] =
\int_{\Omega} |X_i(\omega)| d\pr(\omega) < \infty.$
\item $\mathcal{F}_0 \subseteq \mathcal{F}_1 \subseteq \ldots $
(this sequence is called a filtration).
\item For all $i \in \naturals$, $X_{i-1}
= \expectation[ X_i | \mathcal{F}_{i-1}]$ almost surely (a.s.).
\end{enumerate}
In this case, it is written that $\{X_i,
\mathcal{F}_i\}_{i=0}^{\infty}$ or $\{X_i, \mathcal{F}_i\}_{i \in
\naturals_0}$ (with $\naturals_0 \triangleq \naturals \cup \{0\}$)
is a martingale sequence (the inclusion of $X_\infty$ and
$\mathcal{F}_{\infty}$ in the martingale is not required here).
\label{definition: Doob's martingales}
\end{definition}

\begin{remark} Since $\{\mathcal{F}_i\}_{i=0}^{\infty}$ forms a
filtration, then it follows from the tower principle for
conditional expectations that a.s.
\begin{equation*}
X_j = \expectation[X_i | \mathcal{F}_j],  \quad \forall \, i>j.
\end{equation*}
Also for every $i \in \naturals$,
$
\expectation[X_i] = \expectation \bigl[ \expectation[X_i |
\mathcal{F}_{i-1}] \bigr] = \expectation[X_{i-1}],
$
so the expectation of a martingale sequence stays constant.
\end{remark}

\begin{remark} One can generate martingale sequences by the
following procedure: Given a RV $X \in \LL^1(\Omega, \mathcal{F},
\pr)$ and an arbitrary filtration of sub $\sigma$-algebras
$\{\mathcal{F}_i\}_{i=0}^{\infty}$, let
\begin{equation*}
X_i = \expectation[X | \mathcal{F}_i], \quad \, \forall \, i \in
\{0, 1, \ldots\}.
\end{equation*}
Then, the sequence $X_0, X_1, \ldots$ forms a martingale since
\begin{enumerate}
\item The RV $X_i = \expectation[X |
\mathcal{F}_i]$ is $\mathcal{F}_i$-measurable, and also
$\expectation[|X_i|] \leq \expectation[|X|] < \infty$ (since
conditioning reduces the expectation of the absolute value).
\item By construction $\{\mathcal{F}_i\}_{i=0}^{\infty}$ is a filtration.
\item For every $i \in \naturals$
\begin{eqnarray*}
&& \hspace*{-1.8cm} \expectation[X_i | \mathcal{F}_{i-1}]
= \expectation \bigl[ \expectation[ X | \mathcal{F}_i] |
\mathcal{F}_{i-1} \bigr] \\
&& = \expectation[X | \mathcal{F}_{i-1}]  \quad (\text{since} \,
\mathcal{F}_{i-1} \subseteq \mathcal{F}_i) \\
&& = X_{i-1} \; \; \; \text{a.s.}
\end{eqnarray*}
\end{enumerate}
\label{remark: construction of martingales}
\end{remark}

\begin{remark}
In continuation to Remark~\ref{remark: construction of martingales},
one can choose
$\mathcal{F}_0 = \{\emptyset, \Omega\}$ and $\mathcal{F}_n = \mathcal{F}$,
so that $X_0, X_1, \ldots, X_n$ is a martingale sequence where
\begin{eqnarray*}
&& \hspace*{-0.8cm} X_0 = \expectation[X | \mathcal{F}_0] =
\expectation[X] \quad
\text{(since $X$ is independent of $\mathcal{F}_0$)} \\
&& \hspace*{-0.8cm} X_n = \expectation[X | \mathcal{F}_n] = X \; \;
\text{a.s.} \quad \text{(since $X$ is $\mathcal{F}$-measurable)}.
\end{eqnarray*}
In this case, one gets a martingale sequence where the first element
is the expected value of $X$, and the last element of the sequence
is $X$ itself (a.s.). This has the following interpretation: At the
beginning, one doesn't know anything about $X$, so it is initially
estimated by its expectation. At each step more and more information
about $X$ is revealed until one is able to specify it exactly (a.s.).
\label{remark: construction of martingales (cont.)}
\end{remark}

\subsection{Azuma's Inequality} \label{subsection:
Azuma's inequality}
Azuma's inequality\footnote{Azuma's inequality is also known as
the Azuma-Hoeffding inequality. Since this inequality is referred
numerous times in this paper, it will be named from this point
as Azuma's inequality for the sake of brevity.}
forms a useful concentration inequality for
bounded-difference martingales \cite{Azuma}. In the following,
this inequality is introduced. The reader is referred to
\cite[Chapter~11]{AlonS_tpm3}, \cite[Chapter~2]{Chung_LU2006},
\cite{survey2006} and \cite{McDiarmid_tutorial} for surveys
on concentration inequalities for (sub/ super) martingales.

\begin{theorem}{\bf[Azuma's inequality]}
Let $\{X_k, \mathcal{F}_k\}_{k=0}^{\infty}$
be a discrete-parameter real-valued martingale sequence
such that for every $k \in \naturals$, the condition
$ |X_k - X_{k-1}| \leq d_k$
holds a.s. for some non-negative constants $\{d_k\}_{k=1}^{\infty}$. Then
\begin{equation}
\pr( | X_n - X_0 | \geq r) \leq 2 \exp\left(-\frac{r^2}{2
\sum_{k=1}^n d_k^2}\right) \, \quad \forall \, r \geq 0.
\label{eq: Azuma's concentration inequality - general case}
\end{equation}
\label{theorem: Azuma's concentration inequality}
\end{theorem}

The concentration inequality stated in Theorem~\ref{theorem:
Azuma's concentration inequality} was proved in \cite{Hoeffding}
for independent bounded random variables, followed by a discussion
on sums of dependent random variables; this inequality was later
derived in \cite{Azuma} for bounded-difference martingales. For a
proof of Theorem~\ref{theorem: Azuma's concentration inequality}
see, e.g., \cite[Chapter~2]{Chung_LU2006},
\cite[Chapter~2.4]{Dembo_Zeitouni} and
\cite[Chapter~12]{MitzenmacherU_book}). It will be revisited in
the next section for the derivation of some refined versions of
Azuma's inequality.

\begin{remark}
In \cite[Theorem~3.13]{McDiarmid_tutorial}, Azuma's inequality is
stated as follows: Let $\{Y_k, \mathcal{F}_k\}_{k=0}^{\infty}$ be a
martingale-difference sequence with $Y_0=0$ (i.e., $Y_k$ is
$\mathcal{F}_k$-measurable, $\expectation[|Y_k|] < \infty$ and
$\expectation[Y_k| \mathcal{F}_{k-1}]=0$ a.s. for every $k \in
\naturals$). Assume that, for every $k \in \naturals$, there
exist numbers $a_k, b_k \in \reals$ such that a.s.
$a_k \leq Y_k \leq b_k$. Then, for every $r \geq 0$,
\begin{equation}
\pr \left( \bigg|\sum_{k=1}^n Y_k\bigg| \geq r \right) \leq 2
\exp\left(-\frac{2r^2}{\sum_{k=1}^n (b_k-a_k)^2} \right).
\label{eq: concentration inequality for a martingale-difference
sequence (McDiarmid's tutorial)}
\end{equation}
Hence, consider a discrete-parameter
real-valued martingale sequence $\{X_k,
\mathcal{F}_k\}_{k=0}^{\infty}$ where $a_k \leq X_k -
X_{k-1} \leq b_k$ a.s. for every $k \in \naturals$. Let $Y_k
\triangleq X_k - X_{k-1}$ for every $k \in \naturals$. This
implies that $\{Y_k, \mathcal{F}_k\}_{k=0}^{\infty}$ is a
martingale-difference sequence. From \eqref{eq: concentration
inequality for a martingale-difference sequence (McDiarmid's
tutorial)}, it follows that for every $r \geq 0$,
\begin{equation}
\pr \left( |X_n - X_0| \geq r \right) \leq 2
\exp\left(-\frac{2r^2}{\sum_{k=1}^n (b_k-a_k)^2} \right).
\label{eq: concentration inequality for a martingale sequence
(McDiarmid's tutorial)}
\end{equation}
Note that 
according to the setting in
Theorem~\ref{theorem: Azuma's concentration inequality}, $a_k
= -d_k$ and $b_k = d_k$ for every $k \in \naturals$, which
implies the equivalence between
\eqref{eq: Azuma's concentration inequality - general case}
and \eqref{eq: concentration inequality for a martingale sequence
(McDiarmid's tutorial)}.
\end{remark}

\vspace*{0.1cm} As a special case of Theorem~\ref{theorem: Azuma's
concentration inequality}, let $\{X_k,
\mathcal{F}_k\}_{k=0}^{\infty}$ be a martingale sequence, and
assume that there exists a constant $d>0$ such that a.s., for
every $k \in \naturals$, $ |X_k - X_{k-1}| \leq d.$ Then, for
every $n \in \naturals$ and $\alpha \geq 0$,
\begin{equation}
\pr( | X_n - X_0 | \geq \alpha \sqrt{n}) \leq 2
\exp\left(-\frac{\alpha^2}{2d^2}\right) \, . \label{eq1: Azuma's
concentration inequality - special case}
\end{equation}

\begin{example}
Let $\{Y_i\}_{i=0}^{\infty}$ be i.i.d. binary random variables
which get the values $\pm d$, for some constant $d>0$, with equal
probability. Let $X_k = \sum_{i=0}^k Y_i$ for $k \in \{0, 1,
\ldots, \}$, and define the natural filtration $\mathcal{F}_0
\subseteq \mathcal{F}_1 \subseteq \mathcal{F}_2 \ldots $ where
$$\mathcal{F}_k = \sigma(Y_0, \ldots, Y_k) \, , \quad \forall
\, k \in \{0, 1, \ldots, \}$$ is the $\sigma$-algebra that is
generated by the random variables $Y_0, \ldots, Y_k$. Note that
$\{X_k, \mathcal{F}_k\}_{k=0}^{\infty}$ is a martingale sequence, and
(a.s.) $ |X_k - X_{k-1}| = |Y_k| = d, \, \forall \, k \in
\naturals$. It therefore follows from Azuma's inequality in
\eqref{eq1: Azuma's concentration inequality - special case} that
\begin{equation}
\pr( | X_n - X_0 | \geq \alpha \sqrt{n}) \leq 2
\exp\left(-\frac{\alpha^2}{2d^2}\right). \label{eq: Azuma's
inequality for example1}
\end{equation}
for every $\alpha \geq 0$ and $n \in \naturals$. From the central
limit theorem (CLT), since the RVs
$\{Y_i\}_{i=0}^{\infty}$ are i.i.d. with zero mean and
variance~$d^2$, then
$\frac{1}{\sqrt{n}} (X_n - X_0) = \frac{1}{\sqrt{n}} \sum_{k=1}^n Y_k$
converges in distribution to
$\mathcal{N}(0,d^2)$. Therefore, for every $\alpha \geq 0$,
\begin{equation}
\lim_{n \rightarrow \infty} \pr( | X_n - X_0 | \geq \alpha
\sqrt{n})= 2 \, Q\Bigl(\frac{\alpha}{d}\Bigr) \label{CLT1 - i.i.d.
RVs}
\end{equation}
where
\begin{equation}
Q(x) \triangleq \frac{1}{\sqrt{2\pi}} \, \int_{x}^{\infty}
\exp\Bigl(-\frac{t^2}{2}\Bigr) \mathrm{d}t, \quad \forall \, x \in
\reals \label{eq: Q function}
\end{equation}
is the probability that a zero-mean and unit-variance Gaussian
RV is larger than $x$. Since the following exponential
upper and lower bounds on the Q-function hold
\begin{equation}
\frac{1}{\sqrt{2\pi}} \, \frac{x}{1+x^2} \cdot e^{-\frac{x^2}{2}}
< Q(x) < \frac{1}{\sqrt{2\pi} \, x} \cdot e^{-\frac{x^2}{2}}, \;
\; \forall \, x>0 \label{eq: upper and lower bounds for the Q
function}
\end{equation}
then it follows from \eqref{CLT1 - i.i.d. RVs} that the exponent
on the right-hand side of \eqref{eq: Azuma's inequality for
example1} is the exact exponent in this example. \label{example1}
\end{example}

\begin{example}
In continuation to Example~\ref{example1}, let $\gamma \in (0,1]$,
and let us generalize this example by considering the case where the
i.i.d. binary RVs $\{Y_i\}_{i=0}^{\infty}$ have the
probability law
$$ \pr(Y_i = +d) = \frac{\gamma}{1+\gamma}, \quad \pr(Y_i =
-\gamma d) = \frac{1}{1+\gamma} \; .$$ Hence, it follows that the
i.i.d. random variables $\{Y_i\}$ have zero mean and variance
$\sigma^2 = \gamma d^2$ as in Example~\ref{example1}. Let $\{X_k,
\mathcal{F}_k\}_{k=0}^{\infty}$ be defined similarly to
Example~\ref{example1}, so that it forms a martingale sequence.
Based on the CLT $$\frac{X_n - X_0}{\sqrt{n}} = \frac{\sum_{k=1}^n
Y_k}{\sqrt{n}}$$ weakly converges to $\mathcal{N}(0, \gamma d^2)$,
so for every $\alpha \geq 0$
\begin{equation}
\lim_{n \rightarrow \infty} \pr( | X_n - X_0 | \geq \alpha
\sqrt{n})= 2 \, Q\biggl(\frac{\alpha}{\sqrt{\gamma} \, d}\biggr).
\label{CLT2 - i.i.d. RVs}
\end{equation}
From the exponential upper and lower bounds of the Q-function in
\eqref{eq: upper and lower bounds for the Q function}, the
right-hand side of \eqref{CLT2 - i.i.d. RVs} scales exponentially
like $e^{-\frac{\alpha^2}{2 \gamma d^2}}$. Hence, the exponent in
this example is improved by a factor $\frac{1}{\gamma}$ as
compared Azuma's inequality (that is the same as in
Example~\ref{example1} since $|X_k - X_{k-1}| \leq d$ for every $k
\in \naturals$). This indicates on the possible refinement of
Azuma's inequality by introducing an additional constraint on the
second moment. This route was studied extensively in the
probability literature, and it is further studied in
Section~\ref{section: Refined Versions of Azuma's Inequality}.
\label{example2}
\end{example}

Example~\ref{example2} serves to motivate the introduction of an
additional constraint on the conditional variance of a martingale
sequence, i.e., adding an inequality constraint of the form
$$ \text{Var}(X_k \, | \, \mathcal{F}_{k-1}) =
\expectation\bigl[(X_k-X_{k-1})^2 \, | \, \mathcal{F}_{k-1}\bigr]
\leq \gamma d^2$$ where $\gamma \in (0,1]$ is a constant. Note
that since, by assumption $|X_k - X_{k-1}| \leq d$ a.s. for every
$k \in \naturals$, then the additional constraint becomes active
when $\gamma < 1$ (i.e., if $\gamma = 1$, then this additional
constraint is redundant, and it coincides with the
setting of Azuma's inequality with a fixed $d_k$ (i.e., $d_k =
d$).


\section{Refined Versions of Azuma's Inequality}
\label{section: Refined Versions of Azuma's Inequality}

\subsection{First Refinement of Azuma's Inequality}

\begin{theorem}
Let $\{X_k, \mathcal{F}_k\}_{k=0}^{\infty}$ be a discrete-parameter real-valued
martingale. Assume that, for some constants $d, \sigma > 0$, the
following two requirements are satisfied a.s.
\begin{eqnarray*}
&& | X_k - X_{k-1} | \leq d, \\
&& \text{Var} (X_k | \mathcal{F}_{k-1}) = \expectation \bigl[(X_k
- X_{k-1})^2 \, | \, \mathcal{F}_{k-1} \bigr] \leq \sigma^2
\end{eqnarray*}
for every $k \in \{1, \ldots, n\}$. Then, for every $\alpha \geq 0$,
\begin{equation}
\hspace*{-0.2cm} \pr(|X_n-X_0| \geq \alpha n) \leq 2 \exp\left(-n
\, D\biggl(\frac{\delta+\gamma}{1+\gamma} \Big|\Big|
\frac{\gamma}{1+\gamma}\biggr) \right) \label{eq: first refined
concentration inequality}
\end{equation}
where
\begin{equation}
\gamma \triangleq \frac{\sigma^2}{d^2}, \quad \delta \triangleq
\frac{\alpha}{d}  \label{eq: notation}
\end{equation}
and
\begin{equation}
D(p || q) \triangleq p \ln\Bigl(\frac{p}{q}\Bigr) + (1-p)
\ln\Bigl(\frac{1-p}{1-q}\Bigr), \quad \forall \, p, q \in [0,1]
\label{eq: divergence}
\end{equation}
is the divergence (a.k.a. relative entropy or
Kullback-Leibler distance) between the two probability
distributions $(p,1-p)$ and $(q,1-q)$. If $\delta>1$, then the
probability on the left-hand side of \eqref{eq: first refined
concentration inequality} is equal to zero. \label{theorem: first
refined concentration inequality}
\end{theorem}

\begin{remark}
Theorem~\ref{theorem: first refined concentration inequality} is
known in the probability literature (see, e.g.,
\cite[Corollary~2.4.7]{Dembo_Zeitouni}), as is discussed later in
Section~\ref{subsection: relations with Theorem 2}. The reasons
for introducing and proving this theorem here are as follows:
\begin{itemize}
\item The geometric interpretation that is associated with the
proof of Theorem~\ref{theorem: first refined concentration
inequality} provides an insight on the underlying connections
between this theorem and some other results (e.g.,
Theorem~\ref{theorem: inequality based on a parabola intersecting
the exponential function at the endpoints of the interval} and
Azuma's inequality).
\item The technique that is used to derive
Theorem~\ref{theorem: first refined concentration inequality}
serves as a starting point for the derivation of
Theorem~\ref{theorem: second inequality}. Then, it is shown that
under a certain sufficient condition, the exponent in
Theorem~\ref{theorem: second inequality} is better than the one in
Theorem~\ref{theorem: first refined concentration inequality}.
This will be also exemplified numerically.
\item Some of the inequalities obtained along the proof of
Theorem~\ref{theorem: first refined concentration inequality} are
meaningful in their own right. They serve to demonstrate, later in
this work, the underlying connections of Theorem~\ref{theorem:
first refined concentration inequality} with some other
concentration inequalities. These inequalities are also helpful
for some applications discussed in the continuation to this work.
\item The proof of Theorem~\ref{theorem: first refined concentration inequality}
is of interest since it indicates that it is possible to improve
the exponent of inequality~\eqref{eq: first refined concentration
inequality} by imposing some additional assumptions on the
conditional distribution of the jumps $\xi_k \triangleq X_k -
X_{k-1}$ given $\mathcal{F}_{k-1}$ (see the first item in
Section~\ref{subsection: Outlook}).
\item The inclusion of Theorem~\ref{theorem: first refined concentration
inequality} and its proof at this stage makes the
material self-contained, and enables the use of the same set
of notation throughout the paper.
\end{itemize}
\end{remark}

\begin{remark}
From the above conditions then without any loss of generality,
$\sigma^2 \leq d^2$ and therefore $\gamma \in (0,1]$.
\end{remark}

\vspace*{0.1cm}
\begin{IEEEproof}
$
X_n - X_0 = \sum_{k=1}^n \xi_k
$
where $\xi_k = X_k - X_{k-1}$ for $k=1, \ldots, n$. By assumption
$|\xi_k| \leq d$ a.s. for some $d>0$, and also for $k=1, \ldots, n$
\begin{eqnarray}
&& \hspace*{-0.7cm} \expectation\bigl[\xi_k \, | \,
\mathcal{F}_{k-1} \bigr] \nonumber \\
&& \hspace*{-0.7cm} = \expectation\bigl[X_k \, | \,
\mathcal{F}_{k-1}\bigr] - \expectation\bigl[X_{k-1}
\, | \, \mathcal{F}_{k-1}\bigr] \nonumber \\
&& \hspace*{-0.7cm} = \expectation\bigl[X_k \, | \,
\mathcal{F}_{k-1}\bigr] - X_{k-1} \quad
\text{(since $X_{k-1}$ is $\mathcal{F}_{k-1}$-measurable)} \nonumber \\
&& \hspace*{-0.7cm} = X_{k-1} - X_{k-1} = 0. \label{eq: zero
expectation}
\end{eqnarray}
Based on Chernoff's inequality, it follows that for every $\alpha
\geq 0$
\begin{eqnarray}
&& \pr(X_n-X_0 \geq \alpha n) \nonumber \\
&& = \pr \left( \sum_{k=1}^n \xi_k \geq \alpha n \right) \nonumber \\
&& \leq \exp(-\alpha nt) \; \expectation \biggl[ \exp \biggl(t
\sum_{k=1}^n \xi_k \biggr) \biggr], \quad \forall \, t \geq 0.
\label{eq: Chernoff}
\end{eqnarray}
For every $t \geq 0$
\begin{eqnarray}
&& \expectation \biggl[ \exp \biggl(t \sum_{k=1}^n \xi_k \biggr)
\biggr] \nonumber \\
&& = \expectation \Biggl[ \expectation \biggl[ \exp \biggl(t
\sum_{k=1}^n \xi_k \biggr) \, | \, \mathcal{F}_{n-1} \biggr] \Biggr]
\nonumber \\
&& = \expectation \Biggl[ \expectation \biggl[ \exp \biggl(t
\sum_{k=1}^{n-1} \xi_k \biggr) \, \exp(t \xi_n) \, | \,
\mathcal{F}_{n-1} \biggr] \Biggr] \nonumber \\
&& = \expectation \Biggl[ \exp \biggl(t \sum_{k=1}^{n-1} \xi_k
\biggr) \, \expectation \bigl[ \exp(t \xi_n) \, | \,
\mathcal{F}_{n-1} \bigr] \Biggr] \label{eq: smoothing theorem}
\end{eqnarray}
where the last transition holds since $Y = \exp \bigl(t
\sum_{k=1}^{n-1} \xi_k \bigr)$ is $\mathcal{F}_{n-1}$-measurable.
The measurability of $Y$ is due to fact that $\xi_k \triangleq X_k
- X_{k-1}$ is $\mathcal{F}_k$-measurable for every $k \in
\naturals$, and $\mathcal{F}_k \subseteq \mathcal{F}_{n-1}$ for $0
\leq k \leq n-1$ since $\{\mathcal{F}_k\}_{k=0}^{\infty}$ is a
filtration; hence, the RV $\sum_{k=1}^{n-1} \xi_k$ and its
exponentiation $(Y)$ are both $\mathcal{F}_{n-1}$-measurable, and
a.s. $\expectation[ XY | \mathcal{F}_{n-1}] = Y \, \expectation[ X
| \mathcal{F}_{n-1}].$

From Bennett's inequality \cite{Bennett} (see, e.g.,
\cite[Lemma~2.4.1]{Dembo_Zeitouni}), if $X$ is a real-valued
random variable with $\overline{x} = \expectation(X)$ and
$\expectation[(X-\overline{x})^2] \leq \sigma^2$ for some
$\sigma>0$, and $X \leq b$ a.s. for some $b \in \reals$, then for
every $\lambda \geq 0$
\begin{equation}
\expectation\bigl[e^{\lambda X}\bigr] \leq
\frac{e^{\lambda \overline{x}} \left[(b-\overline{x})^2 \exp^{-\frac{\lambda
\sigma^2}{b-\overline{x}}}+\sigma^2
e^{\lambda(b-\overline{x})}\right]}{(b-\overline{x})^2+\sigma^2}.
\label{eq: Bennett's inequality for unconditional expectation}
\end{equation}
Applying Bennett's inequality for the conditional law of $\xi_k$
given the $\sigma$-algebra $\mathcal{F}_{k-1}$, since
$\expectation[\xi_k | \mathcal{F}_{k-1}] = 0$, $\text{Var}[\xi_k |
\mathcal{F}_{k-1}] \leq \sigma^2$ and $\xi_k \leq d$ a.s. for $k
\in \naturals$, then a.s.
\begin{equation}
\expectation \left[ \exp(t \xi_k) \, | \, \mathcal{F}_{k-1}
\right] \leq \frac{\sigma^2 \exp(td) + d^2 \exp\left(-\frac{t
\sigma^2}{d}\right)}{d^2 + \sigma^2}. \label{eq: Bennett's
inequality for the conditional law of xi_k}
\end{equation}
Hence, it follows from \eqref{eq: smoothing theorem} and
\eqref{eq: Bennett's inequality for the conditional law of xi_k}
that, for every $t \geq 0$,
\begin{eqnarray*}
&& \hspace*{-0.7cm} \expectation \biggl[ \exp \biggl(t
\sum_{k=1}^n \xi_k \biggr) \biggr] \\
&& \hspace*{-0.7cm} \leq \left(\frac{\sigma^2 \exp(td) + d^2
\exp\left(-\frac{t \sigma^2}{d}\right)}{d^2 + \sigma^2}\right)
\expectation \biggl[ \exp \biggl(t \sum_{k=1}^{n-1} \xi_k \biggr)
\biggr]
\end{eqnarray*}
and, by induction, it follows that for every $t \geq 0$
\begin{equation*}
\expectation \biggl[ \exp \biggl(t \sum_{k=1}^n \xi_k \biggr)
\biggr] \leq \left(\frac{\sigma^2 \exp(td)
+ d^2 \exp\left(-\frac{t \sigma^2}{d}\right)}{d^2 + \sigma^2}\right)^n.
\end{equation*}
From the definition of $\gamma$ in \eqref{eq: notation}, this
inequality is rewritten as

\vspace{-0.2cm} \small
\begin{equation}
\hspace*{-0.2cm} \expectation \biggl[ \exp \biggl(t \sum_{k=1}^n
\xi_k \biggr) \biggr] \leq \left(\frac{\gamma \exp(td) +
\exp(-\gamma td)}{1+\gamma}\right)^n, \; \forall \, t \geq 0.
\label{eq: important inequality used for the derivation of Theorem 2}
\end{equation}
\normalsize Let $x \triangleq td$ (so $x \geq 0$). Combining
\eqref{eq: Chernoff} with
\eqref{eq: important inequality used for the derivation of Theorem 2}
gives that, for every $\alpha \geq 0$ (based on the definition
of $\delta$ in \eqref{eq: notation}, $\alpha t = \delta x$),
\begin{eqnarray}
&& \hspace*{-1.3cm} \pr(X_n-X_0 \geq \alpha n) \nonumber \\
&& \hspace*{-1.3cm} \leq \left(\frac{\gamma \exp\bigl((1-\delta)
x\bigr) + \exp\bigl(-(\gamma+\delta) x\bigr)}{1+\gamma}\right)^n,
\; \forall \, x \geq 0. \label{eq: first concentration inequality
before the optimization over the non-negative parameter x}
\end{eqnarray}
Consider first the case where $\delta=1$ (i.e., $\alpha = d$),
then \eqref{eq: first concentration inequality before the
optimization over the non-negative parameter x} is particularized to
\begin{equation*}
\pr(X_n-X_0 \geq d n) \leq \left( \frac{\gamma +
\exp\bigl(-(\gamma+1)x\bigr)}{1+\gamma} \right)^n, \quad \forall
\, x \geq 0
\end{equation*}
and the tightest bound within this form is obtained in the limit
where $x \rightarrow \infty$. This provides the inequality
\begin{equation}
\pr(X_n-X_0 \geq d n) \leq \left( \frac{\gamma}{1+\gamma}
\right)^n. \label{eq: first concentration inequality for delta=1}
\end{equation}
Otherwise, if $\delta \in [0, 1)$, the minimization of the base of
the exponent on the right-hand side of \eqref{eq: first
concentration inequality before the optimization over the
non-negative parameter x} w.r.t. the free non-negative parameter
$x$ yields that the optimized value is
\begin{equation}
x = \left(\frac{1}{1+\gamma}\right) \ln
\left(\frac{\gamma+\delta}{\gamma (1-\delta)}\right). \label{eq:
optimized non-negative x for the first refined concentration
inequality}
\end{equation}
and its substitution into the right-hand side of
\eqref{eq: first concentration inequality before
the optimization over the non-negative parameter x} gives that,
for every $\alpha \geq 0$,
\begin{eqnarray}
&& \hspace*{-0.7cm} \pr(X_n-X_0 \geq \alpha n) \nonumber \\
&& \hspace*{-0.7cm} \leq \left[ \left(\frac{\gamma+\delta}
{\gamma}\right)^{-\frac{\gamma+\delta}{1+\gamma}}
(1-\delta)^{-\frac{1-\delta}{1+\gamma}} \right]^n \nonumber \\
&& \hspace*{-0.7cm} = \exp \left\{ -n \left[\left(\frac{\gamma
+\delta}{1+\gamma}\right) \ln\left(\frac{\gamma+\delta}{\gamma}\right)+
\left(\frac{1-\delta}{1+\gamma}\right) \ln(1-\delta) \right] \right\} \nonumber\\
&& \hspace*{-0.7cm} = \exp \left( -n \, D\left(\frac{\delta+\gamma}
{1+\gamma} \Big|\Big| \frac{\gamma}{1+\gamma}\right) \right)
\label{eq: one-sided concentration inequality of the first refined bound}
\end{eqnarray}
and the exponent is equal to infinity if $\delta > 1$ (i.e., if
$\alpha > d$). Applying inequality~\eqref{eq: one-sided
concentration inequality of the first refined bound} to the
martingale $\{-X_k, \mathcal{F}_k\}_{k=0}^{\infty}$, and using the
union bound gives the two-sided concentration inequality in
\eqref{eq: first refined concentration inequality}. This completes
the proof of Theorem~\ref{theorem: first refined concentration
inequality}.
\end{IEEEproof}

\begin{example}
Let $d>0$ and $\varepsilon \in (0, \frac{1}{2}]$ be some
constants. Consider a discrete-time real-valued martingale $\{X_k,
\mathcal{F}_k\}_{k=0}^{\infty}$ where a.s. $X_0 = 0$, and for
every $m \in \naturals$
\begin{eqnarray*}
&& \pr(X_m - X_{m-1} = d
\, | \, \mathcal{F}_{m-1}) = \varepsilon \, , \\[0.1cm]
&& \pr\left(X_m - X_{m-1} = -\frac{\varepsilon d}{1-\varepsilon}
\, \Big| \, \mathcal{F}_{m-1}\right) = 1-\varepsilon \, .
\end{eqnarray*}
This indeed implies that a.s. for every $m \in \naturals$
$$ \expectation[X_m - X_{m-1} \, | \, \mathcal{F}_{m-1}] =
\varepsilon d + \left(-\frac{\varepsilon d}{1-\varepsilon}\right)
(1-\varepsilon) = 0$$ and since $X_{m-1}$ is
$\mathcal{F}_{m-1}$-measurable then a.s.
$$\expectation[X_m \, | \, \mathcal{F}_{m-1}] = X_{m-1}.$$
Since $\varepsilon \in (0, \frac{1}{2}]$ then a.s.
$$|X_m - X_{m-1}| \leq \max \left\{d,
\frac{\varepsilon d}{1-\varepsilon} \right\} = d.$$
From Azuma's inequality, for every $x \geq 0$,
\begin{equation}
\pr(X_k \geq x) \leq \exp\left(-\frac{x^2}{2kd^2}\right)
\label{example: Azuma's inequality}
\end{equation}
independently of the value of $\varepsilon$ (note that $X_0=0$
a.s.). The concentration inequality
in Theorem~\ref{theorem: first refined concentration inequality} enables
one to get a better bound: Since a.s., for every $m \in \naturals$,

\vspace*{-0.2cm} \small \begin{eqnarray*} \expectation \bigl[ (X_m
- X_{m-1})^2 \, | \, \mathcal{F}_{m-1} \bigr] = d^2 \varepsilon +
\Bigl(-\frac{\varepsilon d}{1-\varepsilon}\Bigr)^2 \,
(1-\varepsilon) = \frac{d^2 \varepsilon}{1-\varepsilon}
\end{eqnarray*}
\normalsize then from \eqref{eq: notation} $$\gamma =
\frac{\varepsilon}{1-\varepsilon} \, , \, \quad \delta =
\frac{x}{d}$$ and from \eqref{eq: one-sided concentration
inequality of the first refined bound}, for every $x \geq 0$,
\begin{equation}
\pr(X_k \geq x) \leq
\exp\left(-k \, D\Bigl(\frac{x(1-\varepsilon)}{d} + \varepsilon
\; || \; \varepsilon \Bigr) \right).
\label{example: one-sided concentration inequality in Theorem2}
\end{equation}
Consider the case where $\varepsilon \rightarrow 0$. Then,
for every $x>0$, Azuma's inequality in \eqref{example: Azuma's
inequality} provides an upper bound that stays bounded away from
zero, whereas the one-sided concentration inequality of
Theorem~\ref{theorem: first refined concentration inequality}
implies a bound in \eqref{example: one-sided concentration
inequality in Theorem2} that tends to zero. This exemplifies the
improvement that is obtained by Theorem~\ref{theorem: first
refined concentration inequality} as compared to Azuma's
inequality.
\end{example}

\begin{remark}
As was noted, e.g., in \cite[Section~2]{McDiarmid_tutorial}, all
the concentration inequalities for martingales whose derivation is
based on Chernoff's bound can be strengthened to refer to maxima.
The reason is that since $\{X_k - X_0,
\mathcal{F}_k\}_{k=0}^{\infty}$ is a martingale and $h(x) =
\exp(tx)$ is a monotonic increasing function for $t \geq 0$, then
$\bigl\{\exp(t(X_k - X_0)), \mathcal{F}_k\bigr\}_{k=0}^{\infty}$
is a sub-martingale for every $t \geq 0$. Hence, by applying
Doob's maximal inequality for sub-martingales, then
for every $\alpha \geq 0$
\begin{eqnarray*}
&& \pr\Bigl(\max_{1 \leq k \leq n} X_k - X_0 \geq \alpha n\Bigr)  \\
&& = \pr\Bigl(\max_{1 \leq k \leq n} \exp\left(t(X_k - X_0)\right)
\geq \exp(\alpha nt) \Bigr) \quad \quad t \geq 0 \\
&& \leq \exp(-\alpha n t) \;
\expectation \Bigl[ \exp \bigl(t (X_n - X_0) \bigr) \Bigr] \\
&& = \exp(-\alpha n t) \; \expectation \left[\exp \biggl(t
\sum_{k=1}^n \xi_k \biggr) \right]
\end{eqnarray*}
which coincides with the proof of Theorem~\ref{theorem: first
refined concentration inequality} when started from~\eqref{eq:
Chernoff}. This concept applies to all
the concentration inequalities derived in this paper. \label{remark:
maxima}
\end{remark}

\begin{corollary}
In the setting of Theorem~\ref{theorem: first refined
concentration inequality}, for every $\alpha \geq 0$,
\begin{equation}
\pr(|X_n-X_0| \geq \alpha n) \leq 2 \exp\left(-2n \,
\biggl(\frac{\delta}{1+\gamma}\biggr)^2\right). \label{eq:
loosened version of the first refined concentration inequality}
\end{equation}
\label{corollary: loosened version of the first refined
concentration inequality}
\end{corollary}
\begin{proof}
This concentration inequality is a loosened version of
Theorem~\ref{theorem: first refined concentration inequality}.
From Pinsker's inequality,
\begin{equation}
D(p||q) \geq \frac{V^2}{2}, \quad \forall \, p,q \in [0,1]
\label{eq: Pinsker's inequality}
\end{equation}
where
\begin{equation}
V  \triangleq ||(p,1-p)-(q,1-q)||_1 = 2|p-q|
\label{eq: total variation}
\end{equation}
denotes the $L^1$-variational distance between the two probability
distributions. Hence, for $\gamma, \delta \in [0,1]$
\begin{equation*}
D\biggl(\frac{\delta+\gamma}{1+\gamma} \Big|\Big|
\frac{\gamma}{1+\gamma}\biggr) \geq 2 \left(\frac{\delta}{1+\gamma}\right)^2.
\end{equation*}
\end{proof}
\begin{remark}
As was shown in the proof of Corollary~\ref{corollary: loosened
version of the first refined concentration inequality}, the
loosening of the exponential bound in Theorem~\ref{theorem: first
refined concentration inequality} by using Pinsker's inequality
gives inequality~\eqref{eq: loosened version of the first refined
concentration inequality}. Note that \eqref{eq: loosened version
of the first refined concentration inequality} forms a
generalization of Azuma's inequality in Theorem~\ref{theorem:
Azuma's concentration inequality} for the special case where, for
every $i$, $d_i \triangleq d$ for some $d>0$.
Inequality~\eqref{eq: loosened version of the first refined
concentration inequality} is particularized to Azuma's inequality
when $\gamma=1$, and then
\begin{equation}
\pr(|X_n - X_0| \geq \alpha n) \leq 2 \exp\left(-\frac{n \delta^2}{2} \right).
\label{eq: Azuma's inequality}
\end{equation}
This is consistent with the observation that if $\gamma=1$ then,
from \eqref{eq: notation}, the requirement in
Theorem~\ref{theorem: first refined concentration inequality} for
the conditional variance of the bounded-difference martingale
sequence becomes redundant (since if $|X_k - X_{k-1}| \leq d$ a.s.
then also $\expectation[(X_k - X_{k-1})^2 \, | \,
\mathcal{F}_{k-1}] \leq d^2$). Hence, if $\gamma=1$, the
concentration inequality in Theorem~\ref{theorem: first refined
concentration inequality} is derived under the same setting as of
Azuma's inequality. \label{remark: on Pinsker's inequality and the
connection to Azuma's inequality}
\end{remark}
\begin{remark}
The combination of the exponential bound in Theorem~\ref{theorem:
first refined concentration inequality} with Pinsker's inequality
deserves further attention. Note that Pinker's inequality is
especially loose in the case where the $L^1$-variational distance
in \eqref{eq: total variation} is close to~2 (due to \eqref{eq:
total variation}, $V \in [0,2]$ so $\frac{V^2}{2}$ is upper
bounded by~2, whereas the divergence $D(p||q)$ can be made
arbitrarily large in the limit where $V=2|p-q|$ tends to~2). Let
$P$ and $Q$ be two discrete probability distributions defined on a
common measurable space $(\Omega, \mathcal{F})$, and let
$$V(P,Q) \triangleq \sum_{\omega \in \Omega} |P(\omega) - Q(\omega)|$$
denote the $L^1$-variational distance between the probability
measure $P$ and $Q$. Let $$D \triangleq \inf_{(P,Q): V(P,Q) = V}
D(P||Q)$$ be the infimum value of the information divergence
subject to the constraint where the value of the $L^1$-variational
distance is set to $V \in [0,2]$. A refinement of
Pinsker's inequality was introduced in
\cite[Theorem~7]{FedotovHT_2003}, and it states that
\begin{equation}
D \geq \frac{V^2}{2} + \frac{V^4}{36} + \frac{V^6}{270} + \frac{221 V^8}{340220}.
\label{eq: Topsoe's refined version of Pinsker's inequality}
\end{equation}
However, this lower bound in \eqref{eq: Topsoe's refined version
of Pinsker's inequality} suffers from the same problem that it
stays uniformly bounded for $V \in [0,2]$, and it is therefore
especially loose for values of $V$ that are close to~2. A recent
lower bound on the divergence, subject to a fixed value of the
$L^1$-variational distance, was introduced in \cite{Gilardoni}; it
states that
\begin{equation}
D \geq \ln \left(\frac{2}{2-V} \right) - \frac{2-V}{2}
\ln \left( \frac{2+V}{2} \right)
\label{eq: Gilardoni's refined version of Pinsker's inequality}
\end{equation}
which also has the pleasing property that it tends to infinity as
one lets $V$ tend to 2. Note however that \eqref{eq: Gilardoni's
refined version of Pinsker's inequality} is a looser lower bound
than \eqref{eq: Topsoe's refined version of Pinsker's inequality}
for $V \leq 1.708$. 

In the context of Theorem~\ref{theorem: first refined
concentration inequality}, the $L^1$-variational distance that
corresponds to the divergence (see the exponent on the right-hand
side of \eqref{eq: first refined concentration inequality}) is
equal to $V = \frac{2 \delta}{1+\gamma}$. This implies that at
$\gamma=1$, where $V=\delta$, then the following holds:
\begin{itemize}
\item Since, without any loss of generality,
$\delta \triangleq \frac{\alpha}{d}$ is less than or equal to~1
(as otherwise, the right hand side of \eqref{eq: slightly improved
Azuma's inequality} due to the bounded jumps of the martingale)
then the $L^1$-variational distance ($V$) is upper bounded by~1.
Hence, according to the previous paragraph (where $V \leq 1.708$),
then it follows that the lower bound in \eqref{eq: Topsoe's
refined version of Pinsker's inequality} gives a better lower
bound on the divergence of Theorem~\ref{theorem: first refined
concentration inequality} (see the right-hand side of \eqref{eq:
first refined concentration inequality}) than the lower bound in
\eqref{eq: Gilardoni's refined version of Pinsker's inequality}.
\item Instead of using Pinsker's inequality to
reproduce Azuma's inequality from Theorem~\ref{theorem: first
refined concentration inequality}, it is possible to apply the
lower bound in \eqref{eq: Topsoe's refined version of Pinsker's
inequality} to get a slightly improved concentration inequality.
This gives that, for every $\alpha \geq 0$,
\begin{eqnarray}
&& \hspace*{-0.8cm} \pr(|X_n - X_0| \geq \alpha n) \nonumber \\
&& \hspace*{-0.8cm} \leq 2 \exp\left[-n \left(\frac{\delta^2}{2} +
\frac{\delta^4}{36} + \frac{\delta^6}{270} + \frac{221
\delta^8}{340220}\right) \right]. \label{eq: slightly improved
Azuma's inequality}
\end{eqnarray}
Note that, for $\delta \in [0,1]$, the exponent on the right-hand
side of \eqref{eq: slightly improved Azuma's inequality} improves
the exponent of Azuma's inequality in \eqref{eq: Azuma's inequality}
by a marginal factor of at most 1.064 (at $\delta=1$).
\end{itemize}
\end{remark}

\vspace*{0.2cm}
\begin{corollary}
Let $\{X_k, \mathcal{F}_k\}_{k=0}^{\infty}$ be a discrete-parameter
real-valued martingale, and assume that for some constant $d > 0$
\begin{equation*}
| X_k - X_{k-1} | \leq d
\end{equation*}
a.s. for every $k \in \{1, \ldots, n\}$. Then, for every $\alpha \geq 0$,
\begin{equation}
\pr(|X_n-X_0| \geq \alpha n) \leq 2 \exp \left(-n f(\delta)
\right) \label{eq: the first refined concentration inequality with
no constraint on the conditional variance}
\end{equation}
where
\begin{equation}
f(\delta) = \left\{
\begin{array}{ll}
\ln(2) \Bigl[1 - h_2\left(\frac{1-\delta}{2} \right) \Bigr], \quad &0
\leq \delta \leq 1
\\[0.2cm]
+\infty, \quad & \delta > 1
\end{array}
\right. \label{eq: f}
\end{equation}
and $h_2(x) \triangleq -x \log_2(x) - (1-x) \log_2(1-x)$ for $0
\leq x \leq 1$ denotes the binary entropy function on base~2.
\label{corollary: a tightened version of Azuma's inequality for
martingales with bounded jumps}
\end{corollary}
\begin{IEEEproof}
By substituting $\gamma=1$ in Theorem~\ref{theorem: first refined
concentration inequality} (i.e., since
there is no constraint on the conditional variance, then one can
take $\sigma^2=d^2$), the corresponding exponent in
\eqref{eq: first refined concentration inequality} is equal to
\begin{equation*}
D\left(\frac{1+\delta}{2} \Big|\Big| \frac{1}{2}\right) = f(\delta)
\end{equation*}
since $D(p || \frac{1}{2}) = \ln 2 [1-h_2(p)]$ for every $p \in [0,1]$.
\end{IEEEproof}

\begin{remark}
Based on Remark~\ref{remark: on Pinsker's inequality and the
connection to Azuma's inequality}, and since
Corollary~\ref{corollary: a tightened version of Azuma's
inequality for martingales with bounded jumps} is a special case
of Corollary~\ref{corollary: loosened version of the first refined
concentration inequality} when $\gamma=1$, then it follows that
Corollary~\ref{corollary: a tightened version of Azuma's
inequality for martingales with bounded jumps} is a tightened
version of Azuma's inequality. This can be verified directly, by
comparing the exponents of \eqref{eq: Azuma's inequality} and
\eqref{eq: the first refined concentration inequality with no
constraint on the conditional variance}. To this end, it is
required to show that $f(\delta) > \frac{\delta^2}{2}$ for every
$\delta > 0$. If $\delta > 1$, then it is obvious since $f$ is by
definition infinity, whereas the right-hand side of this
inequality is finite. In order to prove this inequality for
$\delta \in (0,1]$, note that the power series expansion of the
binary entropy function around one-half is equal to
\begin{equation*}
h_2(x) = 1 - \frac{1}{2 \ln 2} \sum_{p=1}^{\infty}
\frac{(1-2x)^{2p}}{p(2p-1)}, \quad 0 \leq x \leq 1
\end{equation*}
so from \eqref{eq: f}, for every $\delta \in [0,1]$,
\begin{equation}
\hspace*{-0.2cm} f(\delta) = \sum_{p=1}^{\infty} \frac{\delta^{2p}}{2p(2p-1)}
= \frac{\delta^2}{2} + \frac{\delta^4}{12} +
\frac{\delta^6}{30} + \frac{\delta^8}{56} + \frac{\delta^{10}}{90} \ldots
\label{eq: power series expansion of f}
\end{equation}
which indeed proves that $f(\delta) > \frac{\delta^2}{2}$ for
$\delta \in (0,1]$. It is shown in Figure~\ref{Figure:
compare_exponents_theorem2} that the two exponents in \eqref{eq:
Azuma's inequality} and \eqref{eq: the first refined concentration
inequality with no constraint on the conditional variance} nearly
coincide for $\delta \leq 0.4$. Also, the improvement in the
exponent of the right-hand side of \eqref{eq: the first refined
concentration inequality with no constraint on the conditional
variance} as compared to the exponent of Azuma's inequality in
\eqref{eq: Azuma's inequality} at the end point where $\delta=1$
is a by a factor $2 \ln 2 \approx 1.386$. This improvement in the
exponent of \eqref{eq: Azuma's inequality} is larger than the
factor of 1.064 obtained by \eqref{eq: slightly improved Azuma's
inequality}. This follows from the use of the lower bound in
\eqref{eq: Topsoe's refined version of Pinsker's inequality} of
the divergence at $\gamma=1$ in Theorem~\ref{theorem: first
refined concentration inequality}, instead of its exact
calculation at $\gamma=1$ that leads to the improved bound in
\eqref{eq: the first refined concentration inequality with no
constraint on the conditional variance}. As a result of this, the
power series on the right-hand side of \eqref{eq: power series
expansion of f} replaces the exponent on the right-hand side of
\eqref{eq: slightly improved Azuma's inequality}. \label{remark:
comparison of exponents}
\end{remark}

\begin{discussion}
Corollary~\ref{corollary: a tightened
version of Azuma's inequality for martingales with bounded jumps}
can be re-derived by
the replacement of Bennett's inequality in \eqref{eq: Bennett's
inequality for the conditional law of xi_k} with the inequality
\begin{equation}
\expectation[\exp(t \xi_k) | \mathcal{F}_{k-1}] \leq \frac{1}{2}
\bigl[e^{td} + e^{-td}\bigr] = \cosh(td) \label{eq: upper bounding
the exponential function over an interval by a line segment}
\end{equation}
that holds a.s. due to the assumption that $|\xi_k| \leq d$ (a.s.)
for every $k$. The geometric interpretation
of this inequality is based on the convexity of the exponential
function, which implies that its curve is below the line segment
that intersects this curve at the two endpoints of the interval $[-d,d]$.
Hence,
\begin{equation}
\exp(t \xi_k) \leq \frac{1}{2} \left(1 + \frac{\xi_k}{d} \right)
e^{td} + \frac{1}{2} \left(1-\frac{\xi_k}{d}\right) e^{-td}
\label{eq: line segment as an upper bound on the exponential
function over an interval}
\end{equation}
a.s. for every $k \in \naturals$ (or vice versa since $\naturals$
is a countable set). Since, by assumption, $\{X_k,
\mathcal{F}_k\}_{k=0}^{\infty}$ is a martingale then
$\expectation[\xi_k | \mathcal{F}_{k-1}]=0$ a.s. for every $k \in
\naturals$, so \eqref{eq: upper bounding the exponential function
over an interval by a line segment} indeed follows from \eqref{eq:
line segment as an upper bound on the exponential function over an
interval}. Combined with Chernoff's inequality, it yields (after
making the substitution $x=td$ where $x \geq 0$) that
\begin{equation}
\hspace*{-0.15cm} \pr(X_n-X_0 \geq \alpha n) \leq
\bigl(\exp(-\delta x) \cosh(x) \bigr)^n, \quad \forall \, x \geq
0. \label{eq: derivation of Azuma's inequality before replacing
the hyperbolic cosine with an exponential upper bound}
\end{equation}
This inequality leads to the derivation of Azuma's inequality. The
difference that makes Corollary~\ref{corollary: a tightened
version of Azuma's inequality for martingales with bounded jumps}
be a tightened version of Azuma's inequality is that in
the derivation of Azuma's inequality, the hyperbolic cosine is
replaced with the bound $\cosh(x) \leq
\exp\bigl(\frac{x^2}{2}\bigr)$ so the inequality in \eqref{eq:
derivation of Azuma's inequality before replacing the hyperbolic
cosine with an exponential upper bound} is loosened, and then the
free parameter $x \geq 0$ is optimized to obtain Azuma's
inequality in Theorem~\ref{theorem: Azuma's concentration
inequality} for the special case where $d_k \triangleq d$ for
every $k \in \naturals$ (note that Azuma's inequality handles the
more general case where $d_k$ is not a fixed value for every $k$).
In the case where $d_k \triangleq d$ for every $k$,
Corollary~\ref{corollary: a tightened
version of Azuma's inequality for martingales with bounded jumps}
is obtained by an optimization of the
non-negative parameter $x$ in \eqref{eq: derivation of Azuma's
inequality before replacing the hyperbolic cosine with an
exponential upper bound}. If $\delta \in [0,1]$, then by setting
to zero the derivative of the logarithm of the right-hand side of
\eqref{eq: derivation of Azuma's inequality before replacing the
hyperbolic cosine with an exponential upper bound}, it follows
that the optimized value is equal to $x = \tanh^{-1}(\delta)$.
Substituting this value into the right-hand side of \eqref{eq:
derivation of Azuma's inequality before replacing the hyperbolic
cosine with an exponential upper bound} provides the concentration
inequality in Corollary~\ref{corollary: a tightened
version of Azuma's inequality for martingales with bounded jumps};
to this end, one needs to rely on the identities
\begin{equation*}
\tanh^{-1}(\delta) = \frac{1}{2} \ln
\left(\frac{1+\delta}{1-\delta} \right), \quad
\cosh(x) = \bigl(1-\tanh^2(x)\bigr)^{-\frac{1}{2}}.
\end{equation*}
\end{discussion}

\begin{figure}[here!]  
\begin{center}
\epsfig{file=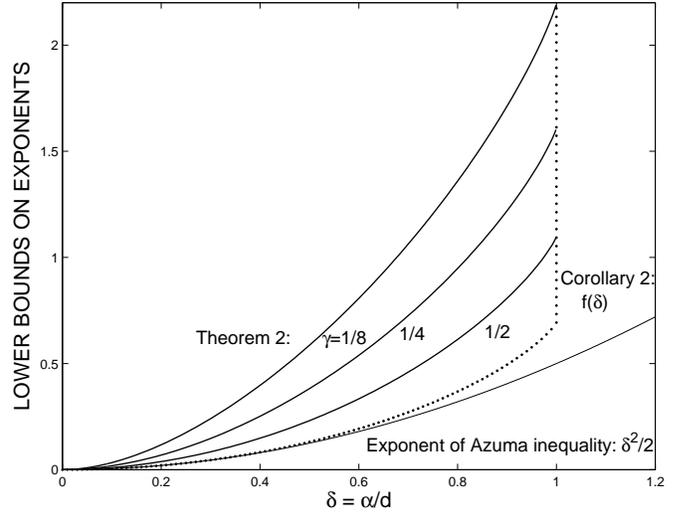,scale=0.5}
\end{center}
\caption{\label{Figure: compare_exponents_theorem2} Plot of the
lower bounds on the exponents from Azuma's inequality in
\eqref{eq: Azuma's inequality} and the refined inequalities in
Theorem~\ref{theorem: first refined concentration inequality} and
Corollary~\ref{corollary: a tightened version of Azuma's
inequality for martingales with bounded jumps} (where $f$
is defined in \eqref{eq: f}). The pointed line
refers to the exponent in Corollary~\ref{corollary: a tightened
version of Azuma's inequality for martingales with bounded jumps},
and the three solid lines for $\gamma = \frac{1}{8}, \frac{1}{4}$
and $\frac{1}{2}$ refer to the exponents in Theorem~\ref{theorem:
first refined concentration inequality}.}
\end{figure}

We obtain in the following a loosened version of
Theorem~\ref{theorem: first refined concentration inequality}.
\begin{lemma}
For every $x, y \in [0,1]$
\begin{equation}
D\left(\frac{x+y}{1+y} \Big|\Big| \frac{y}{1+y} \right)
\geq \frac{x^2}{2y} \; B\Bigl(\frac{x}{y}\Bigr)
\label{eq: lower bound on the divergence}
\end{equation}
where
\begin{equation}
B(u) \triangleq \frac{2[(1+u) \ln(1+u) - u]}{u^2}, \quad \forall
\, u>0. \label{eq: B}
\end{equation}
\label{lemma: lower bound on the divergence}
\end{lemma}
\begin{IEEEproof}
This inequality follows by calculus, and it appears in
\cite[Exercise~2.4.21~(a)]{Dembo_Zeitouni}.
\end{IEEEproof}

\begin{corollary}
Let $\{X_k, \mathcal{F}_k\}_{k=0}^{\infty}$ be a discrete-parameter
real-valued martingale that satisfies the conditions in
Theorem~\ref{theorem: first refined concentration inequality}.
Then, for every $\alpha \geq 0$,
\begin{eqnarray}
&& \hspace*{-1cm} \pr(|X_n-X_0| \geq \alpha n) \nonumber\\
&& \hspace*{-1cm} \leq 2 \exp \left(-n \gamma
\left[\left(1+\frac{\delta}{\gamma}\right)
\ln\left(1+\frac{\delta}{\gamma}\right) - \frac{\delta}{\gamma}
\right] \right)
\label{eq: 3rd corollary}
\end{eqnarray}
where $\gamma, \delta \in [0,1]$ are introduced in \eqref{eq: notation}.
\label{corollary: 3rd corollary}
\end{corollary}
\begin{IEEEproof}
This inequality follows directly by combining
inequalities~\eqref{eq: first refined concentration inequality}
and~\eqref{eq: lower bound on the divergence} with the equality
in \eqref{eq: B}.
\end{IEEEproof}

\subsection{Geometric Interpretation}
The basic inequality that leads to the derivation of Azuma's
inequality (and also its tightened version in
Corollary~\ref{corollary: a tightened
version of Azuma's inequality for martingales with bounded jumps})
relies on the convexity of the exponential function. Hence, this function
is upper bounded over an arbitrary interval by the line segment
that intersects the curve of this exponential function at the two
endpoints of this interval. Under the additional assumption made
in Theorem~\ref{theorem: first refined concentration inequality}
regarding the conditional variance, one may be motivated by the
above geometric viewpoint to improve Azuma's inequality by looking
for a suitable parabola that coincides with the exponential
function at the two endpoints of the interval, and which forms an
improved upper bound to this exponential function over the
considered interval (as compared to the upper bound that is
obtained by referring to the line segment that intersects the curve of the
exponential function at the two endpoints of this interval, see
inequality~\eqref{eq: line segment as an upper bound on the
exponential function over an interval}). The analysis that follows
from this approach leads to the following theorem.
\begin{theorem}
Let $\{X_k, \mathcal{F}_k\}_{k=0}^{\infty}$ be a discrete-parameter
real-valued martingale that satisfies the conditions in
Theorem~\ref{theorem: first refined concentration inequality}
with some constants $d, \sigma > 0$. Then, for every $\alpha \geq 0$,
\begin{equation*}
\pr(|X_n-X_0| \geq \alpha n) \leq 2 \exp\bigl(-n C(\gamma, \delta)
\bigr)
\end{equation*}
where $\gamma$ and $\delta$ are introduced in \eqref{eq:
notation}, and the exponent in this bound is defined as follows:
\begin{itemize}
\item If $\delta>1$ then $C(\gamma, \delta) = \infty$.
\item If $\delta=1$ then
\begin{equation*}
C(\gamma, \delta) = \ln \biggl(\frac{4}{1+\gamma}\biggr).
\end{equation*}
\item Otherwise, if $\delta \in [0,1)$, then
\begin{equation*}
C(\gamma, \delta) = -\ln(u+v)
\end{equation*}
where
\begin{eqnarray*}
&& u \triangleq \left(\frac{1+\gamma}{4}\right) \, e^{(1-\delta)x} \\
&& v \triangleq \left(\frac{1}{2} + \frac{(1+2x)(1-\gamma)}{4}
\right) \, e^{-(1+\delta)x}.
\end{eqnarray*}
In the above two equalities, $x \in (0, \infty)$ is given by
\begin{equation*}
x \triangleq -\frac{1+W_{-1}(w)}{2} - \frac{\gamma +
\delta}{(1+\delta)(1-\gamma)}
\end{equation*}
where $W_{-1}$ stands for a branch of the Lambert W function
\cite{Lambert_function}, and
\begin{equation*}
w \triangleq - \frac{(1+\gamma)(1-\delta)}{(1-\gamma)(1+\delta)}
\cdot e^{-1-\frac{2(\gamma+\delta)}{(1+\delta)(1-\gamma)}}.
\end{equation*}
\end{itemize}
\label{theorem: inequality based on a parabola intersecting the
exponential function at the endpoints of the interval}
\end{theorem}

\begin{proof}
See Appendix~\ref{appendix: inequality based on a parabola
intersecting the exponential function at the endpoints of the
interval}.
\end{proof}

As is explained in the following discussion, Theorem~\ref{theorem:
inequality based on a parabola intersecting the exponential
function at the endpoints of the interval} is looser than
Theorem~\ref{theorem: first refined concentration inequality}
(though it improves Corollary~\ref{corollary: a tightened version
of Azuma's inequality for martingales with bounded jumps} and
Azuma's inequality that are independent of $\gamma$). The reason
for introducing Theorem~\ref{theorem: inequality based on a
parabola intersecting the exponential function at the endpoints of
the interval} here is in order to emphasize the geometric
interpretation of the concentration inequalities that were
introduced so far, as is discussed in the following.

\begin{discussion}
A common ingredient in proving Azuma's inequality, and
Theorems~\ref{theorem: first refined concentration inequality}
and~\ref{theorem: inequality based on a parabola intersecting the
exponential function at the endpoints of the interval} is a
derivation of an upper bound on the conditional expectation
$\expectation\bigl[e^{t \xi_k} \, | \, \mathcal{F}_{k-1}\bigr]$
for $t \geq 0$ where $\expectation\bigl[\xi_k \, | \,
\mathcal{F}_{k-1}\bigr] = 0$, $\text{Var}\bigl[\xi_k |
\mathcal{F}_{k-1}\bigr] \leq \sigma^2$, and $|\xi_k| \leq d$ a.s.
for some $\sigma, d>0$ and for every $k \in \naturals$. The
derivation of Azuma's inequality and Corollary~\ref{corollary: a
tightened version of Azuma's inequality for martingales with
bounded jumps} is based on the line segment that connects the
curve of the exponent $y(x) = e^{tx}$ at the endpoints of the
interval $[-d, d]$; due to the convexity of $y$, this chord is
above the curve of the exponential function $y$ over the interval
$[-d,d]$. The derivation of Theorem~\ref{theorem: first refined
concentration inequality} is based on Bennett's inequality which
is applied to the conditional expectation above. The proof of
Bennett's inequality (see, e.g.,
\cite[Lemma~2.4.1]{Dembo_Zeitouni}) is shortly reviewed, while
adopting its proof to our notation, for the continuation of this
discussion. Let $X$ be a random variable with zero mean and
variance $E[X^2]=\sigma^2$, and assume that $X \leq d$ a.s. for
some $d>0$. Let $\gamma \triangleq \frac{\sigma^2}{d^2}$. The
geometric viewpoint of Bennett's inequality is based on the
derivation of an upper bound on the exponential function $y$ over
the interval $(-\infty, d]$; this upper bound on $y$ is a parabola
that intersects $y$ at the right endpoint $(d, e^{td})$ and is
tangent to the curve of $y$ at the point $(-\gamma d, e^{-t \gamma
d})$. As is verified in the proof of
\cite[Lemma~2.4.1]{Dembo_Zeitouni}, it leads to the inequality
$y(x) \leq \varphi(x)$ for every $x \in (-\infty, d]$ where
$\varphi$ is the parabola that satisfies the conditions
\begin{eqnarray*}
&& \varphi(d) = y(d) = e^{td}, \\
&& \varphi(-\gamma d) = y(-\gamma d) = e^{-t\gamma d}, \\
&& \varphi'(-\gamma d) = y'(-\gamma d) = t e^{-t \gamma d}.
\end{eqnarray*}
Calculation shows that this parabola admits the form

\vspace*{-0.2cm}
\small
\begin{equation*}
\varphi(x) = \frac{(x+\gamma d) e^{td} + (d-x) e^{-t
\gamma d}}{(1+\gamma)d} + \frac{\alpha [\gamma d^2 + (1-\gamma)d
\; x - x^2]}{(1+\gamma)^2 d^2}
\end{equation*}
\normalsize
where $\alpha \triangleq \bigl[(1+\gamma) t d + 1 \bigr] e^{-t
\gamma d} - e^{td}$. At this point, since $\expectation[X] = 0$,
$\expectation[X^2] = \gamma d^2$ and $X \leq d$ a.s., then the
following bound holds:
\begin{eqnarray}
&& \hspace*{-1.2cm} \expectation\bigl[e^{t X}\bigr] \leq
\expectation\bigl[\varphi(X)\bigr] \nonumber \\
&& = \frac{\gamma e^{td} + e^{-\gamma t d}}{1+\gamma} + \alpha
\left(\frac{\gamma d^2 - \expectation[X^2]}{(1+\gamma)^2 d^2}
\right) \nonumber \\ 
&& = \frac{\gamma e^{td} + e^{-\gamma t d}}{1+\gamma} \nonumber \\
&& = \frac{\expectation[X^2] e^{td} + d^2 e^{-\frac{t
\expectation[X^2]}{d}}}{d^2+\expectation[X^2]} \nonumber
\end{eqnarray}
which indeed proves Bennett's inequality in the considered
setting, and it also provides a geometric viewpoint to the proof
of this inequality. Note that under the above assumption, the
bound is achieved with equality when $X$ is a RV that
gets the two values $+d$ and $-\gamma d$ with probabilities
$\frac{\gamma}{1+\gamma}$ and $\frac{1}{1+\gamma}$, respectively.
This bound also holds when $\expectation[X^2] \leq \sigma^2$ since
the right-hand side of the inequality is a monotonic non-decreasing
function of $\expectation[X^2]$ (as it was verified in the proof
of \cite[Lemma~2.4.1]{Dembo_Zeitouni}).

\begin{figure}[here!]
\begin{center}
\epsfig{file=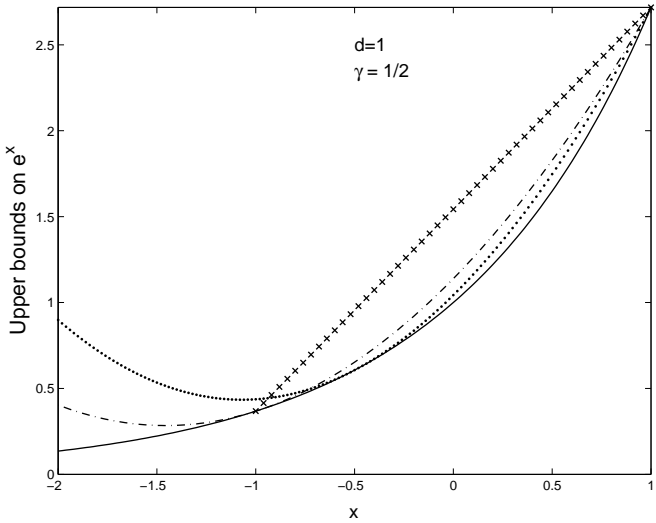,scale=0.5}
\end{center}
\caption{\label{Figure: linear_and_parabolic_bounds_of_exponent}
The function $y = e^x$ (solid line), and the upper bounds on this
function that are used to derive Azuma's inequality and
Corollary~\ref{corollary: a tightened
version of Azuma's inequality for martingales with bounded jumps}
(the dashed line segment intersecting the exponential function at the
endpoints of the interval $[-d,d]$), Theorem~\ref{theorem: first
refined concentration inequality} (the pointed line for the
parabola that coincides with the exponential function at $x=d$ and
is tangent to this function at $x = -\gamma d$), and
Theorem~\ref{theorem: inequality based on a parabola intersecting
the exponential function at the endpoints of the interval} (the
dash-dot line for the parabola that coincides with the exponential
function at $x=d$ and is tangent to this function at $x = -d$).
These parabolas are upper bounds on the exponential function over
$(-\infty, d]$.}
\end{figure}

Applying Bennett's inequality to the conditional law of $\xi_k$
given $\mathcal{F}_{k-1}$ gives \eqref{eq: Bennett's inequality
for the conditional law of xi_k} (with $\gamma$ in \eqref{eq:
notation}). From this discussion, the parabola that serves for the
derivation of Bennett's inequality is the best one in the sense
that it achieves the minimal upper bound on the conditional
expectation $\expectation\bigl[e^{t \xi_k} \, | \,
\mathcal{F}_{k-1}\bigr]$ (where $t \geq 0$) with equality for a
certain conditional probability distribution. In light of this
geometric interpretation, it follows from the proof of
Theorem~\ref{theorem: inequality based on a parabola intersecting
the exponential function at the endpoints of the interval} that
the concentration inequality in this theorem is looser than the
one in Theorem~\ref{theorem: first refined concentration
inequality}. The reason is that the underlying parabola that
serves to get an upper bound on the exponential function in
Theorem~\ref{theorem: inequality based on a parabola intersecting
the exponential function at the endpoints of the interval} is the
parabola that intersects $y$ at $x=d$ and is tangent to the curve
of this exponent at $x=-d$; as is illustrated in
Figure~\ref{Figure: linear_and_parabolic_bounds_of_exponent}, this
parabola forms an upper bound on the exponential function $y$ over
the interval $(-\infty, d]$. On the other hand,
Theorem~\ref{theorem: inequality based on a parabola intersecting
the exponential function at the endpoints of the interval} refines
Azuma's inequality and Corollary~\ref{corollary: a tightened
version of Azuma's inequality for martingales with bounded jumps}
since the chord that connects the curve of the exponential
function at the two endpoints of the interval $[-d, d]$ is
replaced by a tighter upper bound which is the parabola that
coincides with the exponent at the two endpoints of this interval.
Figure~\ref{Figure: linear_and_parabolic_bounds_of_exponent}
compares the three considered upper bounds on the exponential
function that serve for the derivation of Azuma's inequality (and
Corollary~\ref{corollary: a tightened version of Azuma's
inequality for martingales with bounded jumps}), and
Theorems~\ref{theorem: first refined concentration inequality}
and~\ref{theorem: inequality based on a parabola intersecting the
exponential function at the endpoints of the interval}. A
comparison of the resulting bounds on the exponents of these
inequalities and some other bounds that are derived later in this
section is shown in Figure~\ref{Figure:
compare_exponents_theorems_1_2_3_and_corollary_4}; it verifies
that indeed the exponent of Theorem~\ref{theorem: first refined
concentration inequality} is superior over the exponent in
Theorem~\ref{theorem: inequality based on a parabola intersecting
the exponential function at the endpoints of the interval}, but
this difference is reduced by increasing the value of $\gamma \in
(0,1]$ (e.g., for $\gamma = \frac{3}{4}$, this difference is
already marginal). The reason for this observation is that the two
underlying parabolas that serve for the derivation of
Theorems~\ref{theorem: first refined concentration inequality}
and~\ref{theorem: inequality based on a parabola intersecting the
exponential function at the endpoints of the interval} almost
coincide when the value of $\gamma$ is approached to~1 (and they
are exactly the same parabola when $\gamma=1$); in this respect,
note that the left tangent point at $x = -\gamma d$ for the
parabola that refers to the derivation of Theorem~\ref{theorem:
first refined concentration inequality} (via Bennet's inequality)
tends to the left endpoint of the interval $[-d, d]$ as $\gamma
\rightarrow 1$, and therefore the two parabolas almost coincide
for $\gamma$ close to~1.
\end{discussion}

\subsection{Another Approach for the Derivation of a Refinement of Azuma's Inequality}
\label{subsection: Another Approach for the Derivation of a
Refinement of Azuma's Inequality}

\begin{theorem}
Let $\{X_k, \mathcal{F}_k\}_{k=0}^{\infty}$ be a discrete-parameter
real-valued martingale, and let
$m \in \naturals$ be an even number. Assume that the following
conditions hold a.s. for every $k \in \naturals$
\begin{eqnarray*}
&& | X_k - X_{k-1} | \leq d, \\[0.1cm]
&& \Big| \expectation \bigl[(X_k - X_{k-1})^l \, | \,
\mathcal{F}_{k-1} \bigr] \Big| \leq \mu_l, \quad l = 2, \ldots, m
\end{eqnarray*}
for some $d>0$ and non-negative numbers $\{\mu_l\}_{l=2}^m$. Then,
for every $\alpha \geq 0$,
\begin{eqnarray}
&& \hspace*{-0.7cm} \pr(|X_n - X_0| \geq n \alpha) \nonumber \\
&& \hspace*{-0.7cm} \leq 2 \left\{ \inf_{x \geq 0} \, e^{-\delta
x} \left[1 + \sum_{l=2}^{m-1} \frac{(\gamma_l - \gamma_m) x^l}{l!}
+ \gamma_m (e^x-1-x) \right] \right\}^n \nonumber \\[0.1cm]
\label{eq: 2nd concentration inequality}
\end{eqnarray}
where
\begin{equation}
\delta \triangleq \frac{\alpha}{d}, \quad
\gamma_l \triangleq \frac{\mu_l}{d^l}, \; \; \forall \; l = 2,
\ldots, m.
\label{eq: gamma and delta for Theorem 4}
\end{equation}
\label{theorem: second inequality}
\end{theorem}
\begin{IEEEproof}
The starting point of this proof relies on \eqref{eq: Chernoff}
and \eqref{eq: smoothing theorem} that were used for the
derivation of Theorem~\ref{theorem: first refined concentration
inequality}. From this point, we deviate from the proof of
Theorem~\ref{theorem: first refined concentration inequality}. For
every $k \in \naturals$ and $t \geq 0$
\begin{eqnarray}
&& \hspace*{-0.5cm} \expectation \bigl[ \exp(t \xi_k) |
\mathcal{F}_{k-1} \bigr] \nonumber \\
&& \hspace*{-0.5cm} = 1 + t \expectation\bigl[\xi_k |
\mathcal{F}_{k-1}\bigr] + \ldots + \frac{t^{m-1}}{(m-1)!}
\cdot \expectation\bigl[(\xi_k)^{m-1} |
\mathcal{F}_{k-1}\bigr] \nonumber \\
&& \hspace*{-0.2cm} + \expectation \left[
\exp(t \xi_k) - 1 - t \xi_k - \ldots -
\frac{t^{m-1} (\xi_k)^{m-1}}{(m-1)!} \right] \nonumber \\
&& \hspace*{-0.5cm} = 1 + t \expectation\bigl[\xi_k |
\mathcal{F}_{k-1}\bigr] + \ldots + \frac{t^{m-1}}{(m-1)!}
\cdot \expectation\bigl[(\xi_k)^{m-1} | \mathcal{F}_{k-1}\bigr]
\nonumber \\
&& \hspace*{-0.2cm} + \frac{t^m}{m!} \cdot \expectation \bigl[
(\xi_k)^m \varphi_m(t \xi_k)  | \mathcal{F}_{k-1}\bigr] \label{eq:
chain of equalities for the conditional expectation}
\end{eqnarray}
where
\begin{equation}
\varphi_m(y) \triangleq \left\{\begin{array}{ll} \frac{m!}{y^m}
\left(e^y-\sum_{l=0}^{m-1} \frac{y^l}{l!} \right) \quad &
\mbox{if} \; y \neq 0 \\
1 \quad & \mbox{if} \; y=0 \end{array}
\right. .
\label{eq: phi function}
\end{equation}
In order to proceed, we need the following lemma:
\begin{lemma}
Let $m \in \naturals$ be an even number, then the function
$\varphi_m$ has the following properties:
\begin{enumerate}
\item $\lim_{y \rightarrow 0} \varphi_m(y) = 1$, so $\varphi_m$
is a continuous function.
\item $\varphi_m$ is monotonic increasing over the interval
$[0, \infty)$.
\item $0 < \varphi_m(y) < 1$ for every $y<0$.
\item $\varphi_m$ is a non-negative function.
\end{enumerate}
\label{lemma: properties of phi}
\end{lemma}
\begin{IEEEproof}
See Appendix~\ref{appendix: properties of phi}.
\end{IEEEproof}
\begin{remark}
Note that \cite[Lemma~3.1]{Freedman_1975} states that $\varphi_2$
is a monotonic increasing and non-negative function over the real line. In
general, for an even $m \in \naturals$, the properties of
$\varphi_m$ in Lemma~\ref{lemma: properties of phi} are sufficient
for the continuation of the proof.
\end{remark}

From \eqref{eq: chain of equalities for the conditional
expectation} and Lemma~\ref{lemma: properties of phi}, since
$\xi_k \leq d$ a.s. and $m$ is even, then it follows that for an
arbitrary $t \geq 0$
\begin{equation}
\varphi_m(t \xi_k) \leq \varphi_m(td), \quad \forall \, k \in \naturals
\label{eq: basic inequality for the 2nd bound}
\end{equation}
a.s. (to see this, lets separate the two cases where $\xi_k$ is
either non-negative or negative. If $0 \leq \xi_k \leq d$ a.s.
then, for $t \geq 0$, inequality~\eqref{eq: basic inequality for
the 2nd bound} holds (a.s.) due to the monotonicity of $\varphi_m$
over $[0, \infty)$. If $\xi_k < 0$ then the second and third
properties in Lemma~\ref{lemma: properties of phi} yield that, for
$t \geq 0$ and every $k \in \naturals$,
$$\varphi_m(t \xi_k) \leq 1 = \varphi_m(0) \leq \varphi_m(td),$$ so in both
cases inequality~\eqref{eq: basic inequality for the 2nd bound} is
satisfied a.s.). Since $m$ is even then $(\xi_k)^m \geq 0$, and
$$\expectation\bigl[(\xi_k)^m \, \varphi_m(t \xi_k) | \mathcal{F}_{k-1}\bigr]
\leq \varphi_m(t d) \, \expectation\bigl[(\xi_k)^m |
\mathcal{F}_{k-1}\bigr], \; \; \forall \, t \geq 0.$$ Also, since
$\{X_k, \mathcal{F}_k\}_{k=0}^{\infty}$ is a martingale then
$\expectation \bigl[\xi_k | \mathcal{F}_{k-1} \bigr] = 0,$ and
based on the assumptions of this theorem
$$ \expectation[(\xi_k)^l | \mathcal{F}_{k-1}] \leq \mu_l = d^l \gamma_l,
\quad \forall \, l \in \{2, \dots, m\}.$$ By substituting the last
three results on the right-hand side of \eqref{eq: chain of
equalities for the conditional expectation}, it follows that for
every $t \geq 0$ and every $k \in \naturals$

\vspace*{-0.1cm} \small
\begin{equation} \expectation \bigl[ \exp(t \xi_k) |
\mathcal{F}_{k-1} \bigr] \leq 1 + \sum_{l=2}^{m-1} \frac{\gamma_l
\, (td)^l}{l!} + \frac{\gamma_m \, (td)^m \, \varphi_m(td)}{m!}
\label{eq: important inequality for the derivation of Theorem 4}
\end{equation}
\normalsize so from \eqref{eq: smoothing theorem}
\begin{eqnarray}
&& \hspace*{-1.4cm} \expectation\left[\exp\Bigl(t
\sum_{k=1}^n \xi_k\Bigr)\right] \nonumber\\
&& \hspace*{-1.4cm} \leq \left( 1 + \sum_{l=2}^{m-1}
\frac{\gamma_l \, (td)^l}{l!} + \frac{\gamma_m \, (td)^m \,
\varphi_m(td)}{m!} \right)^n, \; \; \forall \, t \geq 0.
\label{eq: upper bound on the exponent in the second approach}
\end{eqnarray}
From \eqref{eq: Chernoff}, if $\alpha \geq 0$ is arbitrary, then
for every $t \geq 0$
\begin{eqnarray*}
&& \hspace*{-0.7cm} \pr(X_n - X_0 \geq \alpha n) \\
&& \hspace*{-0.7cm} \leq \exp(-\alpha n t) \left(1+
\sum_{l=2}^{m-1} \frac{\gamma_l \, (td)^l}{l!} + \frac{\gamma_m \,
(td)^m \, \varphi_m(td)}{m!} \right)^n \; .
\end{eqnarray*}
Let $x \triangleq td$. Then, based on \eqref{eq: notation} and
\eqref{eq: phi function}, for every $\alpha \geq 0$
\begin{eqnarray}
&& \hspace*{-0.7cm} \pr(X_n - X_0 \geq \alpha n) \nonumber \\
&& \hspace*{-0.7cm} \leq \left\{ \inf_{x \geq 0} e^{-\delta x}
\left( 1 + \sum_{l=2}^{m-1} \frac{\gamma_l \, x^l}{l!}
+ \frac{\gamma_m x^m \, \varphi_m(x)}{m!} \right) \right\}^n \nonumber \\[0.1cm]
&& \hspace*{-0.7cm} = \left\{ \inf_{x \geq 0} e^{-\delta x}
\left[1 + \sum_{l=2}^{m-1} \frac{\gamma_l \, x^l}{l!} + \gamma_m
\left(e^x-\sum_{l=0}^{m-1} \frac{x^l}{l!} \right)
\right] \right\}^n \nonumber \\[0.1cm]
&& \hspace*{-0.7cm} = \left\{ \inf_{x \geq 0} e^{-\delta x}
\left[1 + \sum_{l=2}^{m-1} \frac{(\gamma_l-\gamma_m) \, x^l}{l!} +
\gamma_m \left(e^x-1-x \right) \right] \right\}^n. \nonumber \\
\label{eq: one-sided inequality in the 2nd approach}
\end{eqnarray}
The two-sided concentration inequality in \eqref{eq: 2nd
concentration inequality} follows by applying the above
one-sided inequality to the martingale
$\{-X_k, \mathcal{F}_k\}_{k=0}^{\infty}$, and using the union bound.
\end{IEEEproof}

\vspace*{0.1cm}
\begin{remark}
Without any loss of generality, it is assumed that $\alpha \in [0,
d]$ (as otherwise, the considered probability is zero for
$\alpha>d$). Based on the above conditions, it is also assumed
that $\mu_l \leq d^l$ for every $l \in \{2, \ldots, m\}$. Hence,
$\delta \in [0, 1]$, and $\gamma_l \in [0,1]$ for all values of
$l$. Note that, from  \eqref{eq: notation}, $\gamma_2 = \gamma$.
\end{remark}

\vspace*{0.1cm}
\begin{remark}
From the proof of Theorem~\ref{theorem: second inequality}, it
follows that the one-sided inequality~\eqref{eq: one-sided
inequality in the 2nd approach} is satisfied if the martingale
$\{X_k, \mathcal{F}_k\}_{k=0}^n$ fulfills the following conditions
a.s.
\begin{eqnarray*}
&& X_k - X_{k-1} \leq d, \\[0.1cm]
&& \expectation \bigl[(X_k - X_{k-1})^l \, | \,
\mathcal{F}_{k-1} \bigr] \leq \mu_l, \quad l = 2, \ldots, m
\end{eqnarray*}
for some $d>0$ and non-negative numbers $\{\mu_l\}_{l=2}^m$. Note
that these conditions are weaker than those that are stated in
Theorem~\ref{theorem: second inequality}. Under these weaker
conditions, $\gamma_l \triangleq \frac{\mu_l}{d^l}$ may be larger
than~1. This remark will be helpful later in this paper.
\label{remark: on the one-sided inequality version of Theorem 4}
\end{remark}

\vspace*{0.1cm}
\subsubsection{Specialization of Theorem~\ref{theorem: second inequality} for $m=2$}
Theorem~\ref{theorem: second inequality}
with $m=2$ (i.e., when the same conditions as
of Theorem~\ref{theorem: first refined concentration inequality}
hold) is expressible in closed form, as follows:

\begin{corollary}
Let $\{X_k, \mathcal{F}_k\}_{k=0}^{\infty}$ be a discrete-parameter
real-valued martingale that satisfies a.s.
the conditions in Theorem~\ref{theorem: first refined
concentration inequality}. Then, for every $\alpha \geq
0$,
\begin{equation*}
\pr(|X_n-X_0| \geq \alpha n) \leq 2 \exp\bigl(-n C(\gamma, \delta)
\bigr)
\end{equation*}
where $\gamma$ and $\delta$ are introduced in \eqref{eq: notation},
and the exponent in this upper bound gets the following form:
\begin{itemize}
\item If $\delta>1$ then $C(\gamma, \delta) = \infty$.
\item If $\delta=1$ then
\begin{equation*}
C(\gamma, \delta) = \frac{1}{\gamma} -
\ln \Bigl( \gamma \bigl( e^{\frac{1}{\gamma}} - 1 \bigr) \Bigr).
\end{equation*}
\item Otherwise, if $\delta \in (0,1)$, then
\begin{equation*}
C(\gamma, \delta) = \delta x - \ln \bigl(1+\gamma (e^x-1-x)\bigr)
\end{equation*}
where $x \in \bigl(0, \frac{1}{\gamma}\bigr)$ is given by
\begin{equation}
x = \frac{1}{\gamma} + \frac{1}{\delta}-1 - W_0
\left(\frac{(1-\delta) e^{\frac{1}{\gamma} +
\frac{1}{\delta}-1}}{\delta} \right) \label{eq: solution of the
optimized x}
\end{equation}
and $W_0$ denotes the principal branch of the Lambert W
function \cite{Lambert_function}.
\end{itemize}
\label{corollary: specialization of the second inequality}
\end{corollary}
\begin{proof}
See Appendix~\ref{appendix: proof of the corollary that
specializes the second inequality for m equal to 2}.
\end{proof}

\begin{proposition}
If $\gamma < \frac{1}{2}$ then
Corollary~\ref{corollary: specialization of the second inequality}
gives a stronger result than Corollary~\ref{corollary: a tightened
version of Azuma's inequality for martingales with bounded jumps}
(and, hence, it is also better than Azuma's inequality).
\label{proposition: a sufficient condition where the
specialization of the second inequality is better than Azuma's
inequality}
\end{proposition}
\begin{IEEEproof}
See Appendix~\ref{appendix: a sufficient condition where the
specialization of the second inequality is better than Azuma's
inequality}.
\end{IEEEproof}

It is of interest to compare the tightness of
Theorem~\ref{theorem: first refined concentration inequality} and
Corollary~\ref{corollary: specialization of the second
inequality}. This leads to the following conclusion:
\begin{proposition}
The concentration inequality in Corollary~\ref{corollary:
specialization of the second inequality} is looser than
Theorem~\ref{theorem: first refined concentration inequality}.
\label{proposition: Theorem 2 gives a stronger result than
Corollary 4}
\end{proposition}
\begin{IEEEproof}
See Appendix~\ref{appendix: Proof of Proposition 2}.
\end{IEEEproof}

The statements in Propositions~\ref{proposition: a sufficient
condition where the specialization of the second inequality is
better than Azuma's inequality} and~\ref{proposition: Theorem 2
gives a stronger result than Corollary 4} are illustrated in
Figure~\ref{Figure:
compare_exponents_theorems_1_2_3_and_corollary_4}. Sub-plots~(a)
and (b) in Figure~\ref{Figure:
compare_exponents_theorems_1_2_3_and_corollary_4} refer to $\gamma
\leq \frac{1}{2}$ where the statement in
Proposition~\ref{proposition: a sufficient condition where the
specialization of the second inequality is better than Azuma's
inequality} holds. On the other hand, sub-plots~(c) and (d) in
Figure~\ref{Figure:
compare_exponents_theorems_1_2_3_and_corollary_4} refer to higher
values of $\gamma$, and therefore the statement in
Proposition~\ref{proposition: a sufficient condition where the
specialization of the second inequality is better than Azuma's
inequality} does not apply to these values of $\gamma$.

\begin{figure}[here!]
\begin{center}
\epsfig{file=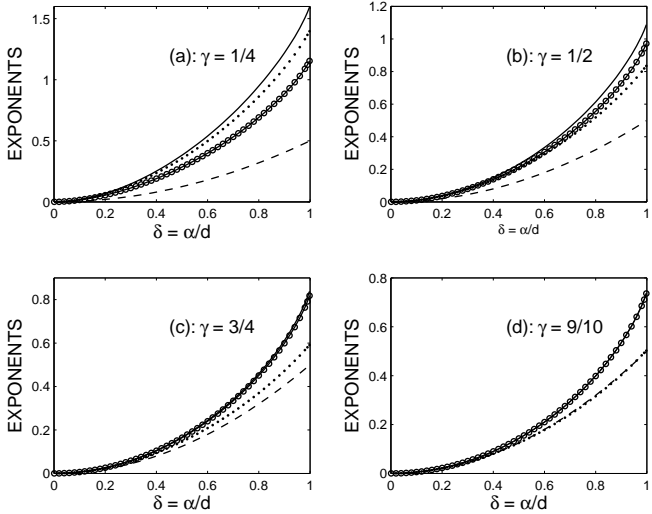,scale=0.5}
\end{center}
\caption{\label{Figure:
compare_exponents_theorems_1_2_3_and_corollary_4} Plots of the
exponents $c \triangleq c(\gamma, \delta)$ for bounds of the form
$\pr\{|X_n - X_0| \geq n \alpha\} \leq e^{-n c(\gamma, \delta)}$
for an arbitrary $\alpha \geq 0$. The sequence $\{X_k,
\mathcal{F}_k\}_{k=0}^{\infty}$ is a discrete-parameter martingale
that satisfies the conditions in Theorem~\ref{theorem: first
refined concentration inequality} for some positive constants $d$
and $\sigma$ (without loss of generality, $\sigma \leq d$), and
where $\gamma \in (0,1]$ and $\delta \in [0, 1]$ are introduced in
\eqref{eq: notation}. The plots show lower bounds on the exponents
according to Azuma's inequality in \eqref{eq: Azuma's inequality},
the bounds in Theorems~\ref{theorem: first refined concentration
inequality} and~\ref{theorem: inequality based on a parabola
intersecting the exponential function at the endpoints of the
interval} Corollary~\ref{corollary: specialization of the second
inequality}. The plots are depicted for a fixed value of $\gamma
\triangleq \frac{\sigma^2}{d^2}$; the horizontal axes refer to the
parameter $\delta \triangleq \frac{\alpha}{d}$, and the vertical
axes refer to the lower bounds on the exponents $c(\gamma,
\delta)$. The dashed lines refer to Azuma's inequality, the solid
lines refer to Theorem~\ref{theorem: first refined concentration
inequality}, the circled lines refer to Theorem~\ref{theorem:
inequality based on a parabola intersecting the exponential
function at the endpoints of the interval}, and the pointed lines
refer to Corollary~\ref{corollary: specialization of the second
inequality}. The subplots (a)-(d) correspond to values of $\gamma
= \frac{1}{4}, \frac{1}{2}, \frac{3}{4}$ and $\frac{9}{10}$,
respectively.}
\end{figure}

\vspace*{0.1cm}
\subsubsection{Exploring the Dependence of the Bound in
Theorem~\ref{theorem: second inequality} in Terms of $m$} In the
previous sub-section, a closed-form expression of
Theorem~\ref{theorem: second inequality} was obtained for the
special case where $m=2$ (see Corollary~\ref{corollary:
specialization of the second inequality}), but also
Proposition~\ref{proposition: Theorem 2 gives a stronger result
than Corollary 4} demonstrated that this special case is looser
than Theorem~\ref{theorem: first refined concentration inequality}
(which is also given as a closed-form expression). Hence, it is
natural to enquire how does the bound in Theorem~\ref{theorem:
second inequality} vary in terms of $m$ (where $m \geq 2$ is
even), and if there is any chance to improve Theorem~\ref{theorem:
first refined concentration inequality} for larger values of $m$.
Also, in light of the closed-form expression that was given in
Corollary~\ref{corollary: specialization of the second inequality}
for the special case where $m=2$, it would be also pleasing to get
an inequality that is expressed in closed form for a general even
number $m \geq 2$. The continuation of the study in this
sub-section is outlined as follows:
\begin{itemize}
\item A loosened version of Theorem~\ref{theorem: second inequality}
is introduced, and it is shown to provide an inequality whose
tightness consistently improves by increasing the value of $m$.
For $m=2$, this loosened version coincides with
Theorem~\ref{theorem: second inequality}. Hence, it follows (by
introducing this loosened version) that $m=2$ provides the weakest
bound in Theorem~\ref{theorem: second inequality}.
\item Inspired by the closed-form expression of the bound in
Corollary~\ref{corollary: specialization of the second
inequality}, we derive a closed-form inequality (i.e., a bound
that is not subject to numerical optimization) by either loosening
Theorem~\ref{theorem: second inequality} or further loosening
its looser version from the previous item.
As will be exemplified numerically in Section~\ref{section: Applications},
the closed-form expression of the new bound causes to a marginal
loosening of Theorem~\ref{theorem: second inequality}. Also, for
$m=2$, it is exactly Theorem~\ref{theorem: second inequality}.
\item A necessary and sufficient condition is derived for the case
where, for an even $m \geq 4$, Theorem~\ref{theorem: second
inequality} provides a bound that is exponentially advantageous
over Theorem~\ref{theorem: first refined concentration
inequality}. Note however that, when $m \geq 4$ in
Theorem~\ref{theorem: second inequality}, one needs to calculate
conditional moments of the martingale differences that are of
higher orders than~2; hence, an improvement in
Theorem~\ref{theorem: second inequality} is obtained at the
expense of the need to calculate higher-order conditional moments.
Saying this, note that the derivation of Theorem~\ref{theorem:
second inequality} deviates from the proof of
Theorem~\ref{theorem: first refined concentration inequality} at
an early stage, and it cannot be considered as a generalization of
Theorem~\ref{theorem: first refined concentration inequality} when
higher-order moments are available (as is also evidenced in
Proposition~\ref{proposition: Theorem 2 gives a stronger result
than Corollary 4} which demonstrates that, for $m=2$,
Theorem~\ref{theorem: second inequality} is weaker than
Theorem~\ref{theorem: first refined concentration inequality}).
\item Finally, this sufficient condition is particularized in
the asymptotic case where $m \rightarrow \infty$. It is of
interest since the tightness of the loosened version of
Theorem~\ref{theorem: second inequality} from the first item is
improved by increasing the value of $m$.
\end{itemize}
The analysis that is related to the above outline is presented in
the following. Then, following this analysis, numerical results
that are related to the comparison of Theorems~\ref{theorem: first
refined concentration inequality} and~\ref{theorem: second
inequality} are relegated to Section~\ref{section: Applications}
(while considered in a certain communication-theoretic context).

\begin{corollary}
Let $\{X_k, \mathcal{F}_k\}_{k=0}^n$ be a discrete-parameter
real-valued martingale, and let
$m \in \naturals$ be an even number. Assume that
$| X_k - X_{k-1} | \leq d$ holds a.s.
for every $k \in \naturals$, and
that there exists a (non-negative) sequence
$\{\mu_l\}_{l=2}^m$ so that for every $k \in \naturals$
\begin{equation}
\mu_l = \expectation[ |X_k - X_{k-1}|^l \, | \, \mathcal{F}_{k-1}
], \quad \forall \, l=2, \ldots, m. \label{eq: mu for the first
loosened version of Theorem 4}
\end{equation}
Then, inequality~\eqref{eq: 2nd concentration inequality} holds
with the notation in \eqref{eq: gamma and delta for Theorem 4}.
\label{corollary: first loosened version of Theorem 4}
\end{corollary}
\begin{IEEEproof}
This corollary is a consequence of Theorem~\ref{theorem: second inequality}
since
$$ | \expectation[ (X_k - X_{k-1})^l \, | \, \mathcal{F}_{k-1} ] |
\leq \expectation[ |X_k - X_{k-1}|^l \, | \, \mathcal{F}_{k-1} ].$$
\end{IEEEproof}

\begin{proposition}
Theorem~\ref{theorem: second inequality} and
Corollary~\ref{corollary: first loosened version of Theorem 4}
coincide for $m=2$ (hence, Corollary~\ref{corollary: first
loosened version of Theorem 4} provides in this case the result
stated in Corollary~\ref{corollary: specialization of the second
inequality}). Furthermore, the bound in Corollary~\ref{corollary:
first loosened version of Theorem 4} improves as the even value of
$m \in \naturals$ is increased.
\label{proposition: first proposition on the loosened version of
Theorem 4}
\end{proposition}
\begin{IEEEproof}
See Appendix~\ref{appendix: proof of the first proposition on the
loosened version of Theorem 4}.
\end{IEEEproof}

\vspace*{0.1cm} Inspired by the closed-form inequality that
follows from Theorem~\ref{theorem: second inequality} for $m=2$
(see Corollary~\ref{corollary: specialization of the second
inequality}), a closed-form inequality is suggested in the
following by either loosening Theorem~\ref{theorem: second
inequality} or Corollary~\ref{corollary: first loosened version of
Theorem 4}. It generalizes the result in Corollary~\ref{corollary:
specialization of the second inequality}, and it
coincides with Theorem~\ref{theorem: second inequality} and
Corollary~\ref{corollary: first loosened version of Theorem 4} for
$m=2$.

\begin{corollary}
Under the conditions of Corollary~\ref{corollary: first loosened
version of Theorem 4} then, for every $\alpha \geq 0$,
\begin{eqnarray}
&& \hspace*{-0.7cm} \pr(X_n - X_0 \geq n \alpha) \nonumber \\
&& \hspace*{-0.7cm} \leq \, \left\{e^{-\delta
x} \left[1 + \sum_{l=2}^{m-1} \frac{(\gamma_l - \gamma_m) x^l}{l!}
+ \gamma_m (e^x-1-x) \right] \right\}^n \nonumber \\[0.1cm]
\label{eq: 2nd looser version of Theorem 4}
\end{eqnarray}
where $\{\gamma_l\}_{l=2}^m$ and $\delta$ are introduced in
\eqref{eq: gamma and delta for Theorem 4},
\begin{equation}
x = \frac{a+b}{c} - W_0 \left(\frac{b}{c} \cdot e^{\frac{a+b}{c}} \right)
\label{eq: sub-optimal x}
\end{equation}
with $W_0$ that denotes the principal branch of the Lambert W
function \cite{Lambert_function}, and
\begin{eqnarray}
&& a \triangleq \frac{1}{\gamma_2}, \quad b
\triangleq \frac{\gamma_m}{\gamma_2} \left(\frac{1}{\delta}-1\right),
\quad c \triangleq \frac{1}{\delta} - b.
\label{eq: a, b, c for the sub-optimal x}
\end{eqnarray}
\label{corollary: second loosened version of Theorem 4}
\end{corollary}
\begin{IEEEproof}
See Appendix~\ref{appendix: proof of 2nd looser version of Theorem 4}.
\end{IEEEproof}

\begin{remark}
It is exemplified numerically in Section~\ref{section:
Applications} that the replacement of the infimum over $x \geq 0$
on the right-hand side of \eqref{eq: 2nd concentration inequality}
with the sub-optimal choice of the value of $x$ that is given in
\eqref{eq: sub-optimal x} and \eqref{eq: a, b, c for the
sub-optimal x} implies a marginal loosening in the exponent of the
bound. Note also that, for $m=2$, this value of $x$ is optimal
since it coincides with the exact value in \eqref{eq: solution of
the optimized x}.
\end{remark}

\begin{corollary}
Under the assumptions of Theorem~\ref{theorem: first refined
concentration inequality} then, for every $\alpha \geq 0$,
\begin{equation}
\pr(X_n - X_0 \geq n \alpha) \leq e^{-n E}
\label{eq: one-sided inequality}
\end{equation}
where
\begin{equation}
E = E_2(\gamma_2, \delta) \triangleq D\left(\frac{\delta
+\gamma_2}{1+\gamma_2} \Big| \Big| \frac{\gamma_2}{1+\gamma_2} \right).
\label{eq: exponent for Theorem 2}
\end{equation}
Also, under the assumptions of Theorem~\ref{theorem: second
inequality} or Corollary~\ref{corollary: first loosened version of
Theorem 4} then \eqref{eq: one-sided inequality} holds for every
$\alpha \geq 0$ with

\vspace*{-0.2cm}
\small
\begin{eqnarray}
&& \hspace*{-0.6cm} E = E_4(\{\gamma_l\}_{l=2}^m, \delta) \nonumber \\
&& \hspace*{-0.2cm} \triangleq \sup_{x \geq 0} \left\{ \delta x -
\ln \left( 1 + \sum_{l=2}^{m-1} \frac{(\gamma_l - \gamma_m) x^l}{l!}
+ \gamma_m (e^x-1-x) \right) \right\} \nonumber \\
\label{eq: exponent for Theorem 4}
\end{eqnarray}
\normalsize where $m \geq 2$ is an arbitrary even number. Hence,
Theorem~\ref{theorem: second inequality} or
Corollary~\ref{corollary: first loosened version of Theorem 4} are
better exponentially than Theorem~\ref{theorem: first refined
concentration inequality} if and only if $E_4 > E_2$.
\end{corollary}
\begin{IEEEproof}
The proof follows directly from \eqref{eq: one-sided concentration
inequality of the first refined bound} and \eqref{eq: one-sided
inequality in the 2nd approach}.
\end{IEEEproof}

\vspace*{0.1cm}
\begin{remark}
In order to avoid the operation of taking the supermum over $x \in
[0, \infty)$, it is sufficient to first check if $\widetilde{E}_4
> E_2$ where
$$\widetilde{E}_4 \triangleq \delta x - \ln \left( 1 +
\sum_{l=2}^{m-1} \frac{(\gamma_l - \gamma_m) x^l}{l!} + \gamma_m
(e^x-1-x) \right)$$ with the value of $x$ in \eqref{eq:
sub-optimal x} and \eqref{eq: a, b, c for the sub-optimal x}. This
sufficient condition is exemplified later in Section~\ref{section:
Applications}. \label{reemark: sufficient condition when Theorem 4
is exponentially better than Theorem 2}
\end{remark}

\subsection{Concentration Inequalities for Small Deviations}
\label{Concentration Inequalities Referring to Small Deviations}
In the following, we consider the probability of the events
$\{|X_n-X_0| \geq \alpha \sqrt{n}\}$ for an arbitrary $\alpha \geq
0$. These events correspond to small deviations. This is in
contrast to events of the form $\{|X_n - X_0| \geq \alpha n\}$,
whose probabilities were analyzed earlier in this section, and
which correspond to large deviations.

\begin{proposition}
Let $\{X_k, \mathcal{F}_k\}_{k=0}^{\infty}$ be a discrete-parameter
real-valued martingale. Then,
Theorem~\ref{theorem: first refined concentration inequality} and
\ref{theorem: inequality based on a parabola intersecting the
exponential function at the endpoints of the interval}, and also
Corollaries~\ref{corollary: 3rd corollary}
and~\ref{corollary: specialization of the second inequality}
imply that, for every $\alpha \geq 0$,
\begin{equation}
\pr(|X_n-X_0| \geq \alpha \sqrt{n}) \leq 2
\exp\Bigl(-\frac{\delta^2}{2\gamma}\Bigr)
\Bigl(1+ O\bigl(n^{-\frac{1}{2}}\bigr)\Bigr).
\label{eq: concentration1}
\end{equation}
Also, under the conditions of Theorem~\ref{theorem: second
inequality}, inequality~\eqref{eq: concentration1} holds
for every even $m \geq 2$ (so the conditional
moments of higher order than~2 do not improve, via Theorem~\ref{theorem:
second inequality}, the scaling of the upper bound in \eqref{eq:
concentration1}).
\label{proposition: a similar scaling of the
concentration inequalities}
\end{proposition}
\begin{IEEEproof}
See Appendix~\ref{appendix: proof of the statement
about the similar scaling of the concentration inequalities}.
\end{IEEEproof}

\begin{remark}
From Proposition~\ref{proposition: a similar scaling of the
concentration inequalities}, all the upper bounds on
$\pr(|X_n-X_0| \geq \alpha \sqrt{n})$ (for an arbitrary $\alpha
\geq 0$) improve the exponent of Azuma's inequality by a factor of
$\frac{1}{\gamma}$.
\end{remark}

\subsection{Inequalities for Sub and Super Martingales}
\label{subsection: concentration inequalities for sub and super
martingales} Upper bounds on the probability $\pr( X_n - X_0 \geq
r)$ for $r \geq 0$, earlier derived in this section for
martingales, can be easily adapted to super-martingales (similarly
to, e.g., \cite[Chapter~2]{Chung_LU2006} or
\cite[Section~2.7]{survey2006}). Alternatively, replacing $\{X_k,
\mathcal{F}_k\}_{k=0}^n$ with $\{-X_k, \mathcal{F}_k\}_{k=0}^n$
provides upper bounds on the probability $\pr( X_n - X_0 \leq -r)$
for sub-martingales.

\section{Relations of the Refined Inequalities to Some Classical Results in Probability Theory}
\label{section: relation to previously reported bound}

\subsection{Relation of Theorem~\ref{theorem: first refined concentration inequality}
to the Method of Types} Consider a sequence of i.i.d. RVs $X_1,
X_2, \ldots$ that are Bernoulli$(p)$ distributed (i.e., for every
$i \in \naturals$, $\pr(X_i=1)=p$ and $\pr(X_i=0)=1-p$). According
to the method of types (see, e.g., \cite[Section~11.1]{Cover and
Thomas}), it follows that for every $n \in \naturals$ and $r \geq
p$
\begin{equation}
\frac{e^{-n D(r||p)}}{n+1} \leq \pr \biggl(\frac{1}{n}
\sum_{i=1}^n X_i \geq r \biggr) \leq e^{-n D(r||p)}
\label{eq: method of types for i.i.d. Bernoulli RVs}
\end{equation}
where the divergence $D(r||p)$ is given in \eqref{eq: divergence},
and therefore
\begin{equation}
\lim_{n \rightarrow \infty} \frac{1}{n} \; \ln \; \pr
\biggl(\frac{1}{n} \sum_{i=1}^n X_i \geq r \biggr) =
-D(r||p), \quad \forall \; r \geq p \label{eq: exact exponent via
the method of types}
\end{equation}
gives the exact exponent. This equality can be
obtained as a particular case of Cram\'{e}r's theorem in $\reals$
where the rate function of $X \sim \text{Bernoulli}(p)$ is given
by
$$ I(x) = \left\{ \begin{array}{ll}
                             D(x||p)  & \mbox{if $x \in [0,1]$}\\
                             +\infty  & \mbox{otherwise}
                             \end{array}
                             \right. $$
(for Cram\'{e}r's theorem in $\reals$ see, e.g.,
\cite[Section~2.2.1 and Exercise~2.2.23]{Dembo_Zeitouni}
and \cite[Section~1.3]{Hollander_book_2000}).

In the following, it is shown that Theorem~\ref{theorem: first
refined concentration inequality} gives in the considered setting
the upper bound on the right-hand side of \eqref{eq: method of
types for i.i.d. Bernoulli RVs}, and it therefore provides the
exact exponent in \eqref{eq: exact exponent via the method of
types}. To this end, consider the filtration where $\mathcal{F}_0
= \{\emptyset, \Omega\}$ and
\begin{equation*}
\mathcal{F}_n = \sigma(X_1, \ldots, X_n), \quad \forall \; n \in \naturals
\end{equation*}
and let the sequence of RVs $\{S_n\}_{n=0}^{\infty}$ be defined as
$S_0=0$, and
\begin{equation}
S_n = \sum_{i=1}^n X_i - np, \quad \forall \, n \in \naturals.
\label{eq: un-biased partial sums of i.i.d. Bernoulli RVs}
\end{equation}
It is easy to verify that $\{S_n, \mathcal{F}_n\}_{n=0}^{\infty}$
is a martingale, and for every $n \in \naturals$
\begin{eqnarray*}
&& |S_n - S_{n-1}| = |X_n - p| \leq \max\{p, 1-p\}, \\
&& \text{Var}(S_n | \mathcal{F}_{n-1}) = \expectation[(X_n-p)^2] = p(1-p).
\end{eqnarray*}
Consider the case where $p \leq \frac{1}{2}$. Then, from the
notation of Theorem~\ref{theorem: first refined concentration
inequality}
\begin{equation*}
\sigma^2 = p(1-p), \quad d = 1-p.
\end{equation*}
Therefore, it follows from
Theorem~\ref{theorem: first refined concentration inequality}
that for every $\alpha \geq 0$
\begin{equation}
\pr(S_n \geq n \alpha) \leq \exp\left(-n
\, D\biggl(\frac{\delta+\gamma}{1+\gamma} \Big|\Big|
\frac{\gamma}{1+\gamma}\biggr) \right)
\label{eq: one sided inequality for Bernoulli i.i.d.}
\end{equation}
where
\begin{equation}
\gamma = \frac{p}{1-p}, \quad \delta = \frac{\alpha}{1-p} \; .
\label{eq: gamma and delta for Bernoulli i.i.d.}
\end{equation}
Substituting \eqref{eq: gamma and delta for Bernoulli i.i.d.} into
\eqref{eq: one sided inequality for Bernoulli i.i.d.} gives that
for every $\alpha \geq 0$
\begin{equation}
\pr(S_n \geq n \alpha) \leq \exp \bigl(-n D(\alpha+p \, || \, p) \bigr).
\label{eq: upper bound for Bernoulli i.i.d.}
\end{equation}
Let $r \triangleq \alpha+p$ (where $r \geq p \Longleftrightarrow
\alpha \geq 0$). The substitution of \eqref{eq: un-biased partial
sums of i.i.d. Bernoulli RVs} into the left-hand side of
\eqref{eq: upper bound for Bernoulli i.i.d.} implies  that
\eqref{eq: upper bound for Bernoulli i.i.d.} coincides with the
upper bound on the right-hand side of \eqref{eq: method of types
for i.i.d. Bernoulli RVs}. Hence, Theorem~\ref{theorem: first
refined concentration inequality} gives indeed the exact exponent
in \eqref{eq: exact exponent via the method of types} for the case
of i.i.d. RVs that are Bernoulli$(p)$ distributed with $p \in [0,
\frac{1}{2}]$.

The method of types gives that a similar one-sided version of
inequality~\eqref{eq: method of types for i.i.d. Bernoulli RVs}
holds for every $r \leq p$, and therefore
\begin{equation}
\hspace*{-0.3cm} \lim_{n \rightarrow \infty} \frac{1}{n} \; \ln \;
\pr \biggl(\frac{1}{n} \sum_{i=1}^n X_i \leq r \biggr) = -D(r||p),
\quad \forall \; r \leq p.
\label{eq: exact exponent via the method of types - version 2}
\end{equation}

For the case where $p \geq \frac{1}{2}$, let $Y_i \triangleq
1-X_i$ for every $i \in \naturals$.
From Theorem~\ref{theorem: first refined concentration inequality},
for every $\alpha \geq 0$,
\begin{eqnarray}
&& \pr \left( \sum_{i=1}^n X_i \leq n(p-\alpha) \right) \nonumber \\
&& = \pr \left( \sum_{i=1}^n Y_i \geq n(\alpha+1-p) \right) \nonumber \\
&& \stackrel{\text{(a)}}{\leq} \exp \bigl(-n D(\alpha+1-p \, ||
\, 1-p) \bigr) \nonumber \\
&& \stackrel{\text{(b)}}{=} \exp \bigl(-n D(p-\alpha \, || \, p)
\bigr) \label{eq: inequality for i.i.d. Bernoulli RVs with p
greater than or equal to one-half}
\end{eqnarray}
where inequality~(a) follows from inequality~\eqref{eq: upper
bound for Bernoulli i.i.d.} since the i.i.d. RVs $\{Y_i\}_{i \in
\naturals}$ are $\text{Bernoulli}(1-p)$ distributed $(1-p \leq
\frac{1}{2})$, and equality~(b) is satisfied since $D(1-x \, || \,
1-y) = D(x||y)$ (see \eqref{eq: divergence}). The substitution $r
\triangleq p-\alpha$ (so $r \leq p \Longleftrightarrow \alpha \geq
0$) in \eqref{eq: inequality for i.i.d. Bernoulli RVs with p
greater than or equal to one-half} gives the same exponent as on
the right-hand side of \eqref{eq: exact exponent via the method of
types - version 2}, so Theorem~\ref{theorem: first refined
concentration inequality} also gives the exact exponent in
\eqref{eq: exact exponent via the method of types - version 2} for
i.i.d. RVs that are Bernoulli$(p)$ distributed with $p \in
[\frac{1}{2}, 1]$.

\subsection{Relations of \cite[Corollary~2.4.7]{Dembo_Zeitouni} with
Theorem~\ref{theorem: first refined concentration inequality} and
Proposition~\ref{proposition: a similar scaling of the concentration inequalities}}
\label{subsection: relations with Theorem 2}

According to \cite[Corollary~2.4.7]{Dembo_Zeitouni}, suppose
$v>0$ and a sequence of real-valued RVs $\{Y_n\}_{n=1}^{\infty}$
satisfies a.s.
\begin{itemize}
\item $Y_n \leq 1$ for every $n \in \naturals$.
\item $\expectation[Y_n \, | \, S_{n-1}]=0$ and $\expectation[Y_n^2 \,
| \, S_{n-1}] \leq v$ for $$S_n \triangleq \sum_{j=1}^n Y_j, \,
S_0=0.$$
\end{itemize}
Then, for every $\lambda \geq 0$,
\begin{equation}
\expectation[\exp(\lambda S_n)] \leq \left(\frac{v \exp(\lambda) +
\exp(-\lambda v)}{1+\lambda}\right)^n \, . \label{eq: First bound
in Corollary 2.4.7 from the book by Dembo and Zeitouni}
\end{equation}
Moreover, for every $x \geq 0$
\begin{equation}
\pr \left( \frac{S_n}{n} \geq x \right) \leq \exp \left(-n
D\Bigl(\frac{x+v}{1+v} \Big|\Big| \frac{v}{1+v}\Bigr) \right)
\label{eq: Second bound in Corollary 2.4.7 from the book by Dembo
and Zeitouni}
\end{equation}
and, for every $y \geq 0$,
\begin{equation}
\pr \left( \frac{S_n}{\sqrt{n}} \geq y \right) \leq \exp
\left(-\frac{2y^2}{(1+v)^2} \right). \label{eq: Third bound in
Corollary 2.4.7 from the book by Dembo and Zeitouni}
\end{equation}

In the following, we show that
\cite[Corollary~2.4.7]{Dembo_Zeitouni} is closely related to
Theorem~\ref{theorem: first refined concentration inequality} in
this paper. To this end, let $\{X_k,
\mathcal{F}_k\}_{k=0}^{\infty}$ be a discrete-parameter
real-valued martingale where $X_k - X_{k-1} \leq d$ a.s. for
every $k \in \naturals$. Let us define the martingale-difference
sequence $\{Y_k, \mathcal{F}_k\}_{k=0}^{\infty}$ where $$Y_k
\triangleq \frac{X_k - X_{k-1}}{d}, \quad \forall \, k \in
\naturals$$ and $Y_0 \triangleq 0$. Based on the assumptions in
Theorem~\ref{theorem: first refined concentration inequality}, it
follows from \eqref{eq: notation} that $Y_k \leq 1$ a.s. for every
$k \in \naturals$, and
$$ \expectation[Y_k \, | \, \mathcal{F}_{k-1}] = 0, \quad
\expectation[Y_k^2 \, | \, \mathcal{F}_{k-1}] \leq
\frac{\sigma^2}{d^2} = \gamma.$$ Hence, by definition, $\xi_k
\triangleq X_k - X_{k-1}$ satisfies the equality $\xi_k = d Y_k$
for every $k \in \naturals$. From \eqref{eq: important inequality
used for the derivation of Theorem 2}, with $t =
\frac{\lambda}{d}$ and $\gamma= v$, it follows that for every
$\lambda \geq 0$
\begin{eqnarray*}
&& \expectation \left[\exp \bigl(\lambda S_n \bigr) \right] \\
&& = \expectation \left[\exp \biggl( \frac{\lambda}{d} \,
\sum_{k=1}^n \xi_k \biggr) \right] \\
&& \leq \left( \frac{v \exp(\lambda) + \exp(-v \lambda)}{1+v} \right)^n
\end{eqnarray*}
which then coincides with \eqref{eq: First bound in Corollary
2.4.7 from the book by Dembo and Zeitouni}. It is noted that in
Theorem~\ref{theorem: first refined concentration inequality} it
was required that $|X_k - X_{k-1}| \leq d$ whereas, due to
\cite[Corollary~2.4.7]{Dembo_Zeitouni}, it is enough that $X_k -
X_{k-1} \leq d$. In fact, this relaxation is possible due to the
use of Bennett's inequality which only requires
that $\xi_k \leq d$. The only reason it was stated in
Theorem~\ref{theorem: first refined concentration inequality} with
the absolute value was simply because we wanted to get without any
loss of generality that $\gamma \leq 1$ (due the second
requirement on the conditional variance). Finally, since
$$\frac{S_n}{n} = \frac{X_n - X_0}{nd} \, ,$$
then it follows from Theorem~\ref{theorem: first refined
concentration inequality} that for every $x \geq 0$
\begin{eqnarray}
&& \pr\left(\frac{S_n}{n} \geq x \right) \nonumber \\
&& = \pr(X_n - X_0 \geq n x d) \nonumber \\
&& \leq \exp \left( - n D\Bigl(\frac{x+v}{1+v}
\Big|\Big| \frac{v}{1+v}\Bigr) \right)
\end{eqnarray}
where, from \eqref{eq: notation}, the correspondence between
Theorem~\ref{theorem: first refined concentration inequality} and
\cite[Corollary~2.4.7]{Dembo_Zeitouni} is that $\gamma=v$ and
$\delta=x$. This shows the relation between
Theorem~\ref{theorem: first refined concentration inequality} and
Eqs.~\eqref{eq: First bound in Corollary 2.4.7 from the book by Dembo
and Zeitouni} and \eqref{eq: Second bound in Corollary 2.4.7 from
the book by Dembo and Zeitouni} (respectively, Eqs.~(2.4.8)
and~(2.4.9) in \cite{Dembo_Zeitouni}).

We show in the following that Proposition~\ref{proposition: a
similar scaling of the concentration inequalities} suggests an
improvement over the bound in \eqref{eq: Third bound in Corollary
2.4.7 from the book by Dembo and Zeitouni} (that is introduced in
\cite[Eq.~(2.4.10)]{Dembo_Zeitouni}). To see this, note that from
Proposition~\ref{proposition: a similar scaling of the
concentration inequalities} (see \eqref{eq: concentration1}), then
for every $y \geq 0$,
\begin{eqnarray}
&& \pr \left(\frac{S_n}{\sqrt{n}} \geq y \right) \nonumber \\
&& = \pr \left(X_n - X_0 \geq y d \sqrt{n} \right) \nonumber \\
&& \leq \exp\left(-\frac{y^2}{2v}\right) \;
\left(1+O\Bigl(\frac{1}{\sqrt{n}}\Bigr)\right)
\label{eq: consequence of the proposition related to small deviations}
\end{eqnarray}
where the term on the right-hand side of \eqref{eq: consequence
of the proposition related to small deviations} that scales like
$O\Bigl(\frac{1}{\sqrt{n}}\Bigr)$
is expressed explicitly in terms of $n$ for each concentration
inequality that was derived in Section~\ref{section: Refined
Versions of Azuma's Inequality} (see the proof of
Proposition~\ref{proposition: a similar scaling of the
concentration inequalities} in Appendix~\ref{appendix: proof of
the statement about the similar scaling of the concentration
inequalities}). The improvement of the exponent \eqref{eq:
consequence of the proposition related to small deviations} over
the exponent in \cite[Eq.~(2.4.10)]{Dembo_Zeitouni}) (see
\eqref{eq: Third bound in Corollary 2.4.7 from the book by Dembo
and Zeitouni}) holds since
$$\frac{y^2}{2v} - \frac{2y^2}{(1+v)^2}
= \frac{y^2}{2v} \; \left(\frac{v-1}{v+1}\right)^2 \geq 0$$ with
equality if and only if $v=1$. Note that this improvement is
especially pronounced if $v \ll 1$; in the limit where $v$ tends
to zero then the improved exponent $(\frac{y^2}{2v})$ tends to
$+\infty$, whereas the other exponent $(i.e., \frac{2 y^2}{(1+v)^2})$
stays bounded.

\subsection{Relations of \cite[Execrise~2.4.21(b)]{Dembo_Zeitouni},
\cite[Theorem~1.6]{Freedman_1975} and
\cite[Theorem~1]{Stieger_1969} with Corollary~\ref{corollary: 3rd
corollary} and Proposition~\ref{proposition: a similar scaling of
the concentration inequalities}}
\label{subsection: Relation with an old theorem}
The following theorem was introduced in
\cite[Theorem~1.6]{Freedman_1975} and \cite[Theorem~1]{Stieger_1969}
(and in \cite[Execrise~2.4.21(b)]{Dembo_Zeitouni} with the weaker
condition below).
\begin{theorem}
Let $\{S_n, \mathcal{F}_n\}_{n=0}^{\infty}$ be a
discrete-parameter real-valued martingale such that $S_0=0$, and
$Y_k \triangleq S_k - S_{k-1} \leq 1$ a.s. for every $k \in
\naturals$. Let us define the random variables
\begin{equation}
Q_n \triangleq \sum_{j=1}^n E(Y_j^2 | \mathcal{F}_{j-1})
\label{eq: Q}
\end{equation}
where $Q_0 \triangleq 0$. Then for every $z,r>0$
\begin{equation}
\pr(S_n \geq z, Q_n \leq r) \leq \exp \left(-\frac{z^2}{2r} \cdot
B\left(\frac{z}{r} \right) \right) \label{eq: inequality in
exercise of Amir and Ofer's book}
\end{equation}
where $B$ was introduced in \eqref{eq: B}.
\label{theorem: from the book of Dembo and Zeitouni}
\end{theorem}

\vspace*{0.1cm}
\begin{proposition}
Let $\{X_k, \mathcal{F}_k\}_{k=0}^{\infty}$ be a
discrete-parameter real-valued martingale. Then,
Theorem~\ref{theorem: from the book of Dembo and Zeitouni} implies
the results in Corollary~\ref{corollary: 3rd corollary}
and inequality~\eqref{eq: concentration1} in
Proposition~\ref{proposition: a similar scaling of the
concentration inequalities}.
\label{proposition: Theorem 5 implies Corollary 3 and Proposition 4}
\end{proposition}
\begin{IEEEproof}
See Appendix~\ref{appendix: proposition related to Theorem 5}.
\end{IEEEproof}

\subsection{Relation between the Martingale Central Limit Theorem
(CLT) and Proposition~\ref{proposition: a similar scaling of the
concentration inequalities}} \label{subsection: relation between
the martingale CLT and Proposition 4.3} In this subsection, we
discuss the relation between the martingale CLT and the
concentration inequalities for discrete-parameter martingales in
Proposition~\ref{proposition: a similar scaling of the
concentration inequalities}.

Let $(\Omega, \mathcal{F}, \pr)$ be a probability space. Given a
filtration $\{\mathcal{F}_k\}$, then $\{Y_k, \mathcal{F}_k\}_{k=0}^{\infty}$
is said to be a martingale-difference sequence if, for every $k$,
\begin{enumerate}
\item $Y_k$ is $\mathcal{F}_k$-measurable,
\item $E[|Y_k|] < \infty$,
\item $ \expectation \bigl[Y_k \, | \,
\mathcal{F}_{k-1}\bigr] = 0.$
\end{enumerate}
Let $$S_n = \sum_{k=1}^n Y_k, \quad \forall \, n \in \naturals$$
and $S_0 = 0$, then $\{S_k, \mathcal{F}_k\}_{k=0}^{\infty}$ is a
martingale. Assume that the sequence of RVs $\{Y_k\}$ is bounded,
i.e., there exists a constant $d$ such that $|Y_k| \leq d$ a.s.,
and furthermore, assume that the limit
$$ \sigma^2 \triangleq \lim_{n \rightarrow \infty} \frac{1}{n}
\sum_{k=1}^n \expectation\bigl[Y_k^2 \, | \,
\mathcal{F}_{k-1}\bigr]$$ exists in probability and is positive.
The martingale CLT asserts that, under the above conditions,
$\frac{S_n}{\sqrt{n}}$ converges in distribution (i.e., weakly
converges) to the Gaussian distribution $\mathcal{N}(0,
\sigma^2)$. It is denoted by $\frac{S_n}{\sqrt{n}} \Rightarrow
\mathcal{N}(0, \sigma^2)$. We note that there exist more general
versions of this statement (see, e.g.,
\cite[pp.~475--478]{Billingsley}).

Let $\{X_k, \mathcal{F}_k\}_{k=0}^{\infty}$ be a
discrete-parameter real-valued martingale with bounded jumps, and
assume that there exists a constant $d$ so that a.s. for every $k
\in \naturals$
$$|X_k - X_{k-1}| \leq d, \quad \forall \, k \in \naturals.$$
Define, for every $k \in \naturals$,
$$Y_k \triangleq X_k - X_{k-1}$$ and $Y_0 \triangleq 0$,
so $\{Y_k, \mathcal{F}_k\}_{k=0}^{\infty}$ is a
martingale-difference sequence, and $|Y_k| \leq d$ a.s. for every
$k \in \naturals \cup \{0\}$. Furthermore, for every $n \in
\naturals$,
$$S_n \triangleq \sum_{k=1}^n Y_k = X_n - X_0.$$ Under the
assumptions in Theorem~\ref{theorem: first refined concentration
inequality} and its subsequences, for every $k \in \naturals$, one
gets a.s. that
$$\expectation[Y_k^2 \, | \, \mathcal{F}_{k-1}] =
\expectation[(X_k - X_{k-1})^2 \, | \, \mathcal{F}_{k-1}] \leq
\sigma^2.$$ Lets assume that this inequality holds a.s. with
equality. It follows from the martingale CLT that
$$ \frac{X_n - X_0}{\sqrt{n}} \Rightarrow \mathcal{N}(0,
\sigma^2)$$ and therefore, for every $\alpha \geq 0$,
\begin{equation*}
\lim_{n \rightarrow \infty} \pr(|X_n - X_0| \geq \alpha \sqrt{n})
= 2 \, Q\Bigl(\frac{\alpha}{\sigma}\Bigr)
\end{equation*}
where the $Q$ function is introduced in \eqref{eq: Q}.

Based on the notation in \eqref{eq: notation}, the equality
$\frac{\alpha}{\sigma} = \frac{\delta}{\sqrt{\gamma}}$ holds, and
\begin{equation}
\lim_{n \rightarrow \infty} \pr(|X_n - X_0| \geq \alpha \sqrt{n})
= 2 \, Q\biggl(\frac{\delta}{\sqrt{\gamma}}\biggr). \label{eq:
consequence of the martingale CLT}
\end{equation}
Since, for every $x \geq 0$, $$Q(x) \leq \frac{1}{2} \,
\exp\left(-\frac{x^2}{2}\right)$$ then it follows that for every
$\alpha \geq 0$
\begin{equation*}
\lim_{n \rightarrow \infty} \pr(|X_n - X_0| \geq \alpha \sqrt{n})
\leq \exp\left(-\frac{\delta^2}{2\gamma}\right).
\end{equation*}
This inequality coincides with the asymptotic result of the
inequalities in Proposition~\ref{proposition: a similar scaling of
the concentration inequalities} (see \eqref{eq: concentration1}
in the limit where $n \rightarrow
\infty$), except for the additional factor of~2. Note also that
the proof of the concentration inequalities in
Proposition~\ref{proposition: a similar scaling of the
concentration inequalities} (see Appendix~\ref{appendix: proof of
the statement about the similar scaling of the concentration
inequalities}) provides inequalities that are informative for
finite $n$, and not only in the asymptotic case where $n$ tends to
infinity. Furthermore, due to the exponential upper and lower
bounds of the Q-function in \eqref{eq: upper and lower bounds for
the Q function}, then it follows from \eqref{eq: consequence of
the martingale CLT} that the exponent in the concentration
inequality \eqref{eq: concentration1}
(i.e., $\frac{\delta^2}{2 \gamma}$) cannot be
improved under the above assumptions (unless some more information
is available).

\subsection{Relation between the Law of the Iterated Logarithm
(LIL) and Proposition~\ref{proposition: a similar scaling of the
concentration inequalities}} \label{subsection: relation between
the LIL and Proposition 4.3} In this subsection, we discuss the
relation between the law of the iterated logarithm (LIL) and the
concentration inequalities for discrete-parameter martingales in
Proposition~\ref{proposition: a similar scaling of the
concentration inequalities}.

According to the law of the iterated logarithm (see, e.g.,
\cite[Theorem~9.5]{Billingsley}) if $\{X_k\}_{k=1}^{\infty}$
are i.i.d. real-valued RVs with zero mean and unit variance, and
$S_n = \sum_{i=1}^n X_i$ for every $n \in \naturals$, then
\begin{eqnarray}
\limsup_{n \rightarrow \infty}
\frac{S_n}{\sqrt{2n \ln \ln n}} = 1 \quad \text{a.s.}
\label{eq: LIL1}
\end{eqnarray}
and
\begin{eqnarray}
\liminf_{n \rightarrow \infty}
\frac{S_n}{\sqrt{2n \ln \ln n}} = -1 \quad \text{a.s.}
\label{eq: LIL2}
\end{eqnarray}
Equations~\eqref{eq: LIL1} and \eqref{eq: LIL2} assert,
respectively, that for $\varepsilon > 0$, along almost any
realization
\begin{equation*}
S_n > (1-\varepsilon) \sqrt{2n \ln \ln n}
\end{equation*}
and
\begin{equation*}
S_n < -(1-\varepsilon) \sqrt{2n \ln \ln n}
\end{equation*}
infinitely often (i.o.).

Let $\{X_k\}_{k=1}^{\infty}$ be i.i.d. real-valued RVs, defined
over the probability space $(\Omega, \mathcal{F}, \pr)$, with
$\expectation[X_1] = 0$ and $\expectation[X_1^2] =~1$. Hence $X_k
\in L^2(\Omega, \mathcal{F}, \pr)$, and therefore $X_k \in
L^1(\Omega, \mathcal{F}, \pr)$ for every $k \in \naturals$.

Let us define the natural filtration where $\mathcal{F}_0 = \{
\emptyset, \Omega \}$, and $\mathcal{F}_k = \sigma(X_1, \ldots,
X_k)$ is the $\sigma$-algebra that is generated by the RVs $X_1,
\ldots, X_k$ for every $k \in \naturals$. Let $S_0 = 0$ and $S_n$
be defined as above for every $n \in \naturals$. It is
straightforward to verify by Definition~\ref{definition: Doob's
martingales} that $\{S_n, \mathcal{F}_n\}_{n=0}^{\infty}$ is a
martingale.

In order to apply Proposition~\ref{proposition: a similar scaling
of the concentration inequalities} to the considered case, let us
assume that the RVs $\{X_k\}_{k=1}^{\infty}$ are uniformly
bounded, i.e., it is assumed that there exists a constant $c>0$
such that $|X_k| \leq c$ a.s. for every $k \in \naturals$. This
implies that the martingale $\{S_n,
\mathcal{F}_n\}_{n=0}^{\infty}$ has bounded jumps, and for every
$n \in \naturals$
$$ |S_n - S_{n-1} | \leq c \quad \text{a.s.}$$
Moreover, due to the independence of the RVs
$\{X_k\}_{k=1}^{\infty}$, then
$$\text{Var}(S_n \, | \, \mathcal{F}_{n-1}) =
\expectation(X_n^2 \, | \, \mathcal{F}_{n-1}) =
\expectation(X_n^2) = 1 \quad \text{a.s.} $$ which by
Proposition~\ref{proposition: a similar scaling of the
concentration inequalities} implies that for every
$\alpha \geq 0$
\begin{equation}
\pr( |S_n| \geq \alpha \sqrt{n}) \leq 2
\exp\left(-\frac{\alpha^2}{2}\right)
\left(1 + O\biggl(\frac{1}{\sqrt{n}}\biggr) \right).
\label{eq: consequence of Proposition 4.3}
\end{equation}
(in the setting of Proposition~\ref{proposition: a similar scaling
of the concentration inequalities}, \eqref{eq:
notation} gives that $\gamma = \frac{1}{c^2}$ and $\delta =
\frac{\alpha}{c}$). Note that the exponent on the right-hand side
of \eqref{eq: consequence of Proposition 4.3} is independent of
the value of $c$, and it improves by a factor of $\frac{1}{c}$
(where $c \leq 1$) the exponent of Azuma's inequality. Under the
additional assumption that the RVs $\{X_k\}$ are uniformly bounded
as above, then inequality~\eqref{eq: consequence of Proposition
4.3} provides further information to \eqref{eq: LIL1} and
\eqref{eq: LIL2} where $\sqrt{2n \ln \ln n}$ roughly scales like
the square root of $n$.

\subsection{Relation of Theorems~\ref{theorem: first refined concentration inequality}
and~\ref{theorem: second inequality} with the Moderate Deviations Principle}
\label{subsection: MDP for real-valued i.i.d. RVs}

According to the moderate deviations theorem (see, e.g.,
\cite[Theorem~3.7.1]{Dembo_Zeitouni}) in $\reals$, let
$\{X_i\}_{i=1}^n$ be a sequence of real-valued RVs such that
$\Lambda_X(\lambda) = \expectation[e^{\lambda X_i}] < \infty$ in
some neighborhood of zero, and also assume that $\expectation[X_i]
= 0$ and $\sigma^2 = \text{Var}(X) > 0$. Let
$\{a_n\}_{n=1}^{\infty}$ be a non-negative sequence such that $a_n
\rightarrow 0$ and $n a_n \rightarrow \infty$ as $n \rightarrow
\infty$, and let
\begin{equation}
Z_n \triangleq \sqrt{\frac{a_n}{n}} \sum_{i=1}^n X_i,
\quad \forall \, n \in \naturals.
\label{eq: Z sequence}
\end{equation}
Then, for every measurable set $\Gamma \subseteq \reals$,
\begin{eqnarray}
&& -\frac{1}{2 \sigma^2} \inf_{x \in \Gamma^0} x^2 \leq
\liminf_{n \rightarrow \infty} a_n \ln \pr(Z_n \in \Gamma) \nonumber \\
&& \hspace*{2.2cm} \leq \limsup_{n \rightarrow \infty}
a_n \ln \pr(Z_n \in \Gamma)  \nonumber \\
&& \hspace*{2.2cm} \leq -\frac{1}{2 \sigma^2}
\inf_{x \in \overline{\Gamma}} x^2
\end{eqnarray}
where $\Gamma^0$ and $\overline{\Gamma}$ designate, respectively,
the interior and closure sets of $\Gamma$.

Let $\eta \in (\frac{1}{2}, 1)$ be an arbitrary fixed number, and
let $\{a_n\}_{n=1}^{\infty}$ be the non-negative sequence
$$a_n = n^{1-2\eta}, \quad \forall \, n \in \naturals$$ so that
$a_n \rightarrow 0$ and $n a_n \rightarrow \infty$ as $n
\rightarrow \infty$. Let $\alpha \in \reals^+$, and $\Gamma
\triangleq (-\infty, -\alpha] \cup [\alpha, \infty)$. Note that,
from \eqref{eq: Z sequence},
$$ \pr\left( \Big|\sum_{i=1}^n X_i \Big| \geq \alpha n^{\eta} \right)
= \pr(Z_n \in \Gamma)$$ so from the moderate deviations principle
(MDP)
\begin{equation}
\hspace*{-0.4cm} \lim_{n \rightarrow \infty} n^{1-2\eta} \;
\pr\left( \Big|\sum_{i=1}^n X_i \Big| \geq \alpha n^{\eta} \right)
= -\frac{\alpha^2}{2 \sigma^2}, \quad \forall \, \alpha \geq 0.
\label{eq: MDP for i.i.d. real-valued RVs}
\end{equation}
It is demonstrated in Appendix~\ref{appendix: MDP} that, in
contrast to Azuma's inequality, Theorems~\ref{theorem: first
refined concentration inequality} and~\ref{theorem: second
inequality} (for every even $m \geq 2$ in Theorem~\ref{theorem:
second inequality}) provide upper bounds on the probability
$$\pr\left( \Big|\sum_{i=1}^n X_i \Big| \geq \alpha n^{\eta} \right),
\quad \forall \, n \in \naturals, \; \alpha \geq 0$$ which both
coincide with the correct asymptotic result in \eqref{eq: MDP for
i.i.d. real-valued RVs}. The analysis in Appendix~\ref{appendix:
MDP} provides another interesting link between
Theorems~\ref{theorem: first refined concentration inequality}
and~\ref{theorem: second inequality} and a classical result in
probability theory, which also emphasizes the significance of the
refinements of Azuma's inequality.

\subsection{Relation of \cite[Lemma~2.8]{McDiarmid_tutorial} with
Theorem~\ref{theorem: second inequality} \&
Corollary~\ref{corollary: specialization of the second
inequality}} In \cite[Lemma~2.8]{McDiarmid_tutorial}, it is proved
that if $X$ is a random variable that satisfies
$\expectation[X]=0$ and $X \leq d$ a.s. (for some $d>0$), then
\begin{equation}
\expectation \bigl[e^X \bigr] \leq
\exp \bigl( \varphi(d) \, \text{Var}(X) \bigr)
\label{eq: bound in Lemma 2.8 from Colin's paper}
\end{equation}
where
\begin{equation*}
\varphi(x) = \left\{\begin{array}{ll}
\frac{\exp(x)-1-x}{x^2} \quad &
\mbox{if} \; x \neq 0 \\[0.2cm]
\hspace*{0.7cm} \frac{1}{2} \quad & \mbox{if} \; x=0 \end{array}
\right. .
\end{equation*}
From \eqref{eq: phi function}, it follows that $\varphi(x) =
\frac{\varphi_2(x)}{2}$ for every $x \in \reals$. Based on
\cite[Lemma~2.8]{McDiarmid_tutorial}, it follows that if $\{\xi_k,
\mathcal{F}_k\}$ is a difference-martingale sequence (i.e., for
every $k \in \naturals$,
$$\expectation[\xi_k \, | \, \mathcal{F}_{k-1}] = 0$$
a.s.), and $\xi_k \leq d$ a.s. for some $d>0$, then for an
arbitrary $t \geq 0$
\begin{equation*}
\expectation \bigl[ \exp(t \xi_k) | \mathcal{F}_{k-1} \bigr] \leq
\exp \left(\frac{\gamma \; (td)^2 \, \varphi_2(td)}{2} \right)
\end{equation*}
holds a.s. for every $k \in \naturals$ (the parameter $\gamma$ was
introduced in~\eqref{eq: notation}). The last inequality can be
rewritten as

\vspace*{-0.2cm}
\begin{equation} \small \expectation \bigl[ \exp(t \xi_k) |
\mathcal{F}_{k-1} \bigr] \leq \exp \left( \gamma \;
\bigl(\exp(td)-1-td\bigr) \right), \quad t \geq 0. \label{eq:
looser bound as compared to the inequality obtained in the
derivation of Theorem 4 and Corollary 4}
\end{equation}

\normalsize This forms a looser bound on the conditional
expectation, as compared to \eqref{eq:
important inequality for the derivation of Theorem 4} with $m=2$,
that gets the form

\vspace*{-0.2cm}
\small \begin{equation} \expectation \bigl[
\exp(t \xi_k) | \mathcal{F}_{k-1} \bigr] \leq 1 + \gamma \;
\bigl(\exp(td)-1-td\bigr), \quad t \geq 0. \label{eq: improved
bound used for the derivation of Corollary 4}
\end{equation}
\normalsize The improvement in \eqref{eq: improved bound used for
the derivation of Corollary 4} over \eqref{eq: looser bound as
compared to the inequality obtained in the derivation of Theorem 4
and Corollary 4} follows since $e^x \geq 1+x$ for $x \geq 0$ with
equality if and only if $x=0$. Note that the proof of
\cite[Lemma~2.8]{McDiarmid_tutorial} shows that indeed the
right-hand side of \eqref{eq: improved bound used for the
derivation of Corollary 4} forms an upper bound on the above
conditional expectation, whereas it is loosened to the bound on
the right-hand side of \eqref{eq: looser bound as compared to the
inequality obtained in the derivation of Theorem 4 and Corollary
4} in order to handle the case where
$$ \frac{1}{n} \sum_{k=1}^n \expectation\bigl[(\xi_k)^2
\, | \, \mathcal{F}_{k-1} \bigr] \leq \sigma^2$$ and derive a
closed-form solution of the optimized parameter $t$ in the
resulting concentration inequality (see the proof of
\cite[Theorem~2.7]{McDiarmid_tutorial} for the case of independent
RVs, and also \cite[Theorem~3.15]{McDiarmid_tutorial} for the
setting of martingales with bounded jumps). However, if for every
$k \in \naturals$, the condition $$\expectation\bigl[(\xi_k)^2 \,
| \, \mathcal{F}_{k-1} \bigr] \leq \sigma^2$$ holds a.s., then the
proof of Corollary~\ref{corollary: specialization of the second
inequality} shows that a closed-form solution of the non-negative
free parameter $t$ is obtained. More on the consequence of the
difference between the bounds in \eqref{eq: looser bound as
compared to the inequality obtained in the derivation of Theorem 4
and Corollary 4} and~\eqref{eq: improved bound used for the
derivation of Corollary 4} is considered in the next sub-section.

\subsection{Relation of the Concentration Inequalities for Martingales
to Discrete-Time Markov Chains}

A striking well-known relation between discrete-time Markov chains
and martingales is the following (see, e.g.,
\cite[p.~473]{GrimmettS_book}): Let $\{X_n\}_{n \in \naturals_0}$
($\naturals_0 \triangleq \naturals \cup \{0\}$) be a discrete-time
Markov chain taking values in a countable state space
$\mathcal{S}$ with transition matrix ${\bf{P}}$, and let the
function $\psi: \mathcal{S} \rightarrow \mathcal{S}$ be harmonic,
i.e., $$\sum_{j \in \mathcal{S}} p_{i,j} \psi(j) = \psi(i), \quad
\forall \, i \in \mathcal{S}$$ and consider the case where
$E[|\psi(X_n)|] < \infty$ for every $n$. Then, $\{Y_n,
\mathcal{F}_n\}_{n \in \naturals_0}$ is a martingale where $Y_n
\triangleq \psi(X_n)$ and $\{\mathcal{F}_n\}_{n \in \naturals_0}$
is a the natural filtration. This relation, which follows directly
from the Markov property, enables to apply the concentration
inequalities in Section~\ref{section: Refined Versions of Azuma's
Inequality} for harmonic functions of Markov chains when the
function $\psi$ is bounded (so that the jumps of the martingale
sequence are uniformly bounded). In the special case of i.i.d.
RVs, one obtains Hoeffding's inequality and its refined versions.

We note that relative entropy and exponential deviation bounds for
an important class of Markov chains, called Doeblin chains (which
are characterized by a convergence to the equilibrium
exponentially fast, uniformly in the initial condition) were
derived in \cite{ISIT05}. These bounds were also shown to be
essentially identical to the Hoeffding inequality in the special
case of i.i.d. RVs (see \cite[Remark~1]{ISIT05}).

\subsection{Relations of \cite[Theorem~2.23]{survey2006} with
Corollary~\ref{corollary: specialization of the second inequality}
and Proposition~\ref{proposition: a similar scaling of the
concentration inequalities}} In the following, we consider the
relation between the inequalities in Corollary~\ref{corollary:
specialization of the second inequality} and
Proposition~\ref{proposition: a similar scaling of the
concentration inequalities} to the particularized form of
\cite[Theorem~2.23]{survey2006} (or also
\cite[Theorem~2.23]{Chung_LU2006}) in the setting where $d_k = d$
and $\sigma_k^2 = \sigma^2$ are fixed for every $k \in \naturals$.
The resulting exponents of these concentration inequalities are
also compared.

Let $\alpha \geq 0$ be an arbitrary non-negative number.
\begin{itemize}
\item In the analysis of small deviations, the bound in
\cite[Theorem~2.23]{survey2006} is particularized to
\begin{equation*}
\pr(|X_n-X_0| \geq \alpha \sqrt{n}) \leq 2 \exp
\left(-\frac{\alpha^2 n}{2 n \sigma^2 + \frac{2d \alpha
\sqrt{n}}{3}}\right).
\end{equation*}
From the notation in \eqref{eq: notation} then
$\frac{\alpha^2}{\sigma^2} = \frac{\delta^2}{\gamma}$, and the
last inequality gets the form
\begin{equation*}
\pr(|X_n-X_0| \geq \alpha \sqrt{n}) \leq 2 \exp \left(
-\frac{\delta^2}{2\gamma}\right) \;
\left(1+O\Bigl(\frac{1}{\sqrt{n}}\Bigr) \right).
\end{equation*}
It therefore follows that \cite[Theorem~2.23]{survey2006}
implies a concentration inequality of the form in \eqref{eq:
concentration1}. This shows that Proposition~\ref{proposition: a
similar scaling of the concentration inequalities} can be also
regarded as a consequence of \cite[Theorem~2.23]{survey2006}.
\item In the analysis of large deviations, the bound in
\cite[Theorem~2.23]{survey2006} is particularized to
\begin{equation*}
\pr(|X_n-X_0| \geq \alpha n) \leq 2 \exp\left(-\frac{\alpha^2 n}{2
\sigma^2 + \frac{2 d \alpha}{3}}\right).
\end{equation*}
From the notation in \eqref{eq: notation}, this inequality is
rewritten as
\begin{equation}
\pr(|X_n-X_0| \geq \alpha n) \leq 2 \exp\left(-\frac{\delta^2 n}{2
\gamma + \frac{2 \delta}{3}} \right). \label{eq: Theorem 2.23 in
Chung's book}
\end{equation}
\end{itemize}
It is claimed that the concentration inequality in \eqref{eq:
Theorem 2.23 in Chung's book} is looser than
Corollary~\ref{corollary: specialization of the second
inequality}. This is a consequence of the proof of
\cite[Theorem~2.23]{survey2006} where the derived concentration
inequality is loosened in order to handle the more general case,
as compared to the setting in this paper (see
Theorem~\ref{theorem: first refined concentration inequality}),
where $d_k$ and $\sigma_k^2$ may depend on $k$. In order to show
it explicitly, lets compare between the steps of the derivation of
the bound in Corollary~\ref{corollary: specialization of the
second inequality}, and the particularization of the derivation of
\cite[Theorem~2.23]{survey2006} in the special setting where $d_k$
and $\sigma_k^2$ are independent of $k$. This comparison is
considered in the following. The derivation of the concentration
inequality in Corollary~\ref{corollary: specialization of the
second inequality} follows by substituting $m=2$ in the proof of
Theorem~\ref{theorem: second inequality}. It then follows that,
for every $\alpha \geq 0$,
\begin{eqnarray}
&& \hspace*{-0.5cm} \pr(X_n-X_0 \geq \alpha n) \nonumber \\
&& \hspace*{-0.5cm} \leq e^{-n \delta x} \Bigl(1 + \gamma
\bigl(e^x-1-x\bigr) \Bigr)^n, \quad \forall \, x \geq 0 \label{eq:
step in the derivation of the relevant corollary in this paper}
\end{eqnarray}
which then leads, after an analytic optimization of the free
non-negative parameter $x$ (see Lemma~\ref{lemma: proof of the
existence and uniqueness of the solution} and
Appendix~\ref{appendix: proof of the corollary that specializes
the second inequality for m equal to 2}), to the derivation of
Corollary~\ref{corollary: specialization of the second
inequality}. On the other hand, the specialization of the proof of
\cite[Theorem~2.23]{survey2006} to the case where $d_k=d$ and
$\sigma_k^2 = \sigma^2$ for every $k \in \naturals$ is equivalent
to a further loosening of \eqref{eq: step in the derivation of the
relevant corollary in this paper} to the bound
\begin{eqnarray}
&& \pr(X_n-X_0 \geq \alpha n) \nonumber \\
&& \leq e^{-n \delta x} e^{n \gamma (e^x-1-x)} \label{eq: 1st loosening} \\
&& \leq e^{n \Bigl(-\delta x + \frac{\gamma
x^2}{1-\frac{x}{3}}\Bigr)}, \quad \; \forall \; x \in (0,3)
\label{eq: 2nd loosening}
\end{eqnarray}
and then choosing an optimal $x \in (0,3)$. This indeed shows that
Corollary~\ref{corollary: specialization of the second inequality}
provides a concentration inequality that is more tight than the
bound in \cite[Theorem~2.23]{survey2006}.

In order to compare quantitatively the exponents of the
concentration inequalities in \cite[Theorem~2.23]{survey2006} and
Corollary~\ref{corollary: specialization of the second
inequality}, let us revisit the derivation of the upper bounds on
the probability of the events $\{|X_n - X_0| \geq \alpha n\}$
where $\alpha \geq 0$ is arbitrary. The optimized value of $x$
that is obtained in Appendix~\ref{appendix: proof of the corollary
that specializes the second inequality for m equal to 2} is
positive, and it becomes larger as we let the value of $\gamma \in
(0,1]$ approach zero. Hence, especially for small values of
$\gamma$, the loosening of the bound from \eqref{eq: step in the
derivation of the relevant corollary in this paper} to \eqref{eq:
2nd loosening} is expected to deteriorate more significantly the
resulting bound in \cite[Theorem~2.23]{survey2006} due to the
restriction that $x \in (0,3)$; this is in contrast to the
optimized value of $x$ in Appendix~\ref{appendix: proof of the
corollary that specializes the second inequality for m equal to 2}
that may be above~3 for small values of $\gamma$, and it lies in
general between 0 and $\frac{1}{\gamma}$. Note also that at
$\delta=1$, the exponent in Corollary~\ref{corollary:
specialization of the second inequality} tends to infinity in the
limit where $\gamma \rightarrow 0$, whereas the exponent in
\eqref{eq: Theorem 2.23 in Chung's book} tends in this case
to~$\frac{3}{2}$. To illustrate these differences,
Figure~\ref{Figure:
compare_exponents_corollary_4_and_looser_version} plots the
exponents of the bounds in Corollary~\ref{corollary:
specialization of the second inequality} and \eqref{eq: Theorem
2.23 in Chung's book}, where the latter refers to
\cite[Theorem~2.23]{survey2006}, for $\gamma = 0.01$ and $0.99$.
As is shown in Figure~\ref{Figure:
compare_exponents_corollary_4_and_looser_version}, the difference
between the exponents of these two bounds is indeed more
pronounced when $\gamma$ gets closer to zero.

\begin{figure}[here!]
\begin{center}
\epsfig{file=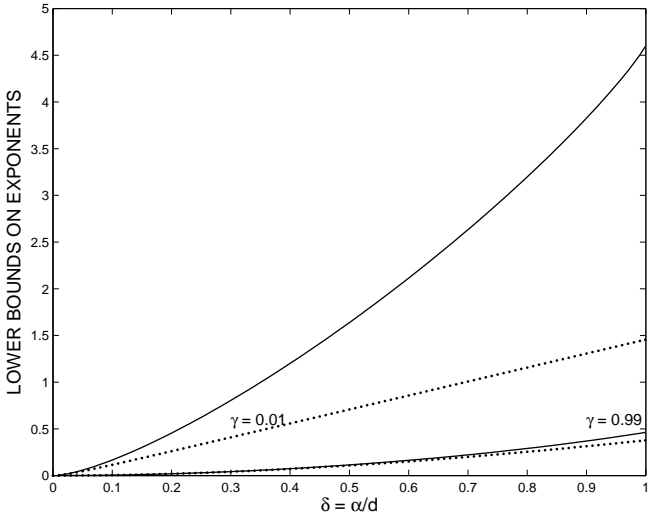,scale=0.50}
\end{center}
\caption{\label{Figure:
compare_exponents_corollary_4_and_looser_version} A comparison of
the exponents of the bound in Corollary~\ref{corollary:
specialization of the second inequality} and the particularized
bound \eqref{eq: Theorem 2.23 in Chung's book} from
\cite[Theorem~2.23]{survey2006}. This comparison is done for both
$\gamma = 0.01$ and $0.99$. The solid curves refer to the
exponents of the bound in Corollary~\ref{corollary: specialization
of the second inequality}, and the dashed curves refer to the
exponents of the looser bound in \eqref{eq: Theorem 2.23 in
Chung's book}. The upper pair of curves refers to the exponents
for $\gamma=0.01$, and the lower pair of curves (that
approximately coincide) refers to the exponents for
$\gamma=0.99$.}
\end{figure}

Consider, on the other hand, the probability of an event $\{|X_n -
X_0| \geq \alpha \sqrt{n}\}$ where $\alpha \geq 0$ is arbitrary.
It was shown in Appendix~\ref{appendix: a sufficient condition
where the specialization of the second inequality is better than
Azuma's inequality} that the optimized value of $x$ for the bound
in Corollary~\ref{corollary: specialization of the second
inequality} (and its generalized version in Theorem~\ref{theorem:
second inequality}) scales like $\frac{1}{\sqrt{n}}$. Hence, it is
approximately zero for $n \gg 1$, and $u \triangleq \gamma
(e^x-1-x) \approx \frac{\gamma x^2}{2}$ scales like
$\frac{1}{n}$. It therefore follows that $(1+u)^n \approx e^{nu}$
for $n \gg 1$. Moreover, the restriction on $x$ to be less than~3
in \eqref{eq: 2nd loosening} does not affect the tightness of the
bound in this case since the optimized value of $x$ is anyway
close to zero. This explains the observation that the two bounds
in Proposition~\ref{proposition: a similar scaling of the
concentration inequalities} and \cite[Theorem~2.23]{survey2006}
essentially scale similarly for small deviations, where
the probability of an event
$\{|X_n - X_0| \geq \alpha \sqrt{n}\}$ for $\alpha \geq 0$
is considered.

\section{Applications in Information Theory and Related Topics}
\label{section: Applications} The refined versions of Azuma's
inequality in Section~\ref{section: Refined Versions of Azuma's
Inequality} are exemplified in this section to hypothesis testing
and information theory, communication and coding.

\subsection{Binary Hypothesis Testing}
Binary hypothesis testing for finite alphabet models was analyzed
via the method of types, e.g., in \cite[Chapter~11]{Cover and
Thomas} and \cite{Csiszar_Shields_FnT}. It is assumed that the
data sequence is of a fixed length $(n)$, and one wishes to make
the optimal decision (based on the Neyman-Pearson ratio test)
based on the received sequence.

Let the RVs $X_1, X_2 ....$ be i.i.d. $\sim Q$, and consider two
hypotheses:
\begin{itemize}
\item $H_1:  Q = P_1$.
\item $H_2:  Q = P_2$.
\end{itemize}
For the simplicity of the analysis, let us assume that the RVs are
discrete, and take their values on a finite alphabet $\mathcal{X}$
where $P_1(x), P_2(x) > 0$ for every $x \in \mathcal{X}$.

In the following, let
\begin{equation*}
L(X_1, \ldots, X_n) \triangleq
\ln \frac{P_1^n(X_1, \ldots, X_n)}{P_2^n(X_1, \ldots, X_n)}
= \sum_{i=1}^n \ln \frac{P_1(X_i)}{P_2(X_i)}
\end{equation*}
designate the log-likelihood ratio.
By the strong law of large number (SLLN), if hypothesis $H_1$ is
true, then a.s.
\begin{equation}
\lim_{n \rightarrow \infty} \frac{L(X_1, \ldots, X_n)}{n} = D(P_1
|| P_2) \label{eq: a.s. limit of the normalized LLR under
hypothesis H1}
\end{equation}
and otherwise, if hypothesis $H_2$ is true, then a.s.
\begin{equation}
\lim_{n \rightarrow \infty} \frac{L(X_1, \ldots, X_n)}{n} = -D(P_2
|| P_1) \label{eq: a.s. limit of the normalized LLR under
hypothesis H2}
\end{equation}
where the above assumptions on the probability mass functions
$P_1$ and $P_2$ imply that the relative entropies, $D(P_1 || P_2)$
and $D(P_2 || P_1)$, are both finite. Consider the case where for
some fixed constants $\overline{\lambda}, \underline{\lambda} \in
\reals$ where $$-D(P_2||P_1) < \underline{\lambda} \leq
\overline{\lambda} < D(P_1||P_2)$$ one decides on hypothesis $H_1$
if $$ L(X_1, \ldots, X_n) > n \overline{\lambda} $$ and on
hypothesis $H_2$ if $$ L(X_1, \ldots, X_n) < n
\underline{\lambda}.$$ Note that if $\overline{\lambda} =
\underline{\lambda} \triangleq \lambda$ then a decision on the two
hypotheses is based on comparing the normalized log-likelihood
ratio (w.r.t. $n$) to a single threshold $(\lambda)$, and deciding
on hypothesis $H_1$ or $H_2$ if this normalized log-likelihood
ratio is, respectively, above or below $\lambda$. If
$\underline{\lambda} < \overline{\lambda}$ then one decides on
$H_1$ or $H_2$ if the normalized log-likelihood ratio is,
respectively, above the upper threshold $\overline{\lambda}$ or
below the lower threshold $\underline{\lambda}$. Otherwise, if the
normalized log-likelihood ratio is between the upper and lower
thresholds, then an erasure is declared and no decision is taken
in this case.

Let
\begin{eqnarray}
&& \alpha_n^{(1)} \triangleq P_1^n \Bigl( L(X_1, \ldots, X_n) \leq
n \overline{\lambda} \Bigr)
\label{eq: error and erasure event under hypothesis H1} \\
&& \alpha_n^{(2)} \triangleq P_1^n \Bigl( L(X_1, \ldots, X_n) \leq
n \underline{\lambda} \Bigr) \label{eq: error event under
hypothesis H1}
\end{eqnarray}
and
\begin{eqnarray}
&& \beta_n^{(1)}  \triangleq P_2^n \Bigl( L(X_1, \ldots, X_n) \geq
n \underline{\lambda} \Bigr)
\label{eq: error and erasure event under hypothesis H2} \\
&& \beta_n^{(2)}  \triangleq P_2^n \Bigl( L(X_1, \ldots, X_n) \geq
n \overline{\lambda} \Bigr) \label{eq: error event under
hypothesis H2}
\end{eqnarray}
then $\alpha_n^{(1)}$ and $\beta_n^{(1)}$ are the probabilities of
either making an error or declaring an erasure under,
respectively, hypotheses $H_1$ and $H_2$; similarly
$\alpha_n^{(2)}$ and $\beta_n^{(2)}$ are the probabilities of
making an error under hypotheses $H_1$ and $H_2$, respectively.

Let $\pi_1, \pi_2 \in (0,1)$ denote the a-priori probabilities of the
hypotheses $H_1$ and $H_2$, respectively, so
\begin{equation}
P_{\text{e}, n}^{(1)} = \pi_1 \alpha_n^{(1)} + \pi_2 \beta_n^{(1)}
\label{eq: overall probability of a mixed error and erasure event}
\end{equation}
is the probability of having either an error or an erasure, and
\begin{equation}
P_{\text{e}, n}^{(2)} = \pi_1 \alpha_n^{(2)} + \pi_2 \beta_n^{(2)}
\label{eq: overall error probability}
\end{equation}
is the probability of error.

\subsubsection{Exact Exponents}
When we let $n$ tend to infinity, the exact exponents of
$\alpha_n^{(j)}$ and $\beta_n^{(j)}$ ($j=1,2$) are derived
via Cram\'{e}r's theorem. The resulting exponents form a
straightforward generalization of, e.g.,
\cite[Theorem~3.4.3]{Dembo_Zeitouni} and
\cite[Theorem~6.4]{Hollander_book_2000} that addresses the case
where the decision is made based on a single threshold of the
log-likelihood ratio. In this particular case where $\overline{\lambda}
= \underline{\lambda} \triangleq \lambda$, the option of erasures
does not exist, and $P_{\text{e}, n}^{(1)} = P_{\text{e}, n}^{(2)}
\triangleq P_{\text{e}, n}$ is the error probability.

In the considered general case with erasures, let $$ \lambda_1
\triangleq -\overline{\lambda}, \quad \lambda_2 \triangleq
-\underline{\lambda} $$ then Cram\'{e}r's theorem on $\reals$
yields that the exact exponents of $\alpha_n^{(1)}$,
$\alpha_n^{(2)}$, $\beta_n^{(1)}$ and $\beta_n^{(2)}$ are given by
\begin{eqnarray}
&& \lim_{n \rightarrow \infty} -\frac{\ln \alpha_n^{(1)}}{n} =
I(\lambda_{1})
\label{eq: exponent of alpha_n1} \\[0.1cm]
&& \lim_{n \rightarrow \infty} -\frac{\ln \alpha_n^{(2)}}{n} =
I(\lambda_{2})
\label{eq: exponent of alpha_n2} \\[0.1cm]
&& \lim_{n \rightarrow \infty} -\frac{\ln \beta_n^{(1)}}{n} =
I(\lambda_2) - \lambda_2 \label{eq: exponent of beta_n1} \\[0.1cm]
&& \lim_{n \rightarrow \infty} -\frac{\ln \beta_n^{(2)}}{n} =
I(\lambda_1) - \lambda_1 \label{eq: exponent of beta_n2}
\end{eqnarray}
where the rate function $I$ is given by
\begin{equation}
I(r) \triangleq \sup_{t \in \reals} \bigl( tr - H(t) \bigr)
\label{eq: rate function}
\end{equation}
and
\begin{equation}
H(t) = \ln \Biggl(\sum_{x \in \mathcal{X}} P_1(x)^{1-t} P_2(x)^t
\Biggr), \quad \forall \, t \in \reals. \label{eq: H}
\end{equation}
The rate function $I$ is convex, lower semi-continuous (l.s.c.)
and non-negative (see, e.g., \cite{Dembo_Zeitouni} and
\cite{Hollander_book_2000}). Note that
$$H(t) = (t-1) D_t(P_2||P_1)$$ where $D_t(P||Q)$ designates R\'{e}yni's
information divergence of order $t$ \cite[Eq.~(3.3)]{Reyni}, and $I$ in
\eqref{eq: rate function} is the Fenchel-Legendre transform of $H$
(see, e.g., \cite[Definition~2.2.2]{Dembo_Zeitouni}).

From \eqref{eq: overall probability of a mixed error and erasure
event}-- \eqref{eq: exponent of beta_n2}, the
exact exponents of $P_{\text{e}, n}^{(1)}$ and
$P_{\text{e}, n}^{(2)}$ are equal to
\begin{equation}
\lim_{n \rightarrow \infty} - \frac{\ln P_{\text{e}, n}^{(1)}}{n}
= \min \Bigl\{ I(\lambda_1), I(\lambda_2) - \lambda_2 \Bigr\}
\label{eq: exact exponent of the overall error and
erasure probability}
\end{equation}
and
\begin{equation}
\lim_{n \rightarrow \infty} - \frac{\ln P_{\text{e}, n}^{(2)}}{n}
= \min \Bigl\{ I(\lambda_2), I(\lambda_1) - \lambda_1 \Bigr\}.
\label{eq: exact error exponent}
\end{equation}

For the case where the decision is based on a single threshold
for the log-likelihood ratio (i.e., $\lambda_1 = \lambda_2
\triangleq \lambda$), then
$P_{\text{e}, n}^{(1)} = P_{\text{e}, n}^{(2)} \triangleq
P_{\text{e}, n}$, and its error exponent is equal to
\begin{equation}
\lim_{n \rightarrow \infty} - \frac{\ln P_{\text{e}, n}}{n} = \min
\Bigl\{ I(\lambda), I(\lambda) - \lambda \Bigr\} \label{eq: exact
exponent of the error probability for a single
threshold}
\end{equation}
which coincides with the error exponent in
\cite[Theorem~3.4.3]{Dembo_Zeitouni} (or
\cite[Theorem~6.4]{Hollander_book_2000}). The optimal threshold
for obtaining the best error exponent of the error
probability $P_{\text{e}, n}$ is equal to zero (i.e.,
$\lambda=0$); in this case, the exact error exponent is
equal to
\begin{eqnarray}
&& I(0) = -\min_{0 \leq t \leq 1} \ln \Biggl( \sum_{x \in
\mathcal{X}} P_1(x)^{1-t} P_2(x)^t \Biggr) \nonumber \\
&&\hspace*{0.7cm} \triangleq C(P_1, P_2) \label{eq: Chernoff
information}
\end{eqnarray}
which is the Chernoff information of the probability measures
$P_1$ and $P_2$ (see \cite[Eq.~(11.239)]{Cover and Thomas}),
and it is symmetric (i.e., $C(P_1, P_2) = C(P_2, P_1)$).
Note that, from \eqref{eq: rate function},
$I(0) = \sup_{t \in \reals}\bigl(-H(t)\bigr) =
-\inf_{t \in \reals}\bigl(H(t)\bigr)$; the minimization in
\eqref{eq: Chernoff information} over the interval $[0,1]$
(instead of taking the infimum of $H$ over $\reals$) is due to the
fact that $H(0) = H(1) = 0$ and the function $H$ in \eqref{eq: H}
is convex, so it is enough to restrict the infimum of $H$ to the
closed interval $[0,1]$ for which it turns to be a minimum.

Paper \cite{Blahut_IT74} forms a classical paper that
considers binary hypothesis testing from an information-theoretic
point of view, and it derives the error exponents of binary hypothesis
testers in analogy to optimum channel codes via the use of relative
entropy measures. We will further explore on this kind of analogy
in the continuation to this section (see later
Sections~\ref{subsubsection: MDP for binary hypothesis testing}
and~\ref{subsunsection: Second-Order Analysis for Binary Hypothesis Testing}
w.r.t. moderate and small deviations analysis of binary hypothesis testing).

\subsubsection{Lower Bound on the Exponents via
Theorem~\ref{theorem: first refined concentration inequality}} In
the following, the tightness of Theorem~\ref{theorem: first
refined concentration inequality} is examined by using it for the
derivation of lower bounds on the error exponent and the exponent
of the event of having either an error or an erasure. These results
will be compared in the next sub-section to the exact exponents
from the previous sub-section.

We first derive a lower bound on the exponent of $\alpha_n^{(1)}$.
Under hypothesis $H_1$, let us construct the martingale sequence
$\{U_k, \mathcal{F}_k\}_{k=0}^n$ where $\mathcal{F}_0 \subseteq
\mathcal{F}_1 \subseteq \ldots \mathcal{F}_n$ is the filtration
$$ \mathcal{F}_0 = \{\emptyset, \Omega\}, \quad \mathcal{F}_k =
\sigma(X_1, \ldots, X_k), \; \; \forall \, k \in \{1, \ldots,
n\}$$ and
\begin{equation}
U_k = \expectation_{P_1^n} \bigl[ L(X_1, \ldots, X_n)
\; | \; \mathcal{F}_k \bigr].
\label{eq: martingale sequence U under hypothesis H1}
\end{equation}
For every $k \in \{0, \ldots, n\}$
\begin{eqnarray*}
&& U_k = \expectation_{P_1^n} \Biggl[  \sum_{i=1}^n
\ln \frac{P_1(X_i)}{P_2(X_i)} \; \Big| \; \mathcal{F}_k  \Biggr] \\
&& \hspace*{0.5cm} =  \sum_{i=1}^k \ln \frac{P_1(X_i)}{P_2(X_i)}
+ \sum_{i=k+1}^n \expectation_{P_1^n} \Biggl[
\ln \frac{P_1(X_i)}{P_2(X_i)} \Biggr] \\
&& \hspace*{0.5cm} =  \sum_{i=1}^k \ln
\frac{P_1(X_i)}{P_2(X_i)} + (n-k) D(P_1 || P_2).
\end{eqnarray*}
In particular
\begin{eqnarray}
&& U_0 = n D(P_1 || P_2), \label{eq: initial value of the
martingale U that is related to the binary hypothesis testing}   \\
&& U_n = \sum_{i=1}^n \ln \frac{P_1(X_i)}{P_2(X_i)} = L(X_1,
\ldots, X_n) \label{eq: final value of the martingale U that is
related to the binary hypothesis testing}
\end{eqnarray}
and, for every $k \in \{1, \ldots, n\}$,
\begin{equation}
U_k - U_{k-1} = \ln \frac{P_1(X_k)}{P_2(X_k)} - D(P_1 || P_2).
\label{eq: jumps of the martingale U that is related to the binary
hypothesis testing}
\end{equation}
Let
\begin{equation}
d_1 \triangleq \max_{x \in \mathcal{X}} \left| \ln
\frac{P_1(x)}{P_2(x)} - D(P_1 || P_2) \right| \label{eq:
d1}
\end{equation}
so $d_1 < \infty$ since by assumption the alphabet set
$\mathcal{X}$ is finite, and $P_1(x), P_2(x) > 0$ for every $x \in
\mathcal{X}$. From \eqref{eq: jumps of the martingale U that is
related to the binary hypothesis testing} and \eqref{eq: d1}
$$|U_k - U_{k-1}| \leq d_1$$ holds a.s. for every $k \in \{1, \ldots,
n\}$, and
\begin{eqnarray}
&& \expectation_{P_1^n} \bigl[ (U_k - U_{k-1})^2 \, | \,
\mathcal{F}_{k-1} \bigr] \nonumber \\
&& = \expectation_{P_1} \left[ \left( \ln \frac{P_1(X_k)}{P_2(X_k)}
- D(P_1 || P_2) \right)^2 \right] \nonumber \\
&& = \sum_{x \in \mathcal{X}} \left\{ P_1(x)
\left( \ln \frac{P_1(x)}{P_2(x)} - D(P_1 || P_2) \right)^2
\right\} \nonumber \\
&& \triangleq \sigma_1^2. \label{eq: sigma1 squared for the jumps
of the martingale U}
\end{eqnarray}

Let
\begin{eqnarray}
&& \hspace*{-1cm} \varepsilon_{1,1} = D(P_1 || P_2) -
\overline{\lambda}, \quad \varepsilon_{2,1} = D(P_2 || P_1) +
\underline{\lambda}
\label{eq: the epsilons introduced for mixed errors and erasures in binary hypothesis testing} \\
&& \hspace*{-1cm} \varepsilon_{1,2} = D(P_1 || P_2) -
\underline{\lambda}, \quad \varepsilon_{2,2} = D(P_2 || P_1) +
\overline{\lambda} \label{eq: the epsilons introduced for errors
in binary hypothesis testing}
\end{eqnarray}
The probability of making an erroneous decision on hypothesis
$H_2$ or declaring an erasure under the hypothesis $H_1$ is equal
to $\alpha_n^{(1)}$, and from Theorem~\ref{theorem: first refined
concentration inequality}
\begin{eqnarray}
&& \alpha_n^{(1)} \triangleq P_1^n \bigl( L(X_1, \ldots, X_n)
\leq n \overline{\lambda} \bigr) \nonumber \\
&& \hspace*{0.7cm} \stackrel{\text{(a)}}{=}
P_1^n(U_n - U_0 \leq -\varepsilon_{1,1} \, n)
\label{eq: intermediate step in the derivation of a bound on alpha1} \\
&& \hspace*{0.7cm} \stackrel{\text{(b)}}{\leq} \exp \left(-n \,
D\Bigl(\frac{\delta_{1,1} + \gamma_1}{1+\gamma_1} \Big|\Big|
\frac{\gamma_1}{1+\gamma_1} \Bigr) \right) \label{eq:
concentration inequality for the first error event}
\end{eqnarray}
where equality~(a) follows from \eqref{eq: initial value of the
martingale U that is related to the binary hypothesis testing},
\eqref{eq: final value of the martingale U that is related to
the binary hypothesis testing} and \eqref{eq: the epsilons
introduced for mixed errors and erasures in binary hypothesis
testing}, and inequality~(b) follows from
Theorem~\ref{theorem: first refined concentration inequality}
with
\begin{equation}
\gamma_1 \triangleq \frac{\sigma_1^2}{d_1^2}, \quad \delta_{1,1}
\triangleq \frac{\varepsilon_{1,1}}{d_1}. \label{eq: gamma1 and delta1,1}
\end{equation}
Note that if $\varepsilon_{1,1} > d_1$ then it follows from
\eqref{eq: jumps of the martingale U that is related to the binary
hypothesis testing} and \eqref{eq: d1} that $\alpha_n^{(1)}$ is zero;
in this case $\delta_{1,1} > 1$, so the divergence in \eqref{eq:
concentration inequality for the first error event} is infinity
and the upper bound is also equal to zero. Hence, it is assumed
without loss of generality that $\delta_{1,1} \in [0, 1]$.

Similarly to \eqref{eq: martingale sequence U under hypothesis H1},
under hypothesis~$H_2$, let us define the martingale sequence
$\{U_k, \mathcal{F}_k\}_{k=0}^n$ with the same filtration and
\begin{equation}
\hspace*{-0.4cm} U_k = \expectation_{P_2^n} \bigl[ L(X_1, \ldots, X_n)
\; | \; \mathcal{F}_k \bigr], \quad \forall \, k \in \{0, \ldots, n\}.
\label{eq: martingale sequence U  under hypothesis H2}
\end{equation}
For every $k \in \{0, \ldots, n\}$
\begin{eqnarray*}
&& U_k = \sum_{i=1}^k \ln \frac{P_1(X_i)}{P_2(X_i)} - (n-k) D(P_2 || P_1)
\end{eqnarray*}
and in particular
\begin{equation}
U_0 = -n D(P_2 || P_1), \quad U_n = L(X_1, \ldots, X_n).
\label{eq: initial and final values of the second martingale sequence}
\end{equation}
For every $k \in \{1, \ldots, n\}$,
\begin{equation}
U_k - U_{k-1} = \ln \frac{P_1(X_k)}{P_2(X_k)} + D(P_2 || P_1).
\label{eq: jumps of the martingale U under hypothesis H2}
\end{equation}
Let
\begin{equation}
d_2 \triangleq \max_{x \in \mathcal{X}} \left| \ln
\frac{P_2(x)}{P_1(x)} - D(P_2 || P_1) \right| \label{eq:
d2}
\end{equation}
then, the jumps of the latter martingale sequence are uniformly
bounded by $d_2$ and, similarly to \eqref{eq: sigma1 squared for
the jumps of the martingale U}, for every $k \in \{1, \ldots, n\}$
\begin{eqnarray}
&& \expectation_{P_2^n} \bigl[ (U_k - U_{k-1})^2 \, | \,
\mathcal{F}_{k-1} \bigr] \nonumber \\
&& = \sum_{x \in \mathcal{X}} \left\{ P_2(x) \left( \ln
\frac{P_2(x)}{P_1(x)} - D(P_2 || P_1) \right)^2 \right\} \nonumber
\\ && \triangleq \sigma_2^2. \label{eq: sigma2 squared for the
jumps of the martingale U}
\end{eqnarray}
Hence, it follows from Theorem~\ref{theorem: first refined
concentration inequality} that
\begin{eqnarray}
&& \beta_n^{(1)} \triangleq P_2^n \bigl( L(X_1, \ldots, X_n)
\geq n \underline{\lambda} \bigr) \nonumber \\
&& \hspace*{0.7cm} = P_2^n(U_n - U_0 \geq \varepsilon_{2,1} \, n)
\label{eq: intermediate step in the derivation of a bound on beta1} \\
&& \hspace*{0.7cm} \leq  \exp \left(-n \,
D\Bigl(\frac{\delta_{2,1} + \gamma_2}{1+\gamma_2} \Big|\Big|
\frac{\gamma_2}{1+\gamma_2} \Bigr) \right) \label{eq:
concentration inequality for the second error event}
\end{eqnarray}
where the equality in \eqref{eq: intermediate step in the
derivation of a bound on beta1} holds due to \eqref{eq:
initial and final values of the second martingale sequence}
and \eqref{eq: the epsilons introduced for mixed errors and
erasures in binary hypothesis testing}, and \eqref{eq:
concentration inequality for the second error event}
follows from Theorem~\ref{theorem: first refined
concentration inequality} with
\begin{equation}
\gamma_2 \triangleq \frac{\sigma_2^2}{d_2^2}, \quad \delta_{2,1}
\triangleq \frac{\varepsilon_{2,1}}{d_2}
\label{eq: gamma2 and delta2,1}
\end{equation}
and $d_2$, $\sigma_2$ are introduced, respectively, in \eqref{eq:
d2} and \eqref{eq: sigma2 squared for the jumps of the martingale
U}.

From \eqref{eq: overall probability of a mixed error and erasure
event}, \eqref{eq: concentration inequality for the first error
event} and \eqref{eq: concentration inequality for the second
error event}, the exponent of the probability of
either having an error or an erasure is lower bounded by
\begin{equation}
\lim_{n \rightarrow \infty} - \frac{\ln P_{\text{e}, n}^{(1)}}{n}
\geq \min_{i=1,2} D\Bigl(\frac{\delta_{i,1} + \gamma_i}{1+\gamma_i}
\Big|\Big| \frac{\gamma_i}{1+\gamma_i} \Bigr). \label{eq: lower
bound on the exponent of mixed errors and erasures
for binary hypothesis testing}
\end{equation}
Similarly to the above analysis, one gets from \eqref{eq: overall
error probability} and \eqref{eq: the epsilons introduced for
errors in binary hypothesis testing} that the error
exponent is lower bounded by
\begin{equation}
\lim_{n \rightarrow \infty} - \frac{\ln P_{\text{e}, n}^{(2)}}{n}
\geq \min_{i=1,2} D\Bigl(\frac{\delta_{i,2} + \gamma_i}{1+\gamma_i}
\Big|\Big| \frac{\gamma_i}{1+\gamma_i} \Bigr) \label{eq: lower
bound on the error exponent for binary hypothesis testing}
\end{equation}
where
\begin{equation}
\delta_{1,2} \triangleq \frac{\varepsilon_{1,2}}{d_1}, \quad
\delta_{2,2} \triangleq \frac{\varepsilon_{2,2}}{d_2}.
\label{eq: delta1,2 and delta2,2}
\end{equation}

For the case of a single threshold (i.e., $\overline{\lambda} =
\underline{\lambda} \triangleq \lambda$) then \eqref{eq: lower
bound on the exponent of mixed errors and erasures for
binary hypothesis testing} and \eqref{eq: lower bound on the
error exponent for binary hypothesis testing} coincide,
and one obtains that the error exponent satisfies
\begin{equation}
\lim_{n \rightarrow \infty} - \frac{\ln P_{\text{e}, n}}{n}
\geq \min_{i=1,2} D\Bigl(\frac{\delta_i + \gamma_i}{1+\gamma_i}
\Big|\Big| \frac{\gamma_i}{1+\gamma_i} \Bigr) \label{eq: lower
bound on the error exponent for binary hypothesis testing with
a single threshold}
\end{equation}
where $\delta_i$ is the common value of $\delta_{i,1}$ and
$\delta_{i,2}$ (for $i=1,2$). In this special case, the zero
threshold is optimal (see, e.g., \cite[p.~93]{Dembo_Zeitouni}),
which then yields that \eqref{eq: lower bound on the error
exponent for binary hypothesis testing with a single threshold}
is satisfied with
\begin{equation}
\delta_1 = \frac{D(P_1 || P_2)}{d_1}, \quad \delta_2 = \frac{D(P_2
|| P_1)}{d_2} \label{eq: delta1 and delta2}
\end{equation}
with $d_1$ and $d_2$ from \eqref{eq: d1} and
\eqref{eq: d2}, respectively. The right-hand side of
\eqref{eq: lower bound on the error exponent for binary
hypothesis testing with a single threshold} forms a lower
bound on Chernoff information which is the exact error
exponent for this special case.

\subsubsection{Comparison of the Lower Bounds on the Exponents with
those that Follow from Azuma's Inequality} The lower bounds on the
error exponent and the exponent of the probability of having either
errors or erasures, that were derived in the previous sub-section
via Theorem~\ref{theorem: first refined concentration inequality},
are compared in the following to the loosened lower bounds on these
exponents that follow from Azuma's inequality.

We first obtain upper bounds on $\alpha_n^{(1)}, \alpha_n^{(2)},
\beta_n^{(1)}$ and $\beta_n^{(2)}$ via Azuma's inequality, and
then use them to derive lower bounds on the exponents of
$P_{\text{e},n}^{(1)}$ and $P_{\text{e},n}^{(2)}$.

From \eqref{eq: jumps of the martingale U that is related to the
binary hypothesis testing}, \eqref{eq: d1}, \eqref{eq: intermediate
step in the derivation of a bound on alpha1}, \eqref{eq: gamma1
and delta1,1}, and Azuma's inequality
\begin{equation}
\alpha_n^{(1)} \leq \exp \biggl(-\frac{\delta_{1,1}^2 n}{2} \biggr)
\label{eq: Azuma's inequality for alpha1}
\end{equation}
and, similarly, from \eqref{eq: jumps of the martingale U under
hypothesis H2}, \eqref{eq: d2}, \eqref{eq: intermediate step
in the derivation of a bound on beta1}, \eqref{eq: gamma2 and
delta2,1}, and Azuma's inequality
\begin{equation}
\beta_n^{(1)} \leq \exp \biggl(-\frac{\delta_{2,1}^2 n}{2} \biggr).
\label{eq: Azuma's inequality for beta1}
\end{equation}
From \eqref{eq: error event under hypothesis H1}, \eqref{eq:
error event under hypothesis H2}, \eqref{eq: the epsilons
introduced for errors in binary hypothesis testing}, \eqref{eq:
delta1,2 and delta2,2} and Azuma's inequality
\begin{eqnarray}
&& \hspace*{-0.5cm} \alpha_n^{(2)} \leq \exp
\biggl(-\frac{\delta_{1,2}^2 n}{2} \biggr)
\label{eq: Azuma's inequality for alpha2} \\
&& \hspace*{-0.5cm} \beta_n^{(2)} \leq \exp
\biggl(-\frac{\delta_{2,2}^2 n}{2} \biggr). \label{eq: Azuma's
inequality for beta2}
\end{eqnarray}
Therefore, it follows from \eqref{eq: overall probability of a mixed error
and erasure event}, \eqref{eq: overall error probability} and
\eqref{eq: Azuma's inequality for alpha1}--\eqref{eq: Azuma's
inequality for beta2} that the resulting lower bounds on the
exponents of $P_{\text{e},n}^{(1)}$ and $P_{\text{e},n}^{(2)}$ are
\begin{equation}
\lim_{n \rightarrow \infty} - \frac{\ln P_{\text{e}, n}^{(j)}}{n}
\geq \min_{i=1,2} \frac{\delta_{i,j}^2}{2}, \quad j = 1, 2
\label{eq: lower bounds on the exponents for binary
hypothesis testing via Azuma inequality}
\end{equation}
as compared to \eqref{eq: lower bound on the exponent
of mixed errors and erasures for binary hypothesis testing} and
\eqref{eq: lower bound on the error exponent for binary
hypothesis testing} which give, for $j=1,2$,
\begin{equation}
\lim_{n \rightarrow \infty} - \frac{\ln P_{\text{e}, n}^{(j)}}{n}
\geq \min_{i=1,2} D\Bigl(\frac{\delta_{i,j} +
\gamma_i}{1+\gamma_i} \Big|\Big| \frac{\gamma_i}{1+\gamma_i}
\Bigr). \label{eq: lower bounds on the exponents for
binary hypothesis testing via Theorem 2}
\end{equation}
For the specific case of a zero threshold, the lower bound
on the error exponent which follows from Azuma's inequality
is given by
\begin{equation}
\lim_{n \rightarrow \infty} - \frac{\ln P_{\text{e}, n}^{(j)}}{n} \geq
\min_{i=1,2} \frac{\delta_i^2}{2}
\label{eq: loosened lower bound on the error exponent for zero threshold}
\end{equation}
with the values of $\delta_1$ and $\delta_2$
in \eqref{eq: delta1 and delta2}.

The lower bounds on the exponents in \eqref{eq: lower bounds on
the exponents for binary hypothesis testing via Azuma inequality}
and \eqref{eq: lower bounds on the exponents for binary hypothesis
testing via Theorem 2} are compared in the following. Note that
the lower bounds in \eqref{eq: lower bounds on the exponents for
binary hypothesis testing via Azuma inequality} are loosened as
compared to those in \eqref{eq: lower bounds on the exponents for
binary hypothesis testing via Theorem 2} since they follow,
respectively, from Azuma's inequality and its improvement in
Theorem~\ref{theorem: first refined concentration inequality}.

The divergence in the exponent of \eqref{eq: lower bounds on the
exponents for binary hypothesis testing via Theorem 2} is equal to

\small \vspace*{-0.2cm}
\begin{eqnarray}
&& \hspace*{-0.8cm} D\Bigl(\frac{\delta_{i,j} +
\gamma_i}{1+\gamma_i}
\Big|\Big| \frac{\gamma_i}{1+\gamma_i} \Bigr) \nonumber \\[0.1cm]
&& \hspace*{-0.8cm} =
\left(\frac{\delta_{i,j}+\gamma_i}{1+\gamma_i} \right) \ln \left(
1 + \frac{\delta_{i,j}}{\gamma_i} \right) +
\left(\frac{1-\delta_{i,j}}{1+\gamma_i}\right)
\ln(1-\delta_{i,j}) \nonumber \\[0.1cm]
&& \hspace*{-0.8cm} = \frac{\gamma_i}{1+\gamma_i} \left[ \left( 1
+ \frac{\delta_{i,j}}{\gamma_i} \right) \ln \Bigl( 1 +
\frac{\delta_{i,j}}{\gamma_i}
\Bigr) + \frac{(1-\delta_{i,j}) \ln(1-\delta_{i,j})}{\gamma_i}\right]. \nonumber \\
\label{eq: equality for the divergence}
\end{eqnarray}

\normalsize
\begin{lemma}
\begin{equation}
(1+u) \ln(1+u) \geq \left\{
\begin{array}{ll}
u + \frac{u^2}{2}, \quad & u \in [-1, 0] \\[0.2cm]
u+\frac{u^2}{2}-\frac{u^3}{6}, \quad & u \geq 0
\end{array}
\right. \label{eq: inequality for lower bounding the divergence}
\end{equation}
where at $u=-1$, the left-hand side is defined to be zero (it is
the limit of this function when $u \rightarrow -1$ from above).
\label{lemma: inequality for lower bounding the divergence}
\end{lemma}
\begin{proof}
The proof follows by elementary calculus. 
\end{proof}

Since $\delta_{i,j} \in [0,1]$, then \eqref{eq: equality
for the divergence} and Lemma~\ref{lemma: inequality for lower
bounding the divergence} imply that
\begin{equation}
D\Bigl(\frac{\delta_{i,j} + \gamma_i}{1+\gamma_i} \Big|\Big|
\frac{\gamma_i}{1+\gamma_i} \Bigr) \geq \frac{\delta_{i,j}^2}{2 \gamma_i} -
\frac{\delta_{i,j}^3}{6 \gamma_i^2 (1+\gamma_i)}.
\label{eq: lower bound on the lower bound of the exponents}
\end{equation}
Hence, by comparing \eqref{eq: lower bounds on the exponents for
binary hypothesis testing via Azuma inequality} with the
combination of \eqref{eq: lower bounds on the exponents for binary
hypothesis testing via Theorem 2} and \eqref{eq: lower bound on
the lower bound of the exponents}, then it follows that (up to a
second-order approximation) the lower bounds on the exponents that
were derived via Theorem~\ref{theorem: first refined concentration
inequality} are improved by at least a factor of $\bigl(\max
\gamma_i\bigr)^{-1}$ as compared to those that follow from Azuma's
inequality.

\begin{example}
Consider two probability measures $P_1$ and $P_2$ where
$$P_1(0) = P_2(1) = 0.4, \quad P_1(1) = P_2(0) = 0.6,$$
and the case of a single threshold of the log-likelihood ratio
that is set to zero (i.e., $\lambda = 0$). The exact error
exponent in this case is Chernoff information that is equal to
$$C(P_1, P_2) = 2.04 \cdot 10^{-2}.$$ The improved lower bound on
the error exponent in \eqref{eq: lower bound on the error exponent
for binary hypothesis testing with a single threshold} and
\eqref{eq: delta1 and delta2} is equal to $1.77 \cdot 10^{-2}$,
whereas the loosened lower bound in \eqref{eq: loosened lower
bound on the error exponent for zero threshold}
is equal to $1.39 \cdot 10^{-2}$. In this case
$\gamma_1 = \frac{2}{3}$ and $\gamma_2 = \frac{7}{9}$, so the
improvement in the lower bound on the error exponent is indeed by
a factor of approximately
$$\left(\max_i \gamma_i \right)^{-1} = \frac{9}{7}.$$
Note that, from \eqref{eq: concentration inequality for the first
error event}, \eqref{eq: concentration inequality for the second
error event} and \eqref{eq: Azuma's inequality for
alpha1}--\eqref{eq: Azuma's inequality for beta2}, these are lower
bounds on the error exponents for any finite block length
$n$, and not only asymptotically in the limit where $n \rightarrow \infty$.
The operational meaning of this example is that
the improved lower bound on the error exponent
assures that a fixed error probability can be obtained based
on a sequence of i.i.d. RVs whose length is reduced by 22.2\%
as compared to the loosened bound which follows from
Azuma's inequality.
\end{example}

\subsubsection{Comparison of the Exact and Lower Bounds on the
Error Exponents, Followed by a Relation to Fisher Information} In
the following, we compare the exact and lower bounds on the error
exponents. Consider the case where there is a single threshold on
the log-likelihood ratio (i.e., referring to the case where the
erasure option is not provided) that is set to zero. The exact
error exponent in this case is given by the Chernoff
information (see \eqref{eq: Chernoff information}), and it will
be compared to the two lower bounds on the error exponents that
were derived in the previous two subsections.

Let $\{P_{\theta}\}_{\theta \in \Theta}$, denote an indexed
family of probability mass functions where $\Theta$ denotes the
parameter set. Assume that $P_{\theta}$ is differentiable in the
parameter $\theta$. Then, the Fisher information is defined as
\begin{equation}
J(\theta) \triangleq \expectation_{\theta} \left[
\frac{\partial}{\partial \theta} \, \ln P_{\theta}(x) \right]^2
\label{eq: Fisher information}
\end{equation}
where the expectation is w.r.t. the probability mass function
$P_{\theta}$. The divergence and Fisher information are
two related information measures, satisfying the equality
\begin{equation}
\lim_{\theta' \rightarrow \theta} \frac{D(P_{\theta} ||
P_{\theta'})}{(\theta - \theta')^2} = \frac{J(\theta)}{2}
\label{eq: relation between the divergence and Fisher information}
\end{equation}
(note that if it was a relative entropy to base~2 then the
right-hand side of \eqref{eq: relation between the divergence and
Fisher information} would have been divided by $\ln 2$, and be
equal to $\frac{J(\theta)}{\ln 4}$ as in \cite[Eq.~(12.364)]{Cover
and Thomas}).
\begin{proposition}
Under the above assumptions,
\begin{itemize}
\item The Chernoff information and Fisher information are related
information measures that satisfy the equality
\begin{equation}
\lim_{\theta' \rightarrow \theta}
\frac{C(P_{\theta}, P_{\theta'})}{(\theta - \theta')^2} = \frac{J(\theta)}{8}.
\label{eq: relation between the Chernoff information and Fisher information}
\end{equation}
\item Let
\begin{equation}
E_{\text{L}}(P_{\theta}, P_{\theta'}) \triangleq
\min_{i=1,2} D\Bigl(\frac{\delta_i + \gamma_i}{1+\gamma_i}
\Big|\Big| \frac{\gamma_i}{1+\gamma_i} \Bigr)
\label{eq: E_L}
\end{equation}
be the lower bound on the error exponent in \eqref{eq: lower bound
on the error exponent for binary hypothesis testing with a single
threshold} which corresponds to $P_1 \triangleq P_{\theta}$ and
$P_2 \triangleq P_{\theta'}$, then also
\begin{equation}
\lim_{\theta' \rightarrow \theta} \frac{E_{\text{L}}(P_{\theta},
P_{\theta'})}{(\theta - \theta')^2} = \frac{J(\theta)}{8}. \label{eq:
relation between the improved lower bound on the error exponent
and Fisher information}
\end{equation}
\item Let
\begin{equation}
\widetilde{E}_{\text{L}}(P_{\theta}, P_{\theta'})
\triangleq \min_{i=1,2} \frac{\delta_i^2}{2}
\label{eq: tilde E_L}
\end{equation}
be the loosened lower bound on the error exponent in \eqref{eq:
loosened lower bound on the error exponent for zero threshold}
which refers to $P_1 \triangleq P_{\theta}$ and $P_2 \triangleq
P_{\theta'}$. Then,
\begin{equation}
\lim_{\theta' \rightarrow \theta} \frac{\widetilde{E}_{\text{L}}(P_{\theta},
P_{\theta'})}{(\theta - \theta')^2} = \frac{a(\theta) \, J(\theta)}{8}
\label{eq: relation between the loosened lower bound on the error
exponent and Fisher information}
\end{equation}
for some deterministic function $a$ bounded in $[0, 1]$,
and there exists an indexed family of probability mass
functions for which $a(\theta)$ can be made arbitrarily close
to zero for any fixed value of $\theta \in \Theta$.
\end{itemize}
\label{proposition: Fisher information}
\end{proposition}
\begin{proof}
See Appendix~\ref{appendix: Fisher information}.
\end{proof}

\vspace*{0.1cm} Proposition~\ref{proposition: Fisher information}
shows that, in the considered setting, the refined lower bound
on the error exponent provides the correct
behavior of the error exponent for a binary hypothesis testing
when the relative entropy between the pair of probability mass
functions that characterize the two hypotheses tends to zero.
This stays in contrast to the
loosened error exponent, which follows from Azuma's inequality,
whose scaling may differ significantly from the
correct exponent (for a concrete example, see the last part of the
proof in Appendix~\ref{appendix: Fisher information}).

\begin{example}
Consider the index family of of probability mass functions defined
over the binary alphabet $\mathcal{X} = \{0,1\}$:
$$ P_\theta(0) = 1-\theta, \; \; P_\theta(1) = \theta, \quad
\forall \, \theta \in (0,1).$$ From \eqref{eq: Fisher
information}, the Fisher information is equal to
$$ J(\theta) = \frac{1}{\theta} + \frac{1}{1-\theta}$$
and, at the point $\theta = 0.5$, $J(\theta) = 4$. Let $\theta_1 =
0.51$ and $\theta_2 = 0.49$, so from \eqref{eq: relation between
the Chernoff information and Fisher information} and \eqref{eq:
relation between the improved lower bound on the error exponent
and Fisher information}
$$ C(P_{\theta_1}, P_{\theta_2}), E_{\text{L}}(P_{\theta_1}, P_{\theta_2}) \approx
\frac{J(\theta) (\theta_1 - \theta_2)^2}{8} = 2.00 \cdot
10^{-4}.$$ Indeed, the exact values of $C(P_{\theta_1},
P_{\theta_2})$ and $E_{\text{L}}(P_{\theta_1}, P_{\theta_2})$ are
$2.000 \cdot 10^{-4}$ and $1.997 \cdot 10^{-4}$, respectively.
\end{example}

\subsubsection{Moderate Deviations Analysis for Binary Hypothesis Testing}
\label{subsubsection: MDP for binary hypothesis testing} So far,
we have discussed large deviations analysis for binary hypothesis
testing, and compared the exact error exponents with lower bounds
that follow from refined versions of Azuma's inequality.

Based on the asymptotic results in \eqref{eq: a.s. limit of the
normalized LLR under hypothesis H1} and \eqref{eq: a.s. limit of
the normalized LLR under hypothesis H2}, which hold a.s. under
hypotheses $H_1$ and $H_2$ respectively, the large deviations
analysis refers to upper and lower thresholds $\overline{\lambda}$
and $\underline{\lambda}$ which are {\em kept fixed} (i.e., these
thresholds do not depend on the block length $n$ of the data
sequence) where
$$ -D(P_2 || P_1) < \underline{\lambda} \leq  \overline{\lambda} <
D(P_1 || P_2).$$ Suppose that instead of having some fixed upper
and lower thresholds, one is interested to set these
thresholds such that as the block length $n$ tends to
infinity, they tend simultaneously to their asymptotic
limits in \eqref{eq: a.s. limit of the normalized LLR under
hypothesis H1} and \eqref{eq: a.s. limit of the normalized LLR
under hypothesis H2}, i.e.,
$$ \lim_{n \rightarrow \infty} \overline{\lambda}^{(n)} = D(P_1 ||
P_2), \quad \lim_{n \rightarrow \infty} \underline{\lambda}^{(n)}
= -D(P_2 || P_1).$$ Specifically, let $\eta \in (\frac{1}{2}, 1)$,
and $\varepsilon_1, \varepsilon_2 > 0$ be arbitrary fixed numbers,
and consider the case where one decides on hypothesis $H_1$ if
\begin{equation*}
L(X_1, \ldots, X_n) > n \overline{\lambda}^{(n)}
\end{equation*}
and on hypothesis $H_2$ if
\begin{equation*}
L(X_1, \ldots, X_n) < n \underline{\lambda}^{(n)}
\end{equation*}
where these upper and lower thresholds are set to
\begin{eqnarray*}
&& \overline{\lambda}^{(n)} = D(P_1 || P_2) - \varepsilon_1
n^{-(1-\eta)} \\
&& \underline{\lambda}^{(n)} = -D(P_2 || P_1) + \varepsilon_2
n^{-(1-\eta)}
\end{eqnarray*}
so that they approach, respectively, the relative entropies $D(P_1
|| P_2)$ and $-D(P_2 || P_1)$ in the asymptotic case where the block
length $n$ of the data sequence
tends to infinity. Accordingly, the conditional probabilities in
\eqref{eq: error and erasure event under hypothesis
H1}--\eqref{eq: error event under hypothesis H2} are modified so
that the fixed thresholds $\overline{\lambda}$ and
$\underline{\lambda}$ are replaced with the above block-length
dependent thresholds $\overline{\lambda}^{(n)}$ and
$\underline{\lambda}^{(n)}$, respectively. The moderate deviations
analysis for binary hypothesis testing studies the probability of
an error event and the probability of a joint error and erasure
event under the two hypotheses, and it studies the interplay
between each of these probabilities, the block length $n$, and the related
thresholds that tend asymptotically to the limits in \eqref{eq:
a.s. limit of the normalized LLR under hypothesis H1} and
\eqref{eq: a.s. limit of the normalized LLR under hypothesis H2}
when the block length tends to infinity.

Before proceeding to the moderate deviations analysis for binary
hypothesis testing, the related literature is reviewed shortly. As
was noted in \cite{AltugW_ISIT2010}, moderate deviations analysis
appears so far in the information theory literature only in two
recent works: Moderate deviations behavior of channel coding for
discrete memoryless channels was studied in
\cite{AltugW_ISIT2010}, with direct and converse results which
explicitly characterize the rate function of the moderate
deviations principle (MDP). In their considered analysis, the
authors of \cite{AltugW_ISIT2010} studied the interplay between
the probability of error, code rate and block length when the
communication takes place over discrete memoryless channels, having
the interest to figure out how the error probability of the
best code scales when simultaneously the block length tends to
infinity and the code rate approaches the channel capacity. The
novelty in the setup of their analysis was the consideration of
the scenario mentioned above, in contrast to the case where the
rate is kept fixed below capacity, and the study is reduced to a
characterization of the dependence between the two remaining
parameters (i.e., the block length $n$ and the average/ maximal
error probability of the best code). As opposed to the latter
case, which corresponds to large deviations analysis and implies a
characterization of error exponents as a function of the fixed
rate, the analysis made in \cite{AltugW_ISIT2010} (via the
introduction of direct and converse theorems) demonstrated a
sub-exponential scaling of the maximal error probability in the
considered moderate deviations regime. In another recent paper
\cite{He_IT09}, the moderate deviations analysis of the
Slepian-Wolf problem was studied, and to the best of our
knowledge, the authors of \cite{He_IT09} were the first to
consider moderate deviations analysis in the information theory
literature. In the probability literature, moderate deviations
analysis was extensively studied (see, e.g.,
\cite[Section~3.7]{Dembo_Zeitouni}), and in particular the MDP was
studied in \cite{Dembo_paper96} in the context of continuous-time
martingales with bounded jumps.

In light of the discussion in Section~\ref{subsection: MDP for
real-valued i.i.d. RVs} on the MDP for i.i.d. RVs and the
discussion of its relation to the concentration inequalities in
Section~\ref{section: Refined Versions of Azuma's Inequality} (see
Appendix~\ref{appendix: MDP}), and
also motivated by the two recent works in \cite{AltugW_ISIT2010} and
\cite{He_IT09}, we proceed to consider in the following moderate
deviations analysis for binary hypothesis testing. Our approach for
this kind of analysis relies on concentration inequalities for
martingales.

In the following, we analyze the probability of a joint error and
erasure event under hypothesis $H_1$, i.e., derive an upper bound
on $\alpha_n^{(1)}$ in \eqref{eq: error and erasure event under
hypothesis H1}. The same kind of analysis can be adapted easily
for the other probabilities in \eqref{eq: error event under
hypothesis H1}--\eqref{eq: error event under hypothesis H2}.
As mentioned earlier, let $\varepsilon_1 > 0$ and $\eta \in
(\frac{1}{2}, 1)$ be arbitrarily fixed numbers. Then, under
hypothesis~$H_1$, it follows that similarly to \eqref{eq:
intermediate step in the derivation of a bound on
alpha1}--\eqref{eq: gamma1 and delta1,1}
\begin{eqnarray}
&& P_1^n\bigl( L(X_1, \ldots, X_n) \leq n \overline{\lambda}^{(n)})
\nonumber \\
&& = P_1^n\bigl( L(X_1, \ldots, X_n) \leq n D(P_1 || P_2) -
\varepsilon_1 n^\eta \bigr) \nonumber \\
&& \leq \exp \left( -n D\biggl(\frac{\delta_1^{(\eta, n)} +
\gamma_1}{1+\gamma_1} \, \big|\big| \, \frac{\gamma_1}{1+\gamma_1}\biggr)
\right) \label{eq:  1st inequality for the moderate deviations
analysis of binary hypothesis testing}
\end{eqnarray}
where
\begin{equation}
\delta_1^{(\eta, n)} \triangleq \frac{\varepsilon_1
n^{-(1-\eta)}}{d_1}, \quad \gamma_1 \triangleq
\frac{\sigma_1^2}{d_1^2} \label{delta1 and gamma1 for the moderate
deviations analysis of binary hypothesis testing}
\end{equation}
with $d_1$ and $\sigma_1^2$ from \eqref{eq: d1} and \eqref{eq:
sigma1 squared for the jumps of the martingale U}. From \eqref{eq:
equality for the divergence}, \eqref{eq: inequality for lower
bounding the divergence} and \eqref{delta1 and gamma1 for the
moderate deviations analysis of binary hypothesis testing}, it
follows that
\begin{eqnarray*}
&& D\biggl(\frac{\delta_1^{(\eta, n)} + \gamma_1}{1+\gamma_1} \,
\big|\big| \, \frac{\gamma_1}{1+\gamma_1}\biggr) \\
&& = \frac{\gamma_1}{1+\gamma_1} \left[ \Bigl(1 +
\frac{\delta_1^{(\eta, n)}}{\gamma_1} \Bigr)
\ln \Bigl(1 + \frac{\delta_1^{(\eta, n)}}{\gamma_1}
\Bigr) \right. \\
&& \hspace*{1.5cm} \left. + \frac{\bigl(1-\delta_1^{(\eta,
n)}\bigr) \ln\bigl(1-\delta_1^{(\eta, n)}\bigr)}{\gamma_1} \right] \\
&& \geq \frac{\gamma_1}{1+\gamma_1} \left[
\biggl(\frac{\delta_1^{(\eta, n)}}{\gamma_1} +
\frac{\bigl(\delta_1^{(\eta, n)}\bigr)^2}{2 \gamma_1^2} -
\frac{\bigl(\delta_1^{(\eta, n)}\bigr)^3}{6 \gamma_1^3}
\biggr) \right. \\
&& \hspace*{1.5cm} \left. + \frac{1}{\gamma_1}
\biggl(-\delta_1^{(\eta, n)} + \frac{(\delta_1^{(\eta, n)})^2}{2}
\biggr) \right] \\
&& = \frac{\bigl(\delta_1^{(\eta, n)}\bigr)^2}{2 \gamma_1} -
\frac{\bigl(\delta_1^{(\eta, n)}\bigr)^3}{6 \gamma_1^2
(1+\gamma_1)} \\[0.1cm]
&& = \frac{\varepsilon_1^2 \, n^{-2(1-\eta)}}{2 \gamma_1 d_1^2}
\left( 1 - \frac{\varepsilon_1}{3 d_1 \gamma_1 (1+\gamma_1)}
\, \frac{1}{n^{1-\eta}} \right) \\[0.1cm]
&& = \frac{\varepsilon_1^2 \, n^{-2(1-\eta)}}{2 \sigma_1^2}
\left( 1 - \frac{\varepsilon_1 d_1}{3 \sigma_1^2 (1+\gamma_1)}
\, \frac{1}{n^{1-\eta}} \right)
\end{eqnarray*}
provided that $\delta_1^{(\eta, n)} < 1$ (which holds for $n \geq
n_0$ for some $n_0 \triangleq n_0(\eta, \varepsilon_1, d_1) \in
\naturals$ that is determined from \eqref{delta1 and gamma1 for
the moderate deviations analysis of binary hypothesis testing}).
By substituting this lower bound on the divergence into \eqref{eq:
1st inequality for the moderate deviations analysis of binary
hypothesis testing}, it follows that
\begin{eqnarray}
&& \hspace*{-1.5cm} P_1^n\bigl( L(X_1, \ldots, X_n) \leq n
D(P_1 || P_2) - \varepsilon_1 n^\eta \bigr) \nonumber \\
&& \hspace*{-1.5cm} \leq \exp \left(-\frac{\varepsilon_1^2 \,
n^{2\eta-1}}{2 \sigma_1^2} \left( 1 - \frac{\varepsilon_1 d_1}{3
\sigma_1^2 (1+\gamma_1)} \, \frac{1}{n^{1-\eta}} \right) \right)
\label{eq: 2nd inequality for the moderate deviations analysis of
binary hypothesis testing}
\end{eqnarray}
so this upper bound has a sub-exponential decay to zero.
In particular, in the limit where $n$ tends to infinity
\begin{eqnarray}
&& \hspace*{-0.5cm} \lim_{n \rightarrow \infty} n^{2\eta-1}
\ln \, P_1^n\bigl( L(X_1, \ldots, X_n) \leq n D(P_1 || P_2) -
\varepsilon_1 n^\eta \bigr) \nonumber \\
&& \hspace*{-0.5cm} \leq -\frac{\varepsilon_1^2}{2 \sigma_1^2}
\label{eq: 3rd inequality for the moderate deviations analysis of
binary hypothesis testing}
\end{eqnarray}
with $\sigma_1^2$ in \eqref{eq: sigma1 squared for the jumps of
the martingale U}, i.e.,
$$\sigma_1^2 \triangleq \sum_{x \in \mathcal{X}} \left\{ P_1(x)
\left( \ln \frac{P_1(x)}{P_2(x)} - D(P_1 || P_2) \right)^2
\right\}.$$

From the analysis in Section~\ref{subsection: MDP for real-valued
i.i.d. RVs} and Appendix~\ref{appendix: MDP}, the following things
hold:
\begin{itemize}
\item The inequality for the asymptotic limit in
\eqref{eq: 3rd inequality for the moderate deviations analysis of
binary hypothesis testing} holds in fact with equality.
\item The same asymptotic result also follows from
Theorem~\ref{theorem: second inequality} for every even-valued $m
\geq 2$ (instead of Theorem~\ref{theorem: first refined
concentration inequality}).
\end{itemize}
To verify these statements, consider the real-valued sequence of i.i.d. RVs
$$ Y_i \triangleq \ln \left( \frac{P_1(X_i)}{P_2(X_i)} \right) - D(P_1 || P_2),
\quad i=1, \dots, n $$ that, under hypothesis $H_1$, have zero
mean and variance $\sigma_1^2$. Since, by assumption, the sequence
$\{X_i\}_{i=1}^n$ are i.i.d., then
\begin{equation}
L(X_1, \ldots, X_n) - n D(P_1 || P_2) = \sum_{i=1}^n Y_i,
\label{eq: equality related to the LLR and the sequence Y}
\end{equation}
and it follows from the one-sided version of the MDP in \eqref{eq:
MDP for i.i.d. real-valued RVs} that indeed \eqref{eq: 3rd
inequality for the moderate deviations analysis of binary
hypothesis testing} holds with equality. Moreover,
Theorem~\ref{theorem: first refined concentration inequality}
provides, via the inequality in \eqref{eq: 2nd inequality for the
moderate deviations analysis of binary hypothesis testing}, a
finite-length result that enhances the asymptotic result for $n
\rightarrow \infty$. The second item above follows from the second
part of the analysis in Appendix~\ref{appendix: MDP} (i.e., the
part of analysis in this appendix that follows from
Theorem~\ref{theorem: second inequality}).

A completely similar analysis w.r.t. moderate deviations for
binary hypothesis testing can be also performed under hypothesis
$H_2$. Note that, in the considered setting of moderate deviations
analysis for binary hypothesis testing, the error probability has
a sub-exponential decay to zero that is similar to the scaling that
was obtained in \cite{AltugW_ISIT2010} by the moderate deviations
analysis for channel coding.

\subsubsection{Second-Order Analysis for Binary Hypothesis Testing}
\label{subsunsection: Second-Order Analysis for Binary Hypothesis Testing}

The moderate deviations analysis in the previous sub-section
refers to deviations that scale like $n^{\eta}$ for $\eta \in
(\frac{1}{2}, 1)$. Let us consider now the case of $\eta =
\frac{1}{2}$ which corresponds to small deviations. To this end,
refer to the real-valued sequence of i.i.d. RVs $\{Y_i\}_{i=1}^n$
with zero mean and variance $\sigma_1^2$ (under hypothesis $H_1$),
and define the partial sums $S_k = \sum_{i=1}^k Y_i$ for $k \in
\{1, \ldots, n\}$ with $S_0 = 0$. This implies that $\{S_k,
\mathcal{F}_k\}_{k=0}^n$ is a martingale-difference sequence. At
this point, it links the current discussion on binary hypothesis
testing to Section~\ref{subsection: relation between the
martingale CLT and Proposition 4.3} which refers to the relation
between the martingale CLT and Proposition~\ref{proposition: a
similar scaling of the concentration inequalities}. Specifically,
since from \eqref{eq: equality related to the LLR and the sequence
Y}, $$S_n - S_0 = L(X_1, \ldots, X_n) - n D(P_1 || P_2)$$ then
from the proof of Proposition~\ref{proposition: a similar scaling
of the concentration inequalities}, one gets an upper bound on the
probability $$ P_1^n\bigl( L(X_1, \ldots, X_n) \leq n D(P_1 || P_2)
- \varepsilon_1 \sqrt{n} \bigr)$$ for a finite block length $n$ (via
an analysis that is either related to Theorem~\ref{theorem: first refined
concentration inequality} or~\ref{theorem: second inequality})
which agrees with the asymptotic result
\begin{eqnarray}
&& \hspace*{-0.5cm} \lim_{n \rightarrow \infty}
\ln \, P_1^n\bigl( L(X_1, \ldots, X_n) \leq n D(P_1 || P_2) -
\varepsilon_1 \sqrt{n} \bigr) \nonumber \\
&& \hspace*{-0.5cm} = -\frac{\varepsilon_1^2}{2 \sigma_1^2}.
\end{eqnarray}

Referring to small deviations analysis and the CLT, it shows a
duality between these kind of results and recent works on
second-order analysis for channel coding (see \cite{Hayashi_IT09},
\cite{Polanskiy_Poor_Verdu_IT2010},
\cite{Polanskiy_Poor_Verdu_IT2011_channel_dispersion_GE_channels}
and \cite{PolanskiyPV_IT paper}, where the variance $\sigma_1^2$
in \eqref{eq: sigma1 squared for the jumps of the martingale U} is
replaced with the channel dispersion that is defined to be the
variance of the mutual information RV between the channel input
and output, and is a property of the communication channel
solely).

\subsection{Pairwise Error Probability for Linear Block Codes over
Binary-Input Output-Symmetric DMCs}

In this sub-section, the tightness of Theorems~\ref{theorem: first
refined concentration inequality} and~\ref{theorem: second
inequality} is studied by the derivation of upper bounds on the
pairwise error probability under maximum-likelihood (ML) decoding
when the transmission takes place over a discrete memoryless channel
(DMC).

Let $\mathcal{C}$ be a binary linear block code of block length
$n$, and assume that the codewords are a-priori equi-probable.
Consider the case where the communication takes place over a
binary-input output-symmetric DMC whose input alphabet is
$\mathcal{X} = \{0, 1\}$, and its output alphabet $\mathcal{Y}$ is
finite.

In the following, boldface letters denote vectors, regular letters
with sub-scripts denote individual elements of vectors, capital
letters represent RVs, and lower-case letters denote individual
realizations of the corresponding RVs. Let
$$ P_{{\bf{Y}} | {\bf{X}}}(\underline{y} | \underline{x}) =
\prod_{i=1}^n P_{Y|X} (y_i | x_i) $$ be the transition probability
of the DMC, where due to the symmetry assumption
$$P_{Y|X}(y|0) = P_{Y|X}(-y|1), \quad \forall \, y \in \mathcal{Y}.$$
It is also assumed in the following that $P_{Y|X}(y|x)> 0$ for
every $(x,y) \in \mathcal{X} \times \mathcal{Y}$. Due to the
linearity of the code and the symmetry of the DMC, the decoding
error probability is independent of the transmitted codeword, so
it is assumed without any loss of generality that the all-zero
codeword is transmitted. In the following, we consider the
pairwise error probability when the competitive codeword
$\underline{x} \in \mathcal{C}$ has a Hamming weight that is equal
to $h$, and denote it by $W_{\text{H}}(\underline{x}) = h$. Let
$P_{\bf{Y}}$ denote the probability distribution of the channel
output.

In order to derive upper bounds on the pairwise error probability,
let us define the following two hypotheses: \vspace*{0.1cm}
\begin{itemize}
\item $H_1: \; P_{\bf{Y}}(\underline{y}) =
\prod_{i=1}^n P_{Y|X} (y_i | 0), \quad
\forall \underline{y} \in \mathcal{Y}^n,$\\
\item $H_2: \; P_{\bf{Y}}(\underline{y}) =
\prod_{i=1}^n P_{Y|X} (y_i | x_i), \quad
\forall \underline{y} \in \mathcal{Y}^n$
\end{itemize}
which correspond, respectively, to the transmission of the
all-zero codeword and the competitive codeword $\underline{x} \in
\mathcal{C}$.

Under hypothesis $H_1$, the considered pairwise error event under
ML decoding occurs if and only if $$ \sum_{i=1}^n \ln \left(
\frac{P_{Y|X} (y_i | x_i)}{P_{Y|X} (y_i | 0)} \right) \geq 0.$$
Let $\{i_k\}_{k=1}^h$ be the $h$ indices of the coordinates of
$\underline{x}$ where $x_i = 1$, ordered such that $1 \leq i_1 <
\ldots < i_h \leq n$. Based on this notation, the log-likelihood
ratio satisfies the equality
\begin{equation}
\sum_{i=1}^n \ln \left( \frac{P_{Y|X}(y_i | x_i)}{P_{Y|X}(y_i |
0)} \right) = \sum_{m=1}^h \ln \left( \frac{P_{Y|X}(y_{i_m} |
1)}{P_{Y|X}(y_{i_m} | 0)} \right). \label{eq: equality for the
LLR}
\end{equation}

For the continuation of the analysis in this sub-section, let us
define the martingale sequence $\{U_k, \mathcal{F}_l\}_{k=0}^n$
with the filtration
\begin{eqnarray*}
&& \mathcal{F}_k = \sigma(Y_{i_1}, \ldots, Y_{i_k}), \quad k=1, \ldots, h \\
&& \mathcal{F}_0 = \{\emptyset, \Omega\}
\end{eqnarray*}
and, under hypothesis $H_1$, let
\begin{equation*}
U_k = \expectation \left[ \sum_{m=1}^h \ln \left( \frac{P_{Y|X}
(Y_{i_m} | 1)}{P_{Y|X} (Y_{i_m} | 0)} \right) \, \Big| \,
\mathcal{F}_k \right], \; \forall \, k \in \{0, 1, \ldots, h\}.
\end{equation*}
Since, under hypothesis $H_1$, the RVs $Y_{i_1}, \ldots, Y_{i_h}$
are statistically independent, then for $k \in \{0, 1, \dots, h\}$
\begin{eqnarray}
&& \hspace*{-0.9cm} U_k = \sum_{m=1}^k \ln \left( \frac{P_{Y|X}(Y_{i_m}
| 1)}{P_{Y|X} (Y_{i_m} | 0)} \right) \nonumber \\
&& + (h-k) \sum_{y \in \mathcal{Y}} P_{Y|X}(y|0) \ln \left(
\frac{P_{Y|X}(y | 1)}{P_{Y|X}(y | 0)} \right) \nonumber \\
&& \hspace*{-0.4cm} = \sum_{m=1}^k \ln \left(
\frac{P_{Y|X}(Y_{i_m}|1)}{P_{Y|X}(Y_{i_m}|0)} \right) \nonumber \\[.1cm]
&& - (h-k) \; D\bigl(P_{Y|X} (\cdot | 0) \, || \, P_{Y|X} (\cdot | 1)\bigr).
\label{eq: martingale sequence U}
\end{eqnarray}
Specifically
\begin{eqnarray}
&& U_0 = -h \, D\bigl(P_{Y|X} (\cdot | 0) \, || \, P_{Y|X} (\cdot | 1)\bigr)
\label{eq: U_0} \\
&& U_h = \sum_{i=1}^n \ln \left( \frac{P_{Y|X}(Y_i | x_i)}{P_{Y|X}
(Y_i | 0)} \right) \label{eq: U_h}
\end{eqnarray}
where the last equality follows from \eqref{eq: equality for the
LLR} and \eqref{eq: martingale sequence U}, and the differences of
the martingale sequence are given by
\begin{eqnarray}
&& \hspace*{-1.5cm} \xi_k \triangleq U_k - U_{k-1} \nonumber \\
&& \hspace*{-1.1cm} = \ln \left( \frac{P_{Y|X}(Y_{i_k} |
1)}{P_{Y|X}(Y_{i_k} | 0)} \right) + D\bigl(P_{Y|X} (\cdot | 0) \, || \,
P_{Y|X} (\cdot | 1)\bigr) \label{eq: differences of the martingale U}
\end{eqnarray}
for every $k \in \{1, \ldots, h\}$. Note that, under hypothesis
$H_1$, indeed $\expectation[\xi_k | \mathcal{F}_{k-1}]=0.$

The probability of a pairwise error event, where the ML decoder
prefers a competitive codeword $\underline{x} \in \mathcal{C}$
($W_{\text{H}}(\underline{x}) = h$) over the transmitted all-zero
codeword, is equal to
\begin{eqnarray}
&& \hspace*{-1.5cm} P_h \triangleq \pr(U_h > 0 \; | \; H_1) \nonumber \\
&& \hspace*{-1.0cm} = \pr\Bigl(U_h - U_0 > h \, D\bigl(P_{Y|X}(\cdot |
0) \, || \, P_{Y|X}(\cdot | 1)\bigr) \; | \; H_1 \Bigr). \label{eq:
pairwise error probability}
\end{eqnarray}
It therefore follows that a.s. for every $k \in \{1, \ldots, h\}$
\begin{eqnarray}
&& \hspace*{-0.8cm} |\xi_k| \leq \max_{y \in \mathcal{Y}} \,
\biggl| \, \ln \left( \frac{P_{Y|X}(y|1)}{P_{Y|X}(y|0)} \right) \biggr|
+ D\bigl(P_{Y|X} (\cdot | 0) \, || \, P_{Y|X} (\cdot | 1)\bigr) \nonumber \\
&& \hspace*{-0.1cm} \triangleq d < \infty \label{eq: upper bound
on the jumps of U}
\end{eqnarray}
which is indeed finite since, by assumption, the alphabet
$\mathcal{Y}$ is finite and $P_{Y|X}(y|x) > 0$ for every $(x,y)
\in \mathcal{X} \times \mathcal{Y}$. Note that, in fact, taking an
absolute value in the maximization of the logarithm on the
right-hand side of \eqref{eq: upper bound on the jumps of U} is
redundant due to the channel symmetry, and also due to the
equality $\sum_y P_{Y|X}(y|0) = \sum_y P_{Y|X}(y|1) = 1$ (so that
it follows, from this equality, that there exists an element $y
\in \mathcal{Y}$ such that $P_{Y|X}(y|1) \geq P_{Y|X}(y|0))$.

As an interim conclusion, $\{U_k, \mathcal{F}_k\}_{k=0}^h$ is a
martingale sequence with bounded jumps, and $|U_k - U_{k-1}| \leq
d$ holds a.s. for every $k \in \{1, \ldots, h\}$. We rely in the
following on the concentration inequalities of
Theorems~\ref{theorem: first refined concentration inequality}
and~\ref{theorem: second inequality} to obtain, via \eqref{eq:
differences of the martingale U}--\eqref{eq: upper bound on the
jumps of U}, upper bounds on the pairwise error probability. The
tightness of these bounds will be examined numerically, and they
will be compared to the Bhattacharyya upper bound.

\subsubsection{Analysis Related to
Theorem~\ref{theorem: first refined concentration inequality}}
From \eqref{eq: differences of the martingale U}, for every $k \in
\{1, \ldots, h\}$
\begin{eqnarray}
&& \hspace*{-2cm} \expectation[\xi_k^2 | \mathcal{F}_{k-1}] \nonumber \\
&& \hspace*{-2cm} = \sum_{y \in \mathcal{Y}} P_{Y|X}(y|0)
\biggl[ \ln \left( \frac{P_{Y|X}(y|1)}{P_{Y|X}(y|0)}
\right) \nonumber \\
&& \hspace*{0.5cm} + D\bigl(P_{Y|X} (\cdot | 0) \, || \,
P_{Y|X} (\cdot | 1)\bigr) \biggr]^2 \nonumber \\
&& \hspace*{-2cm} = \sum_{y \in \mathcal{Y}} P_{Y|X}(y|0)
\biggl[ \ln \left( \frac{P_{Y|X}(y|1)}{P_{Y|X}(y|0)}
\right) \biggr]^2 \nonumber \\
&& \hspace*{-1.7cm} - \Bigl[D\bigl(P_{Y|X} (\cdot | 0) \, || \,
P_{Y|X} (\cdot | 1)\bigr) \Bigr]^2 \triangleq \sigma^2
\label{eq: conditional variance of U}
\end{eqnarray}
holds a.s., where the last equality follows from the definition of
the divergence (relative entropy). Based on \eqref{eq: pairwise
error probability} and the notation in \eqref{eq: notation}, let
\begin{equation}
\gamma = \frac{\sigma^2}{d^2}, \quad \delta \triangleq
\frac{D\bigl(P_{Y|X} (\cdot | 0) \, || \, P_{Y|X} (\cdot | 1)\bigr)}{d}
\label{eq: gamma and delta in the pairwise error probability context}
\end{equation}
where $d$ and $\sigma^2$ are introduced in \eqref{eq: upper bound
on the jumps of U} and \eqref{eq: conditional variance of U},
respectively. Under hypothesis~$H_1$, one gets from \eqref{eq:
pairwise error probability} and Theorem~\ref{theorem: first
refined concentration inequality} that the pairwise error
probability satisfies the upper bound
\begin{equation}
P_h \leq Z_1^h \label{eq: exponential bound on the pairwise error
probability from Theorem 2}
\end{equation}
where
\begin{equation}
Z_1 \triangleq \exp\left( -D\Bigl(\frac{\delta+\gamma}{1+\gamma}
\, \big|\big| \, \frac{\gamma}{1+\gamma}  \Bigr) \right)
\label{eq: base of the exponential bound on the pairwise error
probability from Theorem 2}
\end{equation}
and $\gamma$, $\delta$ are introduced in \eqref{eq: gamma and
delta in the pairwise error probability context}.

In the following, we compare the exponential bound in \eqref{eq:
exponential bound on the pairwise error probability from Theorem
2} with the Bhattacharyya bound
\begin{equation}
P_h \leq Z_{\text{B}}^h \label{eq: Bhattacharyya bound on the
pairwise error probability from Theorem 2}
\end{equation}
where the Bhattacharyya parameter $Z_{\text{B}}$ of the
binary-input DMC is given by
\begin{equation}
Z_{\text{B}} \triangleq \sum_{y \in \mathcal{Y}}
\sqrt{P_{Y|X}(y|0) P_{Y|X}(y|1)} \, .
\label{eq: Bhattacharyya parameter of a binary-input DMC}
\end{equation}

\begin{example}
Consider a binary symmetric channel (BSC) with crossover
probability $p$. The Bhattacharyya parameter which corresponds to
this channel is $Z_{\text{B}} = \sqrt{4p(1-p)}$. In the following,
$Z_1$ from \eqref{eq: base of the exponential bound on the
pairwise error probability from Theorem 2} is calculated for
comparison. Without loss of generality, assume that $p \leq
\frac{1}{2}$. Straightforward calculation shows that
\begin{eqnarray*}
&& d = 2(1-p) \ln \Bigl( \frac{1-p}{p} \Bigr)   \\
&& \sigma^2 = 4p(1-p) \left[ \ln \Bigl( \frac{1-p}{p} \Bigr) \right]^2     \\
&& D\bigl(P_{Y|X} (\cdot | 0) \, || \, P_{Y|X} (\cdot | 1)\bigr)
= (1-2p) \ln \Bigl( \frac{1-p}{p} \Bigr)
\end{eqnarray*}
and therefore \eqref{eq: gamma and delta in the pairwise error
probability context} gives that
\begin{equation*}
\gamma = \frac{p}{1-p}, \quad \delta = \frac{1-2p}{2(1-p)} \; .
\end{equation*}
Substituting $\gamma$ and $\delta$ into \eqref{eq: base of the
exponential bound on the pairwise error probability from Theorem
2} gives that the base of the exponential bound in \eqref{eq:
exponential bound on the pairwise error probability from Theorem
2} is equal to
\begin{equation*}
Z_1 = \exp\left( -D\Bigl(\frac{1}{2} \, \big|\big| \, p \Bigr) \right)
= \sqrt{4p(1-p)}
\end{equation*}
which coincides with the Bhattacharyya parameter for the BSC. This
shows that, for the BSC, Theorem~\ref{theorem: first refined
concentration inequality} implies the Bhattacharyya upper bound on
the pairwise error probability.  \label{example: BSC}
\end{example}

\vspace*{0.1cm} In general, it is observed numerically that $Z_1
\geq Z_{\text{B}}$ for binary-input output-symmetric DMCs with an
equality for the BSC (this will be exemplified after introducing
the bound on the pairwise error probability which follows from
Theorem~\ref{theorem: second inequality}). This implies that
Theorem~\ref{theorem: first refined concentration inequality}
yields in general a looser bound than the Bhattacharyya upper
bound in the context of the pairwise error probability for DMCs.

\subsubsection{Analysis Related to Theorem~\ref{theorem: second inequality}}
In the following, a parallel upper bound on the pairwise error
probability is derived from Remark~\ref{remark: on the one-sided
inequality version of Theorem 4} on Theorem~\ref{theorem: second
inequality}, and the martingale sequence $\{U_k,
\mathcal{F}_k\}_{k=0}^h$. Under hypothesis $H_1$ (i.e., the
assumption that the all-zero codeword is transmitted), \eqref{eq:
differences of the martingale U} implies that the conditional
expectation of $(U_k-U_{k-1})^l$ given $\mathcal{F}_{k-1}$ is
equal (a.s.) to the un-conditional expectation where $l$ is an
arbitrary natural number. Also, it follows from \eqref{eq:
differences of the martingale U} that for every $k \in \{1,
\ldots, h\}$ and $l \in \naturals$

\vspace*{-0.2cm}
\small
\begin{eqnarray*}
&& \hspace*{-0.7cm} \expectation[(U_k - U_{k-1})^l \, | \,
\mathcal{F}_{k-1}] \nonumber\\
&& \hspace*{-0.7cm} = (-1)^l \expectation\left[ \left( \ln \left(
\frac{P_{Y|X}(Y|0)}{P_{Y|X}(Y|1)} \right) - D\bigl(P_{Y|X} (\cdot
|0) \, || \, P_{Y|X} (\cdot |1)\bigr) \right)^l \right]
\end{eqnarray*}
\normalsize and, from the requirement that the sequence
$\{\mu_l\}$ be non-negative, (then, based on Remark~\ref{remark:
on the one-sided inequality version of Theorem 4}) let

\vspace*{-0.2cm}
\small
\begin{eqnarray}
&& \hspace*{-2cm} \mu_l \triangleq \max \left\{0, (-1)^l \expectation
\Biggl[ \biggl( \ln \left( \frac{P_{Y|X}(Y |
0)}{P_{Y|X}(Y | 1)} \right) \right. \nonumber \\
&& \hspace*{0.7cm} \left. \; - D\bigl(P_{Y|X} (\cdot | 0) \, || \,
P_{Y|X} (\cdot | 1)\bigr) \biggr)^l \Biggr] \right\}.
\end{eqnarray}
\normalsize for every $l \in \naturals$ (for even-valued $l$,
there is no need to take the maximization with zero). Based on the
notation used in the context of Remark~\ref{remark: on the
one-sided inequality version of Theorem 4}, let
$$ \gamma_l \triangleq \frac{\mu_l}{d^l}, \quad l = 2, 3, \ldots $$
and $\delta$ be the same parameter as in \eqref{eq: gamma and
delta in the pairwise error probability context}. Note that the
equality $\gamma_2 = \gamma$ holds for the parameter $\gamma$ in \eqref{eq: gamma
and delta in the pairwise error probability context}. Then,
Remark~\ref{remark: on the one-sided inequality version of Theorem
4} on Theorem~\ref{theorem: second inequality} yields that
for every even-valued $m \geq 2$
\begin{equation}
P_h \leq \bigl(Z_2^{(m)}\bigr)^h \label{eq: exponential bound on
the pairwise error probability from Theorem 4}
\end{equation}
where

\vspace*{-0.2cm}
\small
\begin{equation*}
Z_2^{(m)} \triangleq \inf_{x \geq 0} \left\{ e^{-\delta x} \left[
1 + \sum_{l=2}^{m-1} \frac{(\gamma_l - \gamma_m)x^l}{l!} +
\gamma_m (e^x-1-x)  \right]  \right\}.
\end{equation*}
\normalsize

\vspace*{0.1cm}
\begin{example}
In the following example, the bases of the two exponential bounds
on the pairwise error probability in \eqref{eq: exponential bound
on the pairwise error probability from Theorem 2} and \eqref{eq:
exponential bound on the pairwise error probability from Theorem
4} are compared to the corresponding Bhattachryya parameter (see
\eqref{eq: Bhattacharyya parameter of a binary-input DMC}) for
some binary-input output-symmetric DMCs.

For a integer-valued $Q \geq 2$, let $P_{Y|X}^{(Q)}$ be a
binary-input output-symmetric DMC with input alphabet $\mathcal{X}
= \{0, 1\}$ and output alphabet $\mathcal{Y} = \{0, 1, \ldots,
Q-1\}$, characterized by the following probability transitions:
\begin{eqnarray}
&& \hspace*{-0.7cm} P_{Y|X}^{(Q)}(0|0) =
P_{Y|X}^{(Q)}(Q-1|1) = 1-(Q-1)p, \nonumber \\
&& \hspace*{-0.7cm} P_{Y|X}^{(Q)}(1|0) =
\ldots = P_{Y|X}^{(Q)}(Q-1|0) = p \nonumber \\
&& \hspace*{-0.7cm} P_{Y|X}^{(Q)}(0|1) =
\ldots = P_{Y|X}^{(Q)}(Q-2|1) = p
\label{eq: probability transitions of the considered DMCs}
\end{eqnarray}
where $0 < p < \frac{1}{Q-1}$. The considered exponential bounds
are exemplified in the following for the case where $p = 0.04$ and
$Q = 2, 3, 4, 5, 10$. The bases of the exponential bounds in
\eqref{eq: exponential bound on the pairwise error probability
from Theorem 2} and \eqref{eq: exponential bound on the pairwise
error probability from Theorem 4} are compared in
Table~\ref{table1} to the corresponding Bhattachryya parameters of
these five DMCs that, from \eqref{eq: Bhattacharyya parameter of a binary-input DMC},
is equal to
$$Z_{\text{B}} = 2 \sqrt{p \, \bigl[1-(Q-1)p\bigr]} + (Q-2)p.$$

\begin{table}
\begin{center}
\caption{ The bases of the exponential bounds $Z_1$ and
$Z_2^{(m)}$ in \eqref{eq: exponential bound on the pairwise error
probability from Theorem 2} and \eqref{eq: exponential bound on
the pairwise error probability from Theorem 4} (for an even-valued
$m \geq 2$), respectively. The bases of these exponential bounds
are compared to the Bhattachryya parameter $Z_{\text{B}}$
in \eqref{eq: Bhattacharyya parameter of a binary-input DMC} for
the five DMC channels in \eqref{eq: probability transitions of the
considered DMCs} with $p = 0.04$ and $|\mathcal{Y}| = Q = 2, 3, 4, 5, 10$.}
\label{table1} \centering
\begin{tabular}{|c|c|c|c|c|c|} \hline
$Q$ & 2\raisebox{-1.3ex}{\rule{0pt}{4ex}} & 3 & 4 & 5 & 10 \\[0.1cm] \hline \hline
$Z_{\text{B}}$ & 0.3919\raisebox{-1.3ex}{\rule{0pt}{4ex}} & 0.4237
& 0.4552 & 0.4866 & 0.6400 \\[0.1cm] \hline
$Z_1$ & 0.3919\raisebox{-1.3ex}{\rule{0pt}{4ex}} & 0.4424 & 0.4879 & 0.5297 & 0.7012
\\[0.1cm] \hline
$Z_2^{(2)}$ & 0.3967 & 0.4484 & 0.4950 & 0.5377 & 0.7102 \\[0.1cm]
$Z_2^{(4)}$ & 0.3919 & 0.4247 & 0.4570 & 0.4877 & 0.6421 \\[0.1cm]
$Z_2^{(6)}$ & 0.3919 & 0.4237 & 0.4553 & 0.4867 & 0.6400 \\[0.1cm]
$Z_2^{(8)}$ & 0.3919 & 0.4237 & 0.4552 & 0.4866 & 0.6400 \\[0.1cm]
$Z_2^{(10)}$ & 0.3919 & 0.4237 & 0.4552 & 0.4866 & 0.6400 \\[0.1cm]
\hline
\end{tabular}
\end{center}
\end{table}
As is shown in Table~\ref{table1}, the choice of $m=2$ gives the
worst upper bound in Theorem~\ref{theorem: second inequality}
(since $Z_2^{(2)} \geq Z_2^{(m)}$ for every even-valued $m \geq
2$). This is consistent with Corollary~\ref{corollary: 3rd
corollary}. Moreover, the comparison of the third and forth lines
in Theorem~\ref{theorem: second inequality} is consistent with
Proposition~\ref{proposition: Theorem 2 gives a stronger result
than Corollary 4} which indeed assures that Theorem~\ref{theorem:
second inequality} with $m=2$ is looser than Theorem~\ref{theorem:
first refined concentration inequality} (hence, indeed $Z_1 <
Z_2^{(2)}$ for the considered DMCs). Also, from
Example~\ref{example: BSC}, it follows that Theorem~\ref{theorem:
first refined concentration inequality} coincides with the
Battacharyya bound (hence, $Z_1 = Z_{\text{B}}$ for the special
case where $Q=2$, as is indeed verified numerically in
Table~\ref{table1}). It is interesting to realize from
Table~\ref{table1} that for the five considered DMCs, the sequence
$\{Z_2^{(2)}, Z_2^{(4)}, Z_2^{(6)}, \ldots \}$ converges very
fast, and the limit is equal to the Bhattacharyya parameter for
all the examined cases. This stays in contrast to the exponential
base $Z_1$ that was derived from Theorem~\ref{theorem: first
refined concentration inequality}, and which appears to be
strictly larger than the corresponding Bhattacharyya parameter of
the DMC (except for the BSC, where the equality $Z_1 =
Z_{\text{B}}$ holds, as is shown in Example~\ref{example: BSC}).
\label{example: Z1, Z2 vs. ZB for several DMCs}
\end{example}

\vspace{0.1cm} Example~\ref{example: Z1, Z2 vs. ZB for several
DMCs} leads to the following conjecture:
\begin{conjecture}
For the martingale sequence $\{U_k, \mathcal{F}_k\}_{k=0}^h$
introduced in this sub-section,
$$ \lim_{m \rightarrow \infty} Z_2^{(m)} = Z_{\text{B}} $$
and this convergence is quadratic.
\end{conjecture}

\vspace{0.1cm}
\begin{example}
The base $Z_2^{(m)}$ of the exponential bound in \eqref{eq:
exponential bound on the pairwise error probability from Theorem
4} involves an operation of taking an infimum over the interval
$[0, \infty)$. This operation is performed numerically in general,
except for the special case where $m=2$ for which a closed-form
solution exists (see Appendix~\ref{appendix: proof of the
corollary that specializes the second inequality for m equal to 2}
for the proof of Corollary~\ref{corollary: specialization of the
second inequality}).

\begin{table}
\caption{The base $\widetilde{Z}_2^{(m)}$ of the exponential bound in
\eqref{eq: exponential bound on the pairwise error probability
from Theorem 2} and its (tight) upper bound
$\widetilde{Z}_2^{(m)}$ that follows by replacing the infimum
operation by the sub-optimal value in \eqref{eq: sub-optimal x}
and \eqref{eq: a, b, c for the sub-optimal x}. The five DMCs are
the same as in \eqref{eq: probability transitions of the
considered DMCs} and Table~\ref{table1}.} \label{table2} \centering
\begin{tabular}{|c|c|c|c|c|c|} \hline
$Q$ & 2\raisebox{-1.3ex}{\rule{0pt}{4ex}} & 3
& 4 & 5 & 10 \\[0.1cm] \hline
$Z_2^{(10)}$\raisebox{-1.3ex}{\rule{0pt}{4ex}} & 0.3919
& 0.4237 & 0.4552 & 0.4866 & 0.6400 \\[0.1cm]
\hline $\widetilde{Z}_2^{(10)}$\raisebox{-1.3ex}{\rule{0pt}{4ex}}
& 0.3919 & 0.4237 & 0.4553 & 0.4868 & 0.6417 \\[0.1cm] \hline
\end{tabular}
\end{table}
Replacing the infimum over $x \in [0, \infty)$ with the
sub-optimal value of $x$ in \eqref{eq: sub-optimal x} and
\eqref{eq: a, b, c for the sub-optimal x} gives an upper bound on
the respective exponential base of the bound (note that due to the
analysis, this sub-optimal value turns to be optimal in the
special case where $m=2$). The upper bound on $Z_2^{(m)}$ which
follows by replacing the infimum with the sub-optimal value in
\eqref{eq: sub-optimal x} and \eqref{eq: a, b, c for the
sub-optimal x} is denoted by $\widetilde{Z}_2^{(m)}$, and the
difference between the two values is marginal (see
Table~\ref{table2}).
\end{example}

\subsection{Concentration of the Crest-Factor for OFDM Signals}
\label{subsection: Concentration of the Crest-Factor for OFDM Signals}

Orthogonal-frequency-division-multiplexing (OFDM) converts a
high-rate data stream into a number of low-rate steams that are
transmitted over parallel narrow-band channels. This modulation is
used in several international standards related to digital
audio broadcasting, digital video broadcasting, and wireless local
area networks. For a textbook that provides a survey on OFDM, see
e.g. \cite[Chapter~19]{Molisch_book}. One of the problems of OFDM
is that the peak amplitude of the signal can be significantly
higher than the average amplitude. In the following, we consider
the concentration issue of the crest-factor (CF) of OFDM signals.

Given an $n$-length codeword $\{X_i\}_{i=0}^{n-1}$, a single OFDM
baseband symbol is described by \vspace*{-0.2cm}
\begin{equation}
s(t) = \frac{1}{\sqrt{n}} \sum_{i=0}^{n-1} X_i \exp\Bigl(\frac{j
\, 2\pi i t}{T}\Bigr), \quad 0 \leq t \leq T. \label{eq: OFDM
signal}
\end{equation}
Lets assume that $X_0, \ldots, X_{n-1}$ are i.i.d. complex RVs
with $|X_i|=1$. Since the sub-carriers are orthonormal over
$[0,T]$, then
\vspace*{-0.2cm}
\begin{equation}
\frac{1}{T} \int_0^T |s(t)|^2 dt = 1.  \label{eq: energy of OFDM
signal}
\end{equation}
The CF of the signal $s$, whose average power over the interval
$[0,T]$ is~1, is defined as
\vspace*{-0.2cm}
\begin{equation}
\text{CF}(s) \triangleq \max_{0 \leq t \leq T} | s(t) |.
\label{eq: CF}
\end{equation}
From \cite[Section~4]{SalemZ} and \cite{WunderB_IT}, it follows
that the CF scales with high probability like $\sqrt{\ln n}$
for large $n$. In \cite[Theorem~3 and Corollary~5]{LitsynW06},
a concentration inequality was derived for the CF of OFDM signals.
It states that for an arbitrary $\gamma \geq 2.5$
\begin{equation*} \pr \biggl(\Bigl| \text{CF}(s) -
\sqrt{\ln n} \Bigr| < \frac{\gamma \ln \ln n}{\sqrt{\ln n}} \biggr)
= 1 - O\Biggl( \frac{1}{\bigl(\ln n \bigr)^4} \Biggr).
\end{equation*}

\begin{remark}
The analysis used to derive this rather strong concentration
inequality (see \cite[Appendix~C]{LitsynW06}) requires some
assumptions on the distribution of the $X_i$'s (see the two
conditions in \cite[Theorem~3]{LitsynW06} followed by
\cite[Corollary~5]{LitsynW06}). These requirements are not needed
in the following analysis, and the derivation of concentration
inequalities via the martingale-based approach is simple, though
it leads here to a weaker concentration result than in
\cite[Theorem~3]{LitsynW06}. The emphasis here is on the approach
of using Azuma's inequality and some of its refined versions, and
applying these probabilistic tools in the context of OFDM signals.
\end{remark}

\subsubsection{Establishing Concentration of the Crest-Factor via Azuma's
Inequality} In the following, Azuma's inequality is used to derive
another concentration result. Let us define
\begin{equation}
Y_i = \expectation[ \, \text{CF}(s) \, | \, X_0, \ldots, X_{i-1}],
\quad i =0, \ldots, n  \label{eq: martingale sequence for OFDM}
\end{equation}
Based on Remarks~\ref{remark: construction of martingales}
and~\ref{remark: construction of martingales (cont.)}, this
sequence forms indeed a martingale sequence where the associated
filtration of the $\sigma$-algebras $\mathcal{F}_0 \subseteq
\mathcal{F}_1 \subseteq \ldots \subseteq \mathcal{F}_n$ is defined
so that $\mathcal{F}_i$ (for $i=0,1,\ldots,n$) is the
$\sigma$-algebra that is generated by all the first $i$
coordinates $(X_0, \ldots, X_{i-1})$ in \eqref{eq: OFDM signal}.
Moreover, it is a martingale sequence with bounded jumps,
where $$|Y_i - Y_{i-1}| \leq \frac{2}{\sqrt{n}} $$ for $i \in \{1,
\ldots, n\}$, since $\text{CF}(s)$ is defined as in \eqref{eq:
CF}, and therefore revealing the additional $i$-th coordinate
$X_i$ affects the CF by at most $\frac{2}{\sqrt{n}}$ (see first
part of the proof in Appendix~\ref{appendix: OFDM}). Hence, one
obtains from Azuma's inequality that
\begin{eqnarray}
&& \pr( | \text{CF}(s) - \expectation [ \text{CF}(s)] | \geq
\alpha ) \nonumber \\ && \leq 2 \exp\left(-\frac{\alpha^2}{2
\sum_{k=1}^n
\bigl( \frac{2}{\sqrt{n}} \bigr)^2 }\right) \nonumber \\
&& = 2 \exp\Bigl(-\frac{\alpha^2}{8}\Bigr), \quad \forall \,
\alpha > 0 \label{eq: Azuma's inequality for OFDM}
\end{eqnarray}
which demonstrates the concentration of this measure
around its expected value.

\bigskip
\subsubsection{Establishing (via
Proposition~\ref{proposition: a similar scaling of the
concentration inequalities}) an Improved Concentration Inequality
for OFDM Signals with an M-ary PSK Constellation} In the
following, we rely on Proposition~\ref{proposition: a similar
scaling of the concentration inequalities} to derive an improved
concentration result. It is assumed here that each of the i.i.d.
RVs $\{X_i\}$ gets the $M$ values $\exp\left(\frac{j (2k+1)
\pi}{M}\right)$ for $k=0, \ldots, M-1$ with equal probability. For
the martingale sequence $\{Y_i\}_{i=0}^n$ in \eqref{eq: martingale
sequence for OFDM}, it is shown in Appendix~\ref{appendix: OFDM}
that the following properties hold a.s.:
\begin{equation}
|Y_i - Y_{i-1}| \leq \frac{2}{\sqrt{n}}, \quad \quad
\expectation\bigl[(Y_i-Y_{i-1})^2 \, | \, \mathcal{F}_{i-1}\bigr]
\leq \frac{2}{n} \label{eq: properties of Y-sequence for OFDM
signals}
\end{equation}
for every $i \in \{1, \ldots, n\}$ where the conditioning on the
$\sigma$-algebra $\mathcal{F}_{i-1}$ is equivalent to the
conditioning on the values of $X_0, \ldots, X_{i-2}$ (for $i=1$
there is no conditioning).

Let $Z_i = \sqrt{n} Y_i$ for $i \in \{0, \ldots, n\}$, so
\eqref{eq: properties of Y-sequence for OFDM signals} gives that
\begin{equation*}
|Z_i - Z_{i-1}| \leq 2, \quad \quad
\expectation\bigl[(Z_i-Z_{i-1})^2 \, | \, \mathcal{F}_{i-1}\bigr]
\leq 2
\end{equation*}
(i.e., $d=2$ and $\sigma^2=2$ in our notation).
Proposition~\ref{proposition: a similar scaling of the
concentration inequalities} therefore implies that for an
arbitrary $\alpha > 0$
\begin{eqnarray}
&& \pr( | \text{CF}(s) - \expectation [
\text{CF}(s)] | \geq \alpha ) \nonumber\\
&& = \pr( |Y_n - Y_0| \geq \alpha) \nonumber\\
&& = \pr( |Z_n - Z_0| \geq \alpha \sqrt{n}) \nonumber\\
&& \leq 2 \exp \left( -\frac{\delta^2}{2 \gamma} \, \Bigl(1 +
O\Bigl(\frac{1}{\sqrt{n}}\Bigr) \right) \label{eq:
OFDM-inequality1}
\end{eqnarray}
where from \eqref{eq: notation}
\begin{equation*}
\gamma = \frac{\sigma^2}{d^2} = \frac{1}{2}, \quad
\delta = \frac{\alpha}{d} = \frac{\alpha}{2}.
\end{equation*}
Substituting $\gamma, \delta$ in \eqref{eq: OFDM-inequality1}
gives the concentration inequality
\begin{eqnarray}
&& \hspace*{-0.8cm} \pr( | \text{CF}(s) - \expectation [
\text{CF}(s)] | \geq \alpha ) \nonumber \\
&& \hspace*{-0.8cm} \leq 2 \exp \left( -\frac{\alpha^2}{4}
\right)\, \left(1+O\Bigl(\frac{1}{\sqrt{n}}\Bigr)\right)
\label{eq: OFDM-inequality2}
\end{eqnarray}
and its exponent is doubled as compared to the bound in \eqref{eq:
Azuma's inequality for OFDM} that was obtained via Azuma's
inequality. Note that the $O\Bigl(\frac{1}{\sqrt{n}}\Bigr)$ term
on the right-hand side of \eqref{eq: OFDM-inequality2} is
expressed explicitly (in terms of $\delta$ and $\gamma$ that are
calculated above) for a finite value of $n$ (see
Appendix~\ref{appendix: proof of the statement about the similar
scaling of the concentration inequalities}).

\subsection{Concentration of the Cardinality of the Fundamental
System of Cycles for LDPC Code Ensembles} \label{subsection:
Concentration of the Cardinality of the Fundamental System of
Cycles} Low-density parity-check (LDPC) codes are linear block
codes that are represented by sparse parity-check matrices
\cite{Gallager_1962}. A sparse parity-check matrix enables to
represent the corresponding linear block code by a sparse
bipartite graph, and to use this graphical representation for
implementing low-complexity iterative message-passing decoding.
The low-complexity decoding algorithms used for LDPC codes and
some of their variants are remarkable in that they achieve rates
close to the Shannon capacity limit for properly designed code
ensembles (see, e.g., \cite{RiU_book}). As a result of their
remarkable performance under practical decoding algorithms, these
coding techniques have revolutionized the field of channel coding
and they have been incorporated in various digital communication
standards during the last decade.

In the following, we consider ensembles of binary LDPC codes. The
codes are represented by bipartite graphs where the variable nodes
are located on the left side of the graph, and the parity-check
nodes are on the right. The parity-check equations that define the
linear code are represented by edges connecting each check node
with the variable nodes that are involved in the corresponding
parity-check equation. The bipartite graphs representing these
codes are sparse in the sense that the number of edges in the
graph scales linearly with the block length $n$ of the code.
Following standard notation, let $\lambda_i$ and $\rho_i$ denote
the fraction of edges attached, respectively, to variable and
parity-check nodes of degree~$i$. The LDPC code ensemble is
denoted by $\text{LDPC}(n,\lambda,\rho)$ where $n$ is the block
length of the codes, and the pair $\lambda(x) \triangleq \sum_i
\lambda_i x^{i-1}$ and $\rho(x) \triangleq \sum_i \rho_i x^{i-1}$
represents, respectively, the left and right degree distributions
of the ensemble from the edge perspective. For a short summary of
preliminary material on binary LDPC code ensembles see, e.g.,
\cite[Section~II-A]{Sason}.

It is well known that linear block codes which can be represented
by cycle-free bipartite (Tanner) graphs have poor performance even
under ML decoding \cite{Cycle-free codes}. The bipartite graphs of
capacity-approaching LDPC codes should therefore have cycles. For
analyzing this issue, we focused on the notion of "the cardinality
of the fundamental system of cycles of bipartite graphs". For the
required preliminary material, the reader is referred to
\cite[Section~II-E]{Sason}. In \cite{Sason}, we address the
following question:

{\em Question}: Consider an LDPC ensemble whose transmission takes
place over a memoryless binary-input output symmetric channel, and
refer to the bipartite graphs which represent codes from this
ensemble where every code is chosen uniformly at random from the
ensemble. How does the average cardinality of the fundamental
system of cycles of these bipartite graphs scale as a function of
the achievable gap to capacity ?

In light of this question, an information-theoretic lower bound on
the average cardinality of the fundamental system of cycles was
derived in \cite[Corollary~1]{Sason}. This bound was expressed in
terms of the achievable gap to capacity (even under ML decoding)
when the communication takes place over a memoryless binary-input
output-symmetric channel. More explicitly, it was shown that if
$\varepsilon$ designates the gap in rate to capacity, then the
number of fundamental cycles should grow at least like $\log
\frac{1}{\varepsilon}$. Hence, this lower bound remains unbounded
as the gap to capacity tends to zero. Consistently with the study
in \cite{Cycle-free codes} on cycle-free codes, the lower bound on
the cardinality of the fundamental system of cycles in
\cite[Corollary~1]{Sason} shows quantitatively the necessity of
cycles in bipartite graphs which represent good LDPC code
ensembles.
As a continuation to this work, we present in the following a
large-deviations analysis with respect to the cardinality of
the fundamental system of cycles for LDPC code ensembles.

Let the triple $(n, \lambda, \rho)$ represent an LDPC code
ensemble, and let $\mathcal{G}$ be a bipartite graph that
corresponds to a code from this ensemble. Then, the cardinality of
the fundamental system of cycles of $\mathcal{G}$, denoted by
$\beta(\mathcal{G})$, is equal to
\begin{equation*}
\beta(\mathcal{G}) = |E(\mathcal{G})| - |V(\mathcal{G})| +
c(\mathcal{G})
\end{equation*}
where $E(\mathcal{G})$, $V(\mathcal{G})$ and $c(\mathcal{G})$
denote the edges, vertices and components of $\mathcal{G}$,
respectively, and $|A|$ denotes the number of elements of a
(finite) set $A$. Note that for such a bipartite graph
$\mathcal{G}$, there are $n$ variable nodes and $m =
n(1-R_{\text{d}})$ parity-check nodes, so there are in total
$|V(\mathcal{G})| = n(2-R_{\text{d}})$ nodes. Let $a_{\text{R}}$
designate the average right degree (i.e., the average degree of
the parity-check nodes), then the number of edges in $\mathcal{G}$
is given by $|E(\mathcal{G})| = m a_{\text{R}}$. Therefore, for a
code from the $(n, \lambda, \rho)$ LDPC code ensemble, the
cardinality of the fundamental system of cycles satisfies the
equality
\begin{equation}
\beta(\mathcal{G}) = n \bigl[(1-R_{\text{d}}) a_{\text{R}} -
(2-R_{\text{d}}) \bigr] + c(\mathcal{G}) \label{eq: cardinality}
\end{equation}
where
\begin{equation*}
R_{\text{d}} = 1 - \frac{\int_0^1 \rho(x) \; \mathrm{d}x}{\int_0^1
\lambda(x) \; \mathrm{d}x}, \quad a_{\text{R}} = \frac{1}{\int_0^1
\rho(x) \; \mathrm{d}x}
\end{equation*}
denote, respectively, the design rate and average right degree
of the ensemble.

Let
\begin{equation}
E \triangleq |E(\mathcal{G})| = n(1-R_{\text{d}}) a_{\text{R}}
\label{eq: number of edges}
\end{equation}
denote the number of edges of an arbitrary bipartite graph
$\mathcal{G}$ from the ensemble (where we refer interchangeably to
codes and to the bipartite graphs that represent these codes from
the considered ensemble). Let us arbitrarily assign numbers $1,
\ldots, E$ to the $E$ edges of $\mathcal{G}$. Based on
Remarks~\ref{remark: construction of martingales} and~\ref{remark:
construction of martingales (cont.)}, lets construct a martingale
sequence $X_0, \ldots, X_E$ where $X_i$ (for $i=0, 1, \ldots, E$)
is a RV that denotes the conditional expected number of components
of a bipartite graph $\mathcal{G}$, chosen uniformly at random
from the ensemble, given that the first $i$ edges of the graph
$\mathcal{G}$ are revealed. Note that the corresponding filtration
$\mathcal{F}_0 \subseteq \mathcal{F}_1 \subseteq \ldots \subseteq
\mathcal{F}_E$ in this case is defined so that $\mathcal{F}_i$ is
the $\sigma$-algebra that is generated by all the sets of
bipartite graphs from the considered ensemble whose first $i$
edges are fixed. For this martingale sequence $$ X_0 =
\expectation_{\text{LDPC}(n,\lambda,\rho)}[\beta(\mathcal{G})],
\quad X_E = \beta(\mathcal{G})$$ and (a.s.) $|X_k - X_{k-1}| \leq
1$ for $k=1, \ldots, E$ (since by revealing a new edge of
$\mathcal{G}$, the number of components in this graph can change
by at most~1). By Corollary~\ref{corollary: a tightened version of
Azuma's inequality for martingales with bounded jumps}, it follows
that for every $\alpha \geq 0$
\begin{eqnarray}
&& \hspace*{-0.9cm} \pr \left( |c(\mathcal{G}) -
\expectation_{\text{LDPC}(n,\lambda,\rho)}[c(\mathcal{G})]| \geq
\alpha E \right) \leq 2 e^{-f(\alpha) E} \nonumber \\
&& \hspace*{-1.4cm} \Rightarrow \pr \left( |\beta(\mathcal{G}) -
\expectation_{\text{LDPC}(n,\lambda,\rho)}[\beta(\mathcal{G})]|
\geq \alpha E \right) \leq 2 e^{-f(\alpha) E} \label{eq:
Corollary 2 for the cardinality of the set of fundamental system
of cycles}
\end{eqnarray}
where the last transition follows from \eqref{eq: cardinality},
and the function $f$ was defined in \eqref{eq: f}. Hence, for
$\alpha > 1$, this probability is zero (since $f(\alpha) =
+\infty$ for $\alpha > 1$).
Note that, from \eqref{eq: cardinality},
$\expectation_{\text{LDPC}(n,\lambda,\rho)}[\beta(\mathcal{G})]$
scales linearly with $n$.
The combination of Eqs.~\eqref{eq: f}, \eqref{eq: number of edges},
\eqref{eq: Corollary 2 for the cardinality of the set of fundamental
system of cycles} gives the following statement:

\begin{theorem}{\bf[Concentration inequality for the cardinality
of the fundamental system of cycles]}
Let $\text{LDPC}(n,\lambda,\rho)$ be the LDPC code ensemble that
is characterized by a block length $n$, and a pair of degree
distributions (from the edge perspective) of $\lambda$ and $\rho$.
Let $\mathcal{G}$ be a bipartite graph chosen uniformly at random
from this ensemble. Then, for every $\alpha \geq 0$, the
cardinality of the fundamental system of cycles of $\mathcal{G}$
satisfies the following inequality
\begin{equation*}
\pr \left( |\beta(\mathcal{G}) -
\expectation_{\text{LDPC}(n,\lambda,\rho)}[\beta(\mathcal{G})]|
\geq \alpha n \right) \leq 2 \cdot 2^{- \left[ 1 -
h_2\left(\frac{1-\beta}{2}\right) \right] n}
\end{equation*}
where $h_2$ designates the binary entropy function to the base~2,
$ \beta \triangleq \frac{\alpha}{(1-R_{\text{d}}) \,
a_{\text{R}}}$, and $R_{\text{d}}$ and $a_{\text{R}}$ designate,
respectively, the design rate and average right degree of the
of the ensemble. Consequently, if $\beta > 1$,
this probability is zero. \label{theorem: Concentration
result for the cardinality of the fundamental system of cycles}
\end{theorem}

\begin{remark}
The loosened version of Theorem~\ref{theorem: Concentration result
for the cardinality of the fundamental system of cycles}, which
follows from Azuma's inequality, gets the form
\begin{eqnarray*}
&& \pr \left( |\beta(\mathcal{G}) -
\expectation_{\text{LDPC}(n,\lambda,\rho)}[\beta(\mathcal{G})]|
\geq \alpha n \right) \leq 2 e^{-\frac{\beta^2 n}{2}}
\end{eqnarray*}
for every $\alpha \geq 0$, and $\beta$ as defined in
Theorem~\ref{theorem: Concentration result for the cardinality
of the fundamental system of cycles}.
Note, however, that the exponential
decay of the two bounds is similar for values of $\alpha$ close to
zero (see the exponents in Azuma's inequality
and Corollary~\ref{corollary: a tightened version of Azuma's
inequality for martingales with bounded jumps} in Figure~\ref{Figure:
compare_exponents_theorem2}).
\end{remark}

\begin{remark}
For various capacity-achieving sequences of LDPC code ensembles on
the binary erasure channel, the average right degree scales like
$\log \frac{1}{\varepsilon}$ where $\varepsilon$ denotes the
fractional gap to capacity under belief-propagation decoding
(i.e., $R_{\text{d}} = (1-\varepsilon)C$) \cite{LubyMSS_IT01}.
Therefore, for small values of $\alpha$, the exponential decay
rate in the inequality of Theorem~\ref{theorem: Concentration
result for the cardinality of the fundamental system of cycles}
scales like $ \left(\log \frac{1}{\varepsilon} \right)^{-2}$. This
large-deviations result complements the result in
\cite[Corollary~1]{Sason} which provides a lower bound on the
average cardinality of the fundamental system of cycles that
scales like $\log \frac{1}{\varepsilon}$.
\end{remark}

\begin{remark}
Consider small deviations from the expected value that scale like
$\sqrt{n}$. Note that Corollary~\ref{corollary: a tightened
version of Azuma's inequality for martingales with bounded jumps}
is a special case of Theorem~\ref{theorem: first refined
concentration inequality} when $\gamma = 1$ (i.e., when only an
upper bound on the jumps of the martingale sequence is available,
but there is no non-trivial upper bound on the conditional
variance). Hence, it follows from Proposition~\ref{proposition: a
similar scaling of the concentration inequalities} that
Corollary~\ref{corollary: a tightened version of Azuma's
inequality for martingales with bounded jumps} does not provide
any improvement in the exponent of the concentration inequality
(as compared to Azuma's inequality) when small deviations are
considered.
\end{remark}

\section{Summary and Outlook}
This section provides a short summary of this work, followed
by a discussion on some directions for further research as
a continuation to this work.

\label{Section: Summary and Outlook}
\subsection{Summary}
This paper derives some refined versions of the Azuma-Hoeffding
inequality (see \cite{Azuma} and \cite{Hoeffding}) for
discrete-parameter martingales with uniformly
bounded jumps, and it considers some of their applications in
information theory and related topics. The first part is focused
on the derivation of these refined inequalities, followed by a
discussion on their relations to some classical results in
probability theory. Along this discussion, these inequalities
are linked to the method of types, martingale
central limit theorem, law of iterated logarithm, moderate deviations
principle, and to some reported concentration inequalities from the
literature. The second part of this work
exemplifies these refined inequalities in the context of
hypothesis testing and information theory, communication, and
coding theory. The interconnections between the concentration
inequalities that are analyzed in the first part of this work
(including some geometric interpretation w.r.t. some of these inequalities)
are studied, and the conclusions of this study serve for the
discussion on information-theoretic aspects related to these concentration
inequalities in the second
part of this work. Rather than covering a large number of applications,
we chose to exemplify the use of the concentration inequalities by considering
several applications carefully, which also provide some insight on these
concentration inequalities. Several more applications and information-theoretic
aspects are outlined shortly in the next sub-section, as a continuation to
this work. It is our hope that the analysis in this work will stimulate
the use of some refined versions of the Azuma-Hoeffding inequality in
information-theoretic aspects.

\subsection{Topics for Further Research}
\label{subsection: Outlook} We gather here what we consider to be the
most interesting directions for future work as a follow-up to this
paper.
\begin{itemize}
\item {\em Possible refinements of
Theorem~\ref{theorem: first refined concentration inequality}}:
The proof of the concentration inequality in Theorem~\ref{theorem:
first refined concentration inequality} relies on Bennett's
inequality \eqref{eq: Bennett's inequality for unconditional
expectation}. This inequality is applied to a
martingale-difference sequence where it is assumed that the jumps
of the martingale are uniformly upper bounded, and a global upper
bound on their conditional variances is available (see \eqref{eq:
Bennett's inequality for the conditional law of xi_k}). As was
noted in \cite[p.~44]{Bennett} with respect to the derivation of
Bennett's inequality: {\em ``The above analysis may be extended
when more information about the distribution of the component
random variables is available.''} Hence, in the context of the
proof of Theorem~\ref{theorem: first refined concentration
inequality}, consider a martingale-difference sequence $\{\xi_k,
\mathcal{F}_k\}_{k=0}^n$ where, e.g., $\xi_k$ is conditionally
symmetrically distributed around zero given $\mathcal{F}_{k-1}$
(for $k = 1, \ldots, n$); it enables to obtain a tightened version
of Bennett's inequality, and accordingly to improve the exponent
of the concentration inequality in Theorem~\ref{theorem: first
refined concentration inequality} under such an assumption. In
general, under some proper assumptions on the conditional
distribution of $\xi_k$ given $\mathcal{F}_{k-1}$, the exponent in
Theorem~\ref{theorem: first refined concentration inequality} can
be improved by a refinement of the bound in
\eqref{eq: Bennett's inequality for the conditional law of xi_k}.
\item {\em Perspectives on the achievable rates and random
coding error exponents for linear ISI and non-linear Volterra
channels}: Martingale-based concentration inequalities were
recently applied in \cite{Volterra_IT_March11} to obtain lower
bounds on the error exponents, and upper bounds on the achievable
rates of random codes whose transmission takes place over
nonlinear Volterra channels. Performance analysis of random coding
over these channels is of theoretical and practical interest since
various wireless communication channels exhibit non-linear
behavior (e.g., the satellite amplifier operates near its
saturation point, and exhibits highly non-linear characteristics).
For background on digital transmission over non-linear Volterra
channels see, e.g., \cite[Chapter~14]{Benedetto_Biglieri_book}.

The concentration inequalities in Section~\ref{section: Refined
Versions of Azuma's Inequality} can be applied to improve the
recent bounds of the work in
\cite{Volterra_IT_March11}. To this end, note that the jumps and
the conditional variance of the martingale in
\cite{Volterra_IT_March11} are uniformly bounded (see
\cite[Eq.~(22)]{Volterra_IT_March11}, followed by the refined
analysis in \cite[Section~IV]{Volterra_IT_March11}). Hence,
inequality \eqref{eq: important inequality for the derivation of
Theorem 4} for the special case where $m=2$ (serving for the
derivation of Corollary~\ref{corollary: specialization of the
second inequality}) provides an improvement to the analysis in
\cite[Eq.~(38)]{Volterra_IT_March11} (since $e^x > 1+x$ for
$x>0$). Furthermore, based on Proposition~\ref{proposition:
Theorem 2 gives a stronger result than Corollary 4}, a further
improvement to this analysis can be obtained by using, instead of
\eqref{eq: upper bound on the exponent in the second approach}
with $m=2$, the inequality in \eqref{eq: important inequality used
for the derivation of Theorem 2} (that was used to derive
Theorem~\ref{theorem: first refined concentration inequality}).
Based on the analysis in Section~\ref{subsection: Another Approach
for the Derivation of a Refinement of Azuma's Inequality}, a yet
another improvement to this analysis can be obtained by relying on
\eqref{eq: upper bound on the exponent in the second approach} for
even values of $m$ larger than~2.
This provides an improvement to the analysis of the lower bound on
the random coding exponents under ML decoding where the
communication takes place over a non-linear Volterra channel;
respectively, it also improves the upper bounds on the maximal
achievable rates of random coding under ML decoding. As was noted
in \cite[Section~V]{Volterra_IT_March11}, the same kind of
analysis can be applied to the special case where the
communication takes place over a stationary, causal and linear
intersymbol-interference (ISI) channel.
\item {\em Channel Polarization}: Channel polarization was recently
introduced by Arikan \cite{Arikan} to
develop a channel coding scheme called polar codes.
The fundamental concept of channel
polarization is introduced in \cite[Theorem~1]{Arikan}, and it is
proved via the convergence theorem for martingales. This analysis
was strengthened in \cite{Arikan_Emre_ISIT2009} where the key to
this analysis is in \cite[Observation~1]{Arikan_Emre_ISIT2009}; it
states that the random processes that keep track of the mutual
information and Bhattacharyya parameter arising in the course of
the channel polarization are, respectively, a martingale and a
super-martingale. Since both random processes are bounded (so
their jumps are also bounded), it is of interest to
consider the applicability of concentration inequalities for
refining the martingale-based analysis of channel
polarization for finite block-lengths.
\item {\em Message-passing decoding for graph-based codes}:
A great simplification in the analysis of codes defined on graphs
under iterative message-passing decoding is obtained by
considering the asymptotic performance of ensembles instead of the
performance of specific codes. The theoretical justification of
this approach is based on Azuma's concentration inequality and a
definition of a proper martingale that enables to assert that all
except an exponentially (in the block length) small fraction of
codes behave within an arbitrary small $\delta$ from the ensemble
average. This important concentration result was proved by
Richardson and Urbanke (see \cite[pp.~487--490]{RiU_book}). It
implies that for a sufficiently large block length, the ensemble
average is a good indicator for the performance of individual
codes from this ensemble, and it therefore seems a reasonable
route to focus on the design and analysis of capacity-approaching
ensembles (by density evolution \cite{RiU_book}). Some more
concentration inequalities for codes defined on graphs and
iterative decoding algorithms were derived in the coding
literature during the last decade (see \cite{RiU_book} and
references therein). The concentration inequalities which have
been proved in the setting of iterative message-passing decoding
so far rely on Azuma's inequality. They are rather loose, and much
stronger concentration phenomena can be observed in practice for
moderate to large block lengths. Therefore, to date, these
concentration inequalities serve mostly to justify theoretically
the ensemble approach, but they do not provide tight bounds for
finite block lengths. It is of interest to apply martingale-based
concentration inequalities, which improve the exponent of Azuma's
inequality, in order to obtain better concentration results for
finite block lengths. To this end, one needs to tackle the problem
of evaluating the conditional variance or higher conditional
moments for the related martingales that were used for the
derivation of some concentration inequalities that refer to
graph-based code ensembles.
\item {\em Second-order lossless source coding theorems for finite
block length}: Shannon's source coding theorem asserts that the
entropy rate is a fundamental limitation on the asymptotic
compression rate of stationary ergodic sources with finite
alphabets. Shannon's theorem was linked in \cite{Kontoyyanis_IT97}
to the central limit theorem (CLT) and the law of the iterated
logarithm (LIL) to prove asymptotic second-order lossless source
coding theorems for the deviation of the codeword lengths from the
entropy rate of a stationary ergodic source. Due to the relations
of Proposition~\ref{proposition: a similar scaling of the
concentration inequalities} with both the CLT and LIL (see
Sections~\ref{subsection: relation between the martingale CLT and
Proposition 4.3} and~\ref{subsection: relation between the LIL and
Proposition 4.3}), it is of interest to explore further possible
refinements of the asymptotic results presented in
\cite{Kontoyyanis_IT97} via the concentration inequalities in
Proposition~\ref{proposition: a similar scaling of the
concentration inequalities} (note that the terms that appear in
\eqref{eq: concentration1} as $O\bigl(n^{-\frac{1}{2}}\bigr)$ are
expressed explicitly in terms of $n$ along the proof of this
proposition in Appendix~\ref{appendix: proof of the statement
about the similar scaling of the concentration inequalities}).

According to the notation in \cite{Kontoyyanis_IT97}, let
$\mathcal{A}$ be a finite alphabet of a stationary ergodic source,
$L_n: \mathcal{A}^n \rightarrow \naturals$ be an arbitrary
sequence of codeword-length assignments, and
$ D_n \triangleq L_n(X_1^n) - H(X_1^n).$
The suggested direction of study may be done by
introducing the natural filtration where
$ \mathcal{F}_0 = \{ \emptyset, \Omega \}$ and
$\mathcal{F}_n = \sigma(X_1, \ldots, X_n)$ for every $n \in \naturals$.
Then, for fixed value of $n$, let
$Y_k = \expectation[D_n \, | \, \mathcal{F}_k]$ for $k = 0, \ldots, n$.
By Remark~\ref{remark: construction of martingales}, the sequence
$\{Y_k, \mathcal{F}_k \}_{k=0}^n$ is a martingale. By
Remark~\ref{remark: construction of martingales (cont.)},
$ Y_0 = \expectation (D_n)$ and $Y_n = D_n$.

It is of interest to prove sufficient conditions to assert
that this martingale sequence has uniformly bounded jumps, and
to calculate the respective conditional variance. The use
of Proposition~\ref{proposition: a similar scaling of the
concentration inequalities} under this setting is likely to
provide some refined information for finite block length $n$.

As a follow-up to \cite{Kontoyyanis_IT97}, the asymptotic redundancy
of lossless source coding with two codeword lengths was studied in
\cite{FigueroaH_IT05}. Furthermore, an un-published extended
version of this paper \cite{FigueroaH_un-published} relied on
Talagrand's concentration inequalities on product spaces
\cite{Talagrand95} for this study.

\item {\em Martingale-based Inequalities Related to Exponential
Bounds on Error Probability with Feedback}:  As a follow-up
to \cite[Section~3.3]{Nakiboglu_PhD_thesis_2011} and
\cite[Theorem~11]{PolanskiyPV_IT paper}, an analysis that relies on
the refined versions of Azuma's inequality in
Section~\ref{section: Refined Versions of Azuma's Inequality} (with
the standard adaptation of these inequalities to sub-martingales)
has the potential to provide further results in this direction.

\item {\em Some possible extensions}:
Azuma's inequality for discrete-time real-valued martingales was
extended in \cite[Theorem~3.5]{Pinelis} to martingales in Hilbert
spaces; this extension was introduced in
\cite[Lemma~B.1]{Yao_IT_December10} as the Pinelis-Hoeffding
inequality. Similarly, the extension of Corollary~\ref{corollary:
3rd corollary} to martingales in a Hilbert space follows from the
discussion in \cite{Pinelis} (see the remark in
\cite[pp.~1685--1686]{Pinelis}); it was introduced in
\cite[Lemma~B.2]{Yao_IT_December10} as the Pinelis-Bennett
inequality, followed by a looser version of this inequality in
\cite[Corollary~B.3]{Yao_IT_December10} that was introduced as
Pinelis-Bernstein inequality. Extensions of some other
concentration inequalities that are introduced in
Section~\ref{section: Refined Versions of Azuma's Inequality} to
martingales in Hilbert spaces are likely to extend the
applicability of these bounds. For example, some concentration
inequalities for martingales in Hilbert spaces were applied in
\cite{Yao_IT_December10} to a probabilistic analysis of the
convergence rate of an online learning algorithm.

%
%
\end{itemize}

\appendices

\section{Proof of Theorem~\ref{theorem: inequality based on a
parabola intersecting the exponential function at the endpoints of
the interval}} \label{appendix: inequality based on a parabola
intersecting the exponential function at the endpoints of the
interval} Consider a discrete-parameter martingale $\{X_k,
\mathcal{F}_k\}_{k=0}^{\infty}$ that satisfies the assumptions of
Theorem~\ref{theorem: first refined concentration inequality}
(a.s.) for some fixed constants $d, \sigma > 0$. Due to the
convexity of the exponential function, this function is upper
bounded over an arbitrary interval by the line segment that
intersects the curve of the exponential function at the two
endpoints of this interval. The improvement made in the derivation
of Theorem~\ref{theorem: inequality based on a parabola intersecting
the exponential function at the endpoints of the interval}
relies on a specification of the tightest parabola
that coincides with the exponential function at the endpoints of
the interval $[-d, d]$, and is above this exponential function
over the interval $(-d, d)$. Let $\xi_k \triangleq X_k - X_{k-1}$
for every $k \in \naturals$. This correspondingly improves the
upper bound on $\expectation\bigl[\exp(t \xi_k) \, | \,
\mathcal{F}_k\bigr]$ for $t \geq 0$ as compared to
the simple upper bound that refers to the line segment that
connects the exponential function at the endpoints of the interval
$[-d, d]$. The calculation of the considered parabola leads to the
following lemma:
\begin{lemma}
Let $d$ be an arbitrary positive number. Then, for every $x \in
(-\infty,d]$ \small
\begin{equation*}
e^x \leq \left(\frac{d+x}{2d}\right) \, e^d +
\left(\frac{d-x}{2d}\right) e^{-d} - \frac{\sinh(d)-de^{-d}}{2}
\left(1-\Bigl(\frac{x}{d}\Bigr)^2 \right).
\end{equation*}
\normalsize Moreover, this is the tightest parabola that coincides
with the exponential function at the endpoints of the interval
$[-d,d]$, and is above the exponential
function over the interval $(-d,d)$.
\label{lemma: the basic inequality used for the second bound}
\end{lemma}
\begin{IEEEproof}
The proof follows by calculus, and the details are omitted for the sake of brevity.
\end{IEEEproof}
Since by definition $\xi_k = X_k - X_{k-1}$, for every $k \in
\naturals$, then
$X_n - X_0 = \sum_{k=1}^n \xi_k.$
By the first assumption of Theorem~\ref{theorem: first refined
concentration inequality}, $|\xi_k| \leq d$ a.s. for every~$k$. By
Lemma~\ref{lemma: the basic inequality used for the second bound},
for every $t \geq 0$,
\begin{eqnarray}
&& \hspace*{-1cm} e^{t \xi_k} \leq \left(\frac{t d + t
\xi_k}{2td}\right) \, e^{td} +
\left(\frac{t d- t \xi_k}{2t d}\right) \, e^{-t d} \nonumber \\[0.05cm]
&& - \left(\frac{\sinh(t d) - t d e^{-t d}}{2} \right) \left(1 -
\Bigl(\frac{\xi_k}{d}\Bigr)^2 \right) \label{eq: first inequality
for this parabola-based bound}
\end{eqnarray}
a.s. for every $k \in \naturals$. 
The assumptions of
Theorem~\ref{theorem: first refined concentration inequality} on
the martingale sequence $\{X_k, \mathcal{F}_k\}_{k=0}^{\infty}$
yield that a.s.
\begin{eqnarray}
&& \expectation[ \xi_k \, | \, \mathcal{F}_{k-1} ] = 0
\label{eq: martingale property} \\
&& \expectation[ \xi_k^2 \, | \, \mathcal{F}_{k-1} ] \leq \sigma^2
\label{eq: bounded conditional second moment of xi} \\
&& |\xi_k| \leq d \label{eq: xi is bounded a.s.}
\end{eqnarray}
for every $k$, where without any loss of generality $\sigma$ can
be determined such that $\sigma \leq d$. From \eqref{eq: first
inequality for this parabola-based bound} and \eqref{eq: bounded
conditional second moment of xi} then, for every $t \geq 0$,
\begin{equation*}
\expectation \bigl[ e^{t \xi_k} \, | \, \mathcal{F}_{k-1}\bigr]
\leq \cosh(td) - \frac{(1-\gamma)(\sinh(t d) - t d e^{-td})}{2}
\end{equation*}
where $\gamma \triangleq \frac{\sigma^2}{d^2}$ is introduced in
\eqref{eq: notation}. From \eqref{eq: smoothing theorem} and the
last inequality, then for every $t \geq 0$
\begin{eqnarray*}
&& \hspace*{-1cm} \expectation \biggl[ \exp
\Bigl( t \sum_{k=1}^n \xi_k \Bigr) \biggr] \nonumber \\
&& \hspace*{-1cm} \leq \left[ \cosh(td) - \frac{(1-\gamma)
(\sinh(t d) - t d e^{-td})}{2} \right]^n.
\end{eqnarray*}
From \eqref{eq: Chernoff} and the last inequality, then for an
arbitrary $\alpha \geq 0$
\begin{eqnarray*}
&& \hspace*{-0.7cm} \pr(X_n-X_0 \geq \alpha n) \\
&& \hspace*{-0.7cm} \leq \Biggl\{ e^{-\alpha t} \;
\left[ \cosh(td) - \frac{(1-\gamma)
(\sinh(t d) - t d e^{-td})}{2} \right]^n \\
&& \hspace*{-0.7cm} = \left\{ \left(\frac{1+\gamma}{4}\right)
e^{(d-\alpha)t} + \left[\frac{1}{2} + \frac{(1+2td)(1-\gamma)}{4}
\right] e^{-(\alpha+d)t} \right\}^n
\end{eqnarray*}
for every $t \geq 0$. 
%
In the following, the value of the non-negative parameter $t$ that obtains
the tightest exponential bound within this form is calculated. Let $\delta
\triangleq \frac{\alpha}{d}$ as in \eqref{eq: notation}.

If $\delta>1$ then $\pr(X_n-X_0 \geq \alpha n)=0$, and the
exponent of the bound is therefore set to infinity. In the
continuation, we consider the case where $\delta \leq 1$. Based on
the notation in \eqref{eq: notation} and the substitution $x=td$
(where $x \geq 0$ is a free parameter), the last inequality admits
the equivalent form
\begin{eqnarray}
&& \hspace*{-1.2cm} \pr(X_n-X_0 \geq \alpha n) \nonumber \\
&& \hspace*{-1.2cm} \leq \Biggl\{ \frac{(1+\gamma)e^{(1-\delta)x}
+ \bigl[2+(1-\gamma)(1+2x)\bigr] e^{-(1+\delta)x}}{4} \Biggr\}^n
\label{eq: third inequality for this parabola-based bound}
\end{eqnarray}
\normalsize where the non-negative parameter $x$ is subject to
optimization in order to get the tightest bound within this form.

In the particular case where $\delta=1$ (note also that $\gamma
\leq 1$), then the tightest bound in \eqref{eq: third inequality
for this parabola-based bound} is obtained in the limit where we
let $x$ tend to infinity. This gives
\begin{equation}
\pr(X_n-X_0 \geq d n) \leq \Bigl(\frac{1+\gamma}{4}\Bigr)^n.
\label{eq: this parabola-based bound in the case where delta is
equal to 1}
\end{equation}

In the following, we derive a closed-form expression for the
optimized bound in \eqref{eq: third inequality for this
parabola-based bound} for the case where $\delta \in [0,1)$. In
this case, by differentiating the base of the exponential bound on
the right-hand side of \eqref{eq: third inequality for this
parabola-based bound} (w.r.t. the free non-negative parameter $x$)
and setting this derivative to zero, one gets the following
equation:
\begin{eqnarray}
&& \Bigl\{ (1+\delta) \bigl[2+(1+2x)(1-\gamma)] - 2(1-\gamma)
\Bigr\} e^{-2x} \nonumber \\
&& = (1+\gamma) (1-\delta). \label{eq: equation for optimized x of
this parabola-based bound}
\end{eqnarray}
Let us first examine the special case where $\sigma = d$ (i.e.,
$\gamma=1$). In this case, one gets from the assumptions in
Theorem~\ref{theorem: first refined concentration inequality} that
the requirement on the conditional variance is redundant which
then implies the same requirement of a bounded-difference
martingale that was used to derive the tightened Azuma's
inequality in Corollary~\ref{corollary: a tightened version of
Azuma's inequality for martingales with bounded jumps}. Indeed, in
the case where $\gamma=1$, equation \eqref{eq: equation for
optimized x of this parabola-based bound} is simplified, and its
solution is $x = \tanh^{-1}(\delta).$
The substitution of this value of $x$ and $\gamma=1$ into the
right-hand side of \eqref{eq: third inequality for this
parabola-based bound} gives the bound
\begin{eqnarray*}
&& \pr(X_n-X_0 \geq \alpha n) \\
&& \leq \Bigl( \frac{e^{(1-\delta)x}+e^{-(1+\delta)x}}{2} \Bigr)^n
\\
&& = \bigl( e^{-\delta x} \cosh(x) \bigr)^n \\
&& = \exp \left\{ -n \, \ln(2) \biggl[1-h_2\left(
\frac{1-\delta}{2}\right)\biggr] \right\}
\end{eqnarray*}
which indeed coincides with Corollary~\ref{corollary: a tightened
version of Azuma's inequality for martingales with bounded jumps}.

The following lemma asserts the existence and uniqueness of a
non-negative solution of equation \eqref{eq: equation for
optimized x of this parabola-based bound}, and it also provides a
closed-form expression for this solution.
\begin{lemma}
If $\gamma, \delta \in (0,1)$, then there exists a unique
non-negative solution to equation \eqref{eq: equation for
optimized x of this parabola-based bound}, and it is equal to
\begin{equation}
x = -\frac{1+W_{-1}(w)}{2} - \frac{\gamma +
\delta}{(1+\delta)(1-\gamma)} \label{eq: optimized x for 2nd
bound}
\end{equation}
where $W_{-1}$ stands for a branch of the Lambert W function
\cite{Lambert_function}, and
\begin{equation*}
w \triangleq - \frac{(1+\gamma)(1-\delta)}{(1-\gamma)(1+\delta)}
\cdot e^{-1-\frac{2(\gamma+\delta)}{(1+\delta)(1-\gamma)}}.
\end{equation*} \label{lemma: closed-form expression for optimized
x of this parabola-based bound}
\end{lemma}
\begin{IEEEproof}
Equation~\eqref{eq: equation for optimized x of this
parabola-based bound} can be rewritten in the form
\begin{equation}
(a+bx) e^{-2x} = c \label{eq: simplified equation for this
parabola-based bound}
\end{equation}
where
\begin{equation*}
a \triangleq 1+\gamma+(3-\gamma) \delta, \quad b \triangleq 2(1-\gamma)(1+\delta), \quad
c \triangleq (1+\gamma)(1-\delta).
\end{equation*}
Lets define the function
\begin{equation*}
f(x)=(a+bx)e^{-2x}-c, \quad \forall \, x \in \reals^{+}
\end{equation*}
then, since $\gamma, \delta \in (0,1)$,
\begin{equation*}
f(0) = a-c = 4 \delta > 0, \quad \lim_{x \rightarrow \infty} f(x) = -c < 0
\end{equation*}
so, it follows from the intermediate-value theorem that
there exists a solution $x \in (0, \infty)$ to the
equation $f(x)=0$; this value of $x$ is also a solution of
equation~\eqref{eq: equation for optimized x of this
parabola-based bound}. This assures the existence of a positive
solution of equation~\eqref{eq: equation for optimized x of this
parabola-based bound}. In order to prove the uniqueness of this
positive solution, note that
\begin{eqnarray*}
&& f'(x) = \bigl[ (b-2a)-2bx \bigr] e^{-2x} \\
&& \hspace*{0.9cm} = -4 \bigl[\gamma+\delta +
(1+\delta)(1-\gamma)x \bigr] e^{-2x} \\
&& \hspace*{0.9cm} < 0, \quad \forall \, x > 0
\end{eqnarray*}
which yields that $f$ is a monotonic decreasing function over the
interval $(0, \infty)$, so the positive solution of the equation
$f(x)= 0$ (or equivalently equation~\eqref{eq: equation for
optimized x of this parabola-based bound}) is unique.

In order to solve equation~\eqref{eq: equation for optimized x of
this parabola-based bound}, we rely on the equivalent simplified
form in \eqref{eq: simplified equation for this parabola-based
bound}. The substitution $z = -2 \bigl(x+\frac{a}{b}\bigr)$
transforms it to the equation $$z e^z = -\frac{2c}{b} \cdot
e^{-\frac{2a}{b}}.$$ Its solution is, by
definition, expressed in terms of the Lambert W-function
\cite{Lambert_function}:
\begin{eqnarray*}
&& z = W\biggl(-\frac{2c}{b} \cdot e^{-\frac{2a}{b}} \biggr) \\
&& \hspace*{0.3cm} = W(w)
\end{eqnarray*}
where
\begin{equation*}
w \triangleq -\frac{1}{e} \,
\frac{(1+\gamma)(1-\delta)}{(1-\gamma)(1+\delta)} \cdot
e^{-\frac{2(\gamma+\delta)}{(1+\delta)(1-\gamma)}}.
\end{equation*}
The inverse transformation gives
\begin{eqnarray*}
&& x = -\frac{z}{2} - \frac{a}{b}\\
&& \hspace*{0.3cm} = -\frac{1+z}{2} -
\frac{\gamma+\delta}{(1+\delta)(1-\gamma)}.
\end{eqnarray*}
Since $w \in \bigl(-\frac{1}{e}, 0\bigr)$, then the branch $W =
W_{-1}$ of the multi-valued Lambert W-function ensures that $x$ is
a real positive number as required. This completes the proof of
Lemma~\ref{lemma: closed-form expression for optimized x of this
parabola-based bound}.
\end{IEEEproof}
Putting all the pieces of this proof together, it completes the
proof of Theorem~\ref{theorem: inequality based on a parabola
intersecting the exponential function at the endpoints of the
interval}.

\section{Proof of Lemma~\ref{lemma: properties of phi}}
\label{appendix: properties of phi} The first two properties of
$\varphi_m$ in Lemma~\ref{lemma: properties of phi} follow from
the power series expansion of the exponential function, so
\begin{equation*}
\varphi_m(y) = \frac{m!}{y^m} \sum_{l=m}^{\infty} \frac{y^l}{l!} =
\sum_{l=0}^{\infty} \frac{m! y^l}{(m+l)!}, \quad \forall \, y \in \reals.
\end{equation*}
By the absolute convergence of this series, $\lim_{y \rightarrow
0} \varphi_m(y) = 1$, and it follows from this power series expansion
that $\varphi_m$ is strictly monotonic increasing over
the interval $[0, \infty)$. In order to show the third property of
$\varphi_m$, it is proved in the following that for every even $k
\geq 2$
\begin{equation}
1 + y + \ldots + \frac{y^{k-1}}{(k-1)!}
< e^y < 1 + y + \ldots + \frac{y^k}{k !}
\, , \; \forall \, y<0.
\label{eq: bounds on the exponent function for a negative argument}
\end{equation}
\begin{IEEEproof}
The proof is based on mathematical induction. For $k=2$, one needs
to verify that
\begin{equation}
1+y < e^y < 1 + y + \frac{y^2}{2} \, , \quad \forall \, y<0.
\label{eq: special bounds on the exponent function for a negative
argument}
\end{equation}
To this end, let $f_1(y) \triangleq e^y-1-y$, so $f_1(0)=0$, and
$f_1'(y) = e^y-1<0$ for $y<0$, so $f_1$ is monotonic
decreasing over $(-\infty, 0]$ and therefore $f_1(y)>0$ for $y<0$.
This proves the left-hand side of \eqref{eq: special bounds on the
exponent function for a negative argument}. Let $f_2(y) \triangleq
e^y-1-y - \frac{y^2}{2}$, so $f_2'(y) = e^y-1-y>0$ for $y<0$ (from the
proof of the left-hand side of this inequality). Hence, $f_2$ is
monotonic increasing over $(-\infty, 0]$, and $f_2(y) <
f_2(0)=0$ for $y<0$. This proves the right-hand side of \eqref{eq:
special bounds on the exponent function for a negative argument}.

Assume that \eqref{eq: bounds on the exponent function for a
negative argument} is satisfied for a specific even number $k \geq
2$. It will be shown to be valid also for $k+2$. Similarly, define
$f_1(y) \triangleq e^y - \left(1 + y + \ldots +
\frac{y^{k+1}}{(k+1)!} \right)$ so $f_1(0)=0$, and by the
assumption of the induction where \eqref{eq: bounds on the
exponent function for a negative argument} holds for some even $k
\geq 2$, then for every $y < 0$
$$f_1'(y) = e^y - \left(1 + y + \ldots + \frac{y^k}{k!} \right) < 0$$
so $f_1$ is monotonically decreasing over $(-\infty, 0]$. This
implies that $f_1(y)>0$ for every $y<0$, so the left-hand side of
\eqref{eq: bounds on the exponent function for a negative
argument} follows when $k$ is replaced by $k+2$. To prove the
second part of the inequality, let $f_2(y) \triangleq e^y -
\left(1 + y + \ldots + \frac{y^{k+2}}{(k+2)!} \right)$ then
$f_2(0)=0$, and for every $y<0$
$$f_2'(y) = e^y - \left(1 + y + \ldots + \frac{y^{k+1}}{(k+1)!} \right) > 0$$
due to the proof of the first part of this inequality. Hence,
$f_2$ is monotonically increasing over $(-\infty, 0]$, so $f_2(y)
< f_2(0) = 0$ for every $y < 0$. This then implies that the
right-hand side of \eqref{eq: bounds on the exponent function for
a negative argument} holds when $k$ is replaced by
$k+2$. To conclude, \eqref{eq: bounds on the exponent function for
a negative argument} holds for $k=2$,
and also if \eqref{eq: bounds on the exponent function for a
negative argument} holds for a specific even number $k \geq 2$
then it is also valid for $k+2$. Hence, by mathematical
induction, inequality \eqref{eq: bounds on the exponent function for a
negative argument} holds for every even number $k \geq 2$.
\end{IEEEproof}
Based on \eqref{eq: bounds on the exponent function for a negative
argument}, it follows that $0 < \varphi_m(y) < 1$ for every even
$m \geq 2$ and $y<0$, so it completes the proof of the third item
in Lemma~\ref{lemma: properties of phi}. The last property of
$\varphi_m$ in Lemma~\ref{lemma: properties of phi} follows
directly by combining the second and third items of this lemma
with the equality $\varphi_m(0) = 1$.

\section{Proof of Corollary~\ref{corollary: specialization of the second inequality}}
\label{appendix: proof of the corollary that specializes the
second inequality for m equal to 2}

The proof of Corollary~\ref{corollary: specialization of the
second inequality} is based on the specialization of
Theorem~\ref{theorem: second inequality} for $m=2$. This gives
that, for every $\alpha \geq 0$, the following concentration
inequality holds:
\begin{eqnarray}
&& \hspace*{-0.7cm} \pr(|X_n - X_0| \geq n \alpha) \nonumber \\
&& \hspace*{-0.7cm} \leq 2 \left\{ \inf_{x \geq 0} \, e^{-\delta x}
\Bigl[1 + \gamma (e^x-1-x) \Bigr] \right\}^n
\label{eq: second bound before optimization of x for the special case}
\end{eqnarray}
where $\gamma = \gamma_2$ according to the notation in \eqref{eq:
notation}.

By differentiating the logarithm of the right-hand side of
\eqref{eq: second bound before optimization of x for the special
case} w.r.t. $x$ (where $x \geq 0$) and setting this derivative to zero,
it follows that
\begin{equation}
\frac{1-\gamma x}{\gamma (e^x-1)} = \frac{1-\delta}{\delta}.
\label{eq: new equation for x}
\end{equation}

Let us first consider the case where $\delta=1$. In this case,
this equation is satisfied either if $x = \frac{1}{\gamma}$ or in
the limit where $x \rightarrow \infty$. In the former case where
$x = \frac{1}{\gamma}$, the resulting bound in \eqref{eq: second
bound before optimization of x for the special case} is equal to
\begin{equation}
\exp \left[-n \left( \frac{1}{\gamma} - \ln \Bigl( \gamma \bigl(
e^{\frac{1}{\gamma}} - 1 \bigr) \Bigr) \right)\right].
\label{eq: optimized second bound if delta is 1}
\end{equation}
In the latter case where $x \rightarrow \infty$, the resulting
bound in \eqref{eq: second bound before optimization of x for the
special case} when $\delta = 1$ is equal to
\begin{eqnarray*}
&& \lim_{x \rightarrow \infty} e^{-nx} \bigl(1 + \gamma
(e^{x}-1-x) \bigr)^n \\
&& = \lim_{x \rightarrow \infty} \Bigl( e^{-x} + \gamma
\bigl(1-(1+x) e^{-x} \bigr) \Bigr)^n \\
&& = \gamma^n.
\end{eqnarray*}
Hence, since for $\gamma \in (0,1)$
\begin{eqnarray*}
&& \ln \Bigl( \frac{1}{\gamma} \Bigr)
= \frac{1}{\gamma} - \ln \Bigl( \gamma
e^{\frac{1}{\gamma}} \Bigr)\\[0.1cm]
&& \hspace*{1.1cm} < \frac{1}{\gamma} - \ln \Bigl( \gamma \bigl(
e^{\frac{1}{\gamma}} - 1 \bigr) \Bigr)
\end{eqnarray*}
then the optimized value is $x = \frac{1}{\gamma}$, and the
resulting bound in the case where $\delta = 1$ is equal to
\eqref{eq: optimized second bound if delta is 1}.

Let us consider now the case where $0<\delta<1$ (the case where
$\delta=0$ is trivial). In the following lemma, the existence and
uniqueness of a solution of this equation is assured, and a
closed-form expression for this solution is provided.
\begin{lemma}
If $\delta \in (0,1)$, then equation~\eqref{eq: new equation for
x} has a unique solution, and it lies in $(0, \frac{1}{\gamma})$.
This solution is given in \eqref{eq: solution of the optimized x}.
\label{lemma: proof of the existence and uniqueness of the
solution}
\end{lemma}
\begin{IEEEproof}
Consider equation~\eqref{eq: new equation for x}, and note that
the right-hand side of this equation is positive for $\delta \in
(0,1)$. The function
\begin{equation*}
t(x) = \frac{1-\gamma x}{\gamma (e^x-1)}, \quad x \in \reals
\end{equation*}
on the left-hand side of \eqref{eq: new equation for x} is
negative for $x > \frac{1}{\gamma}$ (since the numerator of the
function $t$ is negative and its denominator is positive), and it
is also negative for $x < 0$ (positive numerator and negative
denominator). Since the function $t$ is continuous on the interval
$(0, \frac{1}{\gamma}]$ and
\begin{equation*}
t\left(\frac{1}{\gamma}\right) = 0, \quad \quad \lim_{x
\rightarrow 0^+} t(x) = +\infty
\end{equation*}
then there is a solution $x \in \bigr(0, \frac{1}{\gamma}\bigl)$.
Moreover, the function $t$ is monotonic decreasing in the interval
$\bigr(0, \frac{1}{\gamma}\bigl]$ (the numerator of $t$ is
monotonic decreasing and the denominator of $t$ is monotonic
increasing and both are positive there, hence, their ratio $t$ is
a positive and monotonic decreasing function in this interval).
This implies the existence and uniqueness of the solution, which
lies in the interval $(0, \frac{1}{\gamma})$. In the following, a
closed-form expression of this solution is derived. Note that
Eq.~\eqref{eq: new equation for x} can be expressed in the form
\begin{equation}
\frac{a-x}{e^x-1} = b \label{eq: simplified equation}
\end{equation}
where
\begin{equation}
a \triangleq \frac{1}{\gamma}, \; b \triangleq
\frac{1-\delta}{\delta} \label{eq: a and b}
\end{equation}
are both positive. Eq.~\eqref{eq: simplified equation} is
equivalent to
\begin{equation*}
(a+b-x) e^{-x} = b.
\end{equation*}
The substitution $u = a+b-x$ transforms this equation to
\begin{equation*}
u e^u = b e^{a+b}.
\end{equation*}
The solution of this equation is, by definition, given by
\begin{equation*}
u = W_0\left(b e^{a+b}\right)
\end{equation*}
where $W_0$ denotes the principal branch of the multi-valued
Lambert W function \cite{Lambert_function}. Since $a,b$ are
positive then $b e^{a+b}>0$, so that the principal branch of $W$
is the only one which is a real number. In the following, it will
be confirmed that the selection of this branch also implies that
$x>0$ as required. By the inverse transformation one gets
\begin{eqnarray}
&& x = a+b-u \nonumber \\
&& \hspace*{0.3cm} = a+b - W_0\left(b e^{a+b}\right) \label{eq:
solution of simplified equation}
\end{eqnarray}
Hence, the selection of this branch for $W$ indeed ensures that
$x$ is the positive solution we are looking for (since $a,b>0$,
then it readily follows from the definition of the Lambert W
function that $W_0\left(b e^{a+b}\right) < a+b$ and it was earlier
proved in this appendix that the positive solution $x$ of
\eqref{eq: new equation for x} is unique). Finally, the
substitution of \eqref{eq: a and b} into \eqref{eq: solution of
simplified equation} gives \eqref{eq: solution of the optimized
x}. This completes the proof of Lemma~\ref{lemma: proof of the
existence and uniqueness of the solution}.
\end{IEEEproof}
The bound in \eqref{eq: second bound before optimization of x for
the special case} is given by
\begin{eqnarray}
&& \hspace*{-1.2cm} \pr(|X_n-X_0| \geq \alpha n) \nonumber \\
&& \hspace*{-1.2cm} \leq 2 \exp \biggl(-n \Bigl[\delta x -
\ln\bigl(1+\gamma(e^x-1-x)\bigr) \Bigr] \biggr) \label{eq:
specialization of the second bound before optimization}
\end{eqnarray}
with the value of $x$ in \eqref{eq: solution of the optimized x}.
This completes the proof of Corollary~\ref{corollary:
specialization of the second inequality}.

\section{Proof of Proposition~\ref{proposition: a sufficient condition where the
specialization of the second inequality is better than Azuma's
inequality}} \label{appendix: a sufficient condition where the
specialization of the second inequality is better than Azuma's
inequality} Lets compare the right-hand sides of \eqref{eq:
derivation of Azuma's inequality before replacing the hyperbolic
cosine with an exponential upper bound} and \eqref{eq: second
bound before optimization of x for the special case} that refer to
Corollaries~\ref{corollary: a tightened version of Azuma's
inequality for martingales with bounded jumps} and~\ref{corollary:
specialization of the second inequality}, respectively.
Proposition~\ref{proposition: a sufficient condition where the
specialization of the second inequality is better than Azuma's
inequality} follows by showing that if $\gamma \leq \frac{1}{2}$
\begin{equation}
1 + \gamma (\exp(x)-1-x) < \cosh(x), \quad \forall \, x > 0.
\label{eq: sufficient condition that asserts the claim}
\end{equation}
To this end, define
\begin{equation*}
f(x) \triangleq \cosh(x) - \gamma \bigl(\exp(x)-1-x\bigr), \quad
\forall \, x\geq 0.
\end{equation*}
If $\gamma \leq \frac{1}{2}$, then for every $x>0$
\begin{eqnarray*}
&& f'(x) = \sinh(x)-\gamma \bigl(\exp(x)-1\bigr) \\
&& \hspace*{0.9cm} = \Bigl( \frac{1}{2}-\gamma \Bigr) \exp(x) +
\gamma - \frac{\exp(-x)}{2} \\
&& \hspace*{0.9cm} > \Bigl( \frac{1}{2}-\gamma \Bigr) + \gamma
-\frac{1}{2} = 0
\end{eqnarray*}
so, since $f$ is monotonic increasing on $[0, \infty)$ and
$f(0)=0$, then $f(x)>0$ for every $x>0$. This validates \eqref{eq:
sufficient condition that asserts the claim}, and it therefore
completes the proof of Proposition~\ref{proposition: a sufficient
condition where the specialization of the second inequality is
better than Azuma's inequality}.

\section{Proof of Proposition~\ref{proposition: Theorem 2 gives a stronger result than Corollary 4}}
\label{appendix: Proof of Proposition 2}
\begin{lemma}
For every $\gamma, x > 0$
\begin{equation}
\frac{\gamma e^x + e^{-\gamma x}}{1+\gamma} < 1 + \gamma
(e^x-1-x). \label{eq: an inequality served to compare the
tightness of Theorem 2 and Corollary 4}
\end{equation}
\label{lemma: a lemma served to compare the tightness of Theorem 2
and Corollary 4}
\end{lemma}
\begin{IEEEproof}
Let $\gamma$ be an arbitrary positive number, and define the function
\begin{equation*}
f_{\gamma}(x) \triangleq \frac{\gamma e^x + e^{-\gamma x}}{1+\gamma}
- [1+\gamma(e^x-1-x)] \, , \quad x \geq 0.
\end{equation*}
Then, $f_{\gamma}(0) = 0$, and the first derivative is equal to
\begin{eqnarray*}
&& f_{\gamma}'(x) 
= \gamma \left(1 - \frac{\gamma e^x + e^{-\gamma x}}{1+\gamma} \right).
\end{eqnarray*}
From the convexity of the exponential function $y(u) = e^u$,
then for every $x>0$
\begin{eqnarray*}
&& \frac{\gamma e^x + e^{-\gamma x}}{1+\gamma} =
\left(\frac{\gamma}{1+\gamma}\right) y(x) +
\left(\frac{1}{1+\gamma}\right) y(-\gamma x) \\
&& \hspace*{1.9cm} > y\left(\frac{\gamma}{1+\gamma} \cdot x
+ \frac{1}{1+\gamma} \cdot (-\gamma x) \right) \\
&& \hspace*{1.9cm} = y(0) = 1
\end{eqnarray*}
so, it follows that $f_{\gamma}'(x) < 0$ for every $x > 0$. Since
$f_{\gamma}(0) = 0$ and the first derivative is negative over $(0,
\infty)$, then $f_{\gamma}(x) < 0$ for every $x>0$. This completes
the proof of inequality~\eqref{eq: an inequality served to compare
the tightness of Theorem 2 and Corollary 4}.
\end{IEEEproof}
This claim in Proposition~\ref{proposition: Theorem 2 gives a
stronger result than Corollary 4} follows directly from
Lemma~\ref{lemma: a lemma served to compare the tightness of
Theorem 2 and Corollary 4}, and the two inequalities in \eqref{eq:
important inequality used for the derivation of Theorem 2} and
\eqref{eq: upper bound on the exponent in the second approach}
with $m=2$. In the case where $m=2$, the right-hand side
of~\eqref{eq: upper bound on the exponent in the second approach}
is equal to
$$\bigl(1 + \gamma (e^{td} - 1 - td) \bigr)^n.$$
Note that \eqref{eq: important inequality used for the derivation
of Theorem 2} and \eqref{eq: upper bound on the exponent in the
second approach} with $m=2$ were used to derive, respectively,
Theorem~\ref{theorem: first refined concentration inequality} and
Corollary~\ref{corollary: specialization of the second inequality}
(based on Chernoff's bound). The conclusion follows by
substituting $x \triangleq td$ on the right-hand sides of
\eqref{eq: important inequality used for the derivation of Theorem
2} and \eqref{eq: upper bound on the exponent in the second
approach} with $m=2$ (so that $x \geq 0$ since $t \geq 0$ and
$d>0$, and \eqref{eq: an inequality served to compare the
tightness of Theorem 2 and Corollary 4} turns from an inequality
if $x>0$ into an equality if $x=0$).

\section{Proof of Proposition~\ref{proposition: first proposition on the loosened version of Theorem 4}}
\label{appendix: proof of the first proposition on the loosened
version of Theorem 4} For the case where $m=2$, the conditions in
Theorem~\ref{theorem: second inequality} are identical to
Corollary~\ref{corollary: first loosened version of Theorem 4}.
Hence, since Corollary~\ref{corollary: specialization of the
second inequality} follows as a particular case of
Theorem~\ref{theorem: second inequality} for $m=2$, then
Corollary~\ref{corollary: first loosened version of Theorem 4}
implies the result in Corollary~\ref{corollary: specialization of
the second inequality}.

The sequence $\{\gamma_l\}_{l=2}^m$ is monotonic non-increasing
and non-negative. This follows from the assumption where a.s.
$$|X_k - X_{k-1}| \leq d, \quad \forall \, k \in \naturals$$
and the definition of the sequences $\{\gamma_l\}_{l=2}^m$
and $\{\mu_l\}_{l=2}^m$ in
\eqref{eq: gamma and delta for Theorem 4} and
\eqref{eq: mu for the first loosened version of Theorem 4},
respectively. Hence, for every $l$
\begin{eqnarray*}
&& 0 \leq \gamma_{l+1} = \expectation \left[ \left(
\frac{|X_{k+1}-X_k|}{d} \right)^{l+1} \, \Big| \,
\mathcal{F}_{k-1} \right] \\
&& \hspace*{1.4cm} \leq \expectation \left[
\left( \frac{|X_{k+1}-X_k|}{d} \right)^l \, \Big| \,
\mathcal{F}_{k-1} \right] = \gamma_l \leq 1.
\end{eqnarray*}
Since $\{\gamma_l\}_{l=2}^{\infty}$ is monotonic non-increasing
and non-negative (note also that $\gamma_l$ is independent of the value of
$m$) then it is a convergent sequence. 

Referring to the base of the exponential bound on the right-hand
side of \eqref{eq: 2nd concentration inequality}, for an arbitrary
$x \geq 0$ and an even $m \in \naturals$, we want to show that the
value of the non-negative base of this exponential bound is
decreased by increasing $m$ to $m+2$; hence, it implies that the
bound in Corollary~\ref{corollary: first loosened version of
Theorem 4} is improved as the value of $m$ is increased (where $m
\in \naturals$ is even). To show this, note that for an arbitrary
$x \geq 0$ and an even $m \in \naturals$,

\vspace*{-0.2cm}
\small
\begin{eqnarray*}
&& \hspace*{-0.7cm} \left[ 1 + \sum_{l=2}^{m-1} \frac{(\gamma_l -
\gamma_m)x^l}{l!} + \gamma_m (e^x-1-x) \right] \\[0.1cm]
&& \hspace*{-0.4cm} - \left[ 1 + \sum_{l=2}^{m+1} \frac{(\gamma_l -
\gamma_{m+2})x^l}{l!} + \gamma_{m+2} (e^x-1-x) \right] \\[0.1cm]
&& \hspace*{-0.7cm} = \sum_{l=2}^{m-1} \frac{(\gamma_{m+2} -
\gamma_m)x^l}{l!} - \frac{(\gamma_m - \gamma_{m+2})x^m}{m!} \\
&& \hspace*{-0.4cm} - \frac{(\gamma_{m+1} - \gamma_{m+2})x^{m+1}}{(m+1)!}
+ (\gamma_m - \gamma_{m+2}) (e^x-1-x) \\[0.1cm]
&& \hspace*{-0.7cm} = (\gamma_m - \gamma_{m+2}) \left( e^x -
\sum_{l=0}^m \frac{x^l}{l!} \right) - \frac{(\gamma_{m+1} -
\gamma_{m+2}) x^{m+1}}{(m+1)!} \\[0.1cm]
&& \hspace*{-0.7cm} = (\gamma_m - \gamma_{m+2}) \sum_{l=m+1}^{\infty}
\frac{x^l}{l!} - \frac{(\gamma_{m+1} - \gamma_{m+2}) x^{m+1}}{(m+1)!} \\[0.1cm]
&& \hspace*{-0.7cm} = (\gamma_m - \gamma_{m+2}) \left[ \frac{x^{m+1}}{(m+1)!}
+ \sum_{l=m+2}^{\infty} \frac{x^l}{l!} \right]
- \frac{(\gamma_{m+1} - \gamma_{m+2}) x^{m+1}}{(m+1)!} \\[0.1cm]
&& \hspace*{-0.7cm} = \frac{(\gamma_m - \gamma_{m+1}) x^{m+1}}{(m+1)!} +
(\gamma_m - \gamma_{m+2}) \sum_{l=m+2}^{\infty} \frac{x^l}{l!} \, .
\end{eqnarray*}
\normalsize  Since $\gamma_m - \gamma_{m+1} \geq 0$ and $\gamma_m -
\gamma_{m+2} \geq 0$ (due to the monotonicity of the sequence
$\{\gamma_l\}_{l=2}^{\infty}$) and $x \geq 0$, then the above
difference is non-negative and therefore
\begin{eqnarray*}
&& \hspace*{-1cm} 1 + \sum_{l=2}^{m-1} \frac{(\gamma_l -
\gamma_m)x^l}{l!} + \gamma_m (e^x-1-x) \\[0.1cm]
&& \hspace*{-1cm} \geq 1 + \sum_{l=2}^{m+1} \frac{(\gamma_l -
\gamma_{m+2})x^l}{l!} + \gamma_{m+2} (e^x-1-x).
\end{eqnarray*}
Multiplying the two sides of this inequality by $e^{-\delta x}$
and taking the infimum w.r.t. $x \geq 0$ gives that, for every $m
\in \naturals$ that is even,

\small \vspace*{-0.2cm}
\begin{eqnarray*}
&& \hspace*{-0.8cm} \inf_{x \geq 0} \left\{ e^{-\delta x}
\left[ 1 + \sum_{l=2}^{m-1} \frac{(\gamma_l - \gamma_m)x^l}{l!} +
\gamma_m (e^x-1-x) \right] \right\} \\[0.1cm]
&& \hspace*{-0.8cm} \geq \inf_{x \geq 0} \left\{ e^{-\delta x}
\left[ 1 + \sum_{l=2}^{m+1} \frac{(\gamma_l - \gamma_{m+2})x^l}{l!}
+ \gamma_{m+2} (e^x-1-x) \right] \right\}.
\end{eqnarray*}
\normalsize Note that the right-hand side of this inequality is
non-negative due to inequality~\eqref{eq: 2nd concentration
inequality} where the right-hand side is a non-negative upper
bound. Hence, it follows that the bound in
Corollary~\ref{corollary: first loosened version of Theorem 4}
improves as the value of the even number $m \in \naturals$ is
increased. This completes the proof of
Proposition~\ref{proposition: first proposition on the loosened
version of Theorem 4}.

\section{Proof of Corollary~\ref{corollary: second loosened version of Theorem 4}}
\label{appendix: proof of 2nd looser version of Theorem 4} A
minimization of the logarithm of the exponential bound on the
right-hand side of \eqref{eq: 2nd looser version of Theorem 4}
gives the equation
\begin{equation*}
\frac{\dsum_{l=2}^{m-1} \frac{(\gamma_l - \gamma_m)
x^{l-1}}{(l-1)!} + \gamma_m (e^x-1)}{1 + \dsum_{l=2}^{m-1}
\frac{(\gamma_l - \gamma_m) x^l}{l!} + \gamma_m (e^x-1-x)} =
\delta
\end{equation*}
and after standard algebraic operations, it gives the equation
\begin{eqnarray}
&& \hspace*{-1cm} \gamma_m \left( \frac{1}{\delta}-1 \right) (e^x-1-x)
+ \frac{\gamma_2 x}{\delta} \nonumber \\
&& \hspace*{-1cm} + \sum_{l=1}^{m-1} \left\{ \Bigl[
\frac{\gamma_{l+1}}{\delta} - \gamma_l - \gamma_m
\Bigl(\frac{1}{\delta}-1\Bigr) \Bigr] \frac{x^l}{l!}
\right\} - 1 = 0. \label{eq: equation for optimized x}
\end{eqnarray}
As we have seen in the proof of Corollary~\ref{corollary:
specialization of the second inequality} (see
Appendix~\ref{appendix: proof of the corollary that specializes
the second inequality for m equal to 2}), the solution of this
equation can be expressed in a closed-form for $m=2$, but in general,
a closed-form solution to this equation is not available. A
sub-optimal value of $x$ on the right-hand side of \eqref{eq: 2nd
concentration inequality} is obtained by neglecting the sum that
appears in the second line of this equation (the rationality for
this approximation is that $\{\gamma_l\}$ was observed to converge
very fast, so it was verified numerically that $\gamma_l$ stays
almost constant starting from a small value of $l$). Note that the
operation of $\inf_{x \geq 0}$ can be loosened by taking an
arbitrary non-negative value of $x$; hence, in particular, $x$
will be chosen in the following to satisfy the equation
\begin{equation*}
\gamma_m \left( \frac{1}{\delta}-1 \right) (e^x-1-x)
+ \frac{\gamma_2 x}{\delta} = 1.
\end{equation*}
By dividing both sides of the equation by $\gamma_2$, then it
gives the equation $a+b-cx = be^x$ with $a, b$ and $c$ from
\eqref{eq: a, b, c for the sub-optimal x}. This equation can be
written in the form
\begin{equation*}
\left(\frac{a+b}{c}-x\right) e^{-x} = \frac{b}{c}.
\end{equation*}
Substituting $u \triangleq \frac{a+b}{c}-x$ gives the equation
$$ u e^u = \frac{b}{c} \cdot e^{\frac{a+b}{c}}$$
whose solution is given by
$$u = W_0\left( \frac{b}{c} \cdot e^{\frac{a+b}{c}} \right)$$
where $W_0$ denotes the principal branch of the lambert W function
\cite{Lambert_function}. The inverse transformation back to $x$
gives that $$x = \frac{a+b}{c} - W_0\left( \frac{b}{c} \cdot
e^{\frac{a+b}{c}} \right).$$ This justifies the choice of $x$ in
\eqref{eq: sub-optimal x}, and it provides a loosening of either
Theorem~\ref{theorem: second inequality} or
Corollary~\ref{corollary: first loosened version of Theorem 4} by
replacing the operation of the infimum over the non-negative
values of $x$ on the right-hand side of \eqref{eq: 2nd
concentration inequality} with the value of $x$ that is given in
\eqref{eq: sub-optimal x} and \eqref{eq: a, b, c for the
sub-optimal x}. For $m=2$ where the sum on the left-hand side of
\eqref{eq: equation for optimized x} that was later neglected is
anyway zero, this forms indeed the exact optimal value of $x$ (so
that it coincide with \eqref{eq: optimized x for 2nd bound} in
Corollary~\ref{corollary: specialization of the second
inequality}).

\section{Proof of Proposition~\ref{proposition: a similar scaling
of the concentration inequalities}} \label{appendix: proof of the
statement about the similar scaling of the concentration
inequalities} Let $\{X_k, \mathcal{F}_k\}_{k=0}^{\infty}$ be a
discrete-parameter martingale. We prove in the following that
Theorems~\ref{theorem: first refined concentration inequality}
and~\ref{theorem: second inequality}, and also
Corollaries~\ref{corollary: 3rd corollary} and~\ref{corollary:
specialization of the second inequality} imply that \eqref{eq:
concentration1} holds. Since Theorem~\ref{theorem: inequality
based on a parabola intersecting the exponential function at the
endpoints of the interval} is looser than Theorem~\ref{theorem:
first refined concentration inequality}, and it was introduced in
Section~\ref{section: Refined Versions of Azuma's Inequality} in
order to highlight geometric interpretations, then we skip the
proof that also Theorem~\ref{theorem: inequality based on a
parabola intersecting the exponential function at the endpoints of
the interval} implies the same scaling as in \eqref{eq:
concentration1}.

\subsection{Analysis Related to Theorem~\ref{theorem: first refined concentration inequality}}
\label{appendix: Analysis related to Theorem 2} Let $\{X_k,
\mathcal{F}_k\}_{k=0}^{\infty}$ be a discrete-parameter martingale
that satisfies the conditions in Theorem~\ref{theorem: first
refined concentration inequality}. From \eqref{eq: first refined
concentration inequality}
\begin{eqnarray}
&& \pr(|X_n-X_0| \geq \alpha \sqrt{n}) \nonumber \\
&& \leq 2 \exp\left(-n \,
D\biggl(\frac{\delta'+\gamma}{1+\gamma} \Big|\Big|
\frac{\gamma}{1+\gamma}\biggr) \right)
\label{eq: exponent from first refined version with delta'}
\end{eqnarray}
where from \eqref{eq: notation}
\begin{equation}
\delta' \triangleq \frac{\frac{\alpha}{\sqrt{n}}}{d} =
\frac{\delta}{\sqrt{n}} \, . \label{eq: new delta}
\end{equation}
From the right-hand side of \eqref{eq: exponent from first refined
version with delta'}
\begin{eqnarray}
&& \hspace*{-0.7cm} D\biggl(\frac{\delta'+\gamma}{1+\gamma}
\Big|\Big| \frac{\gamma}{1+\gamma}\biggr) \nonumber \\
&& \hspace*{-0.7cm} = \left(\frac{\frac{\delta}{\sqrt{n}}+
\gamma}{1+\gamma}\right) \ln\left(1+\frac{\delta}{\gamma \sqrt{n}}
\right) + \left(\frac{1-\frac{\delta}{\sqrt{n}}}{1+\gamma} \right)
\ln\left(1-\frac{\delta}{\sqrt{n}} \right) \nonumber \\
&& \hspace*{-0.7cm} = \frac{\gamma}{1+\gamma}
\left[\left(1+\frac{\delta}{\gamma \sqrt{n}} \right)
\ln\left(1+\frac{\delta}{\gamma \sqrt{n}} \right)
\right. \nonumber\\
&& \hspace*{1cm} \left. + \frac{1}{\gamma}
\left(1-\frac{\delta}{\sqrt{n}} \right)
\ln\left(1-\frac{\delta}{\sqrt{n}} \right)
\right]. \label{eq: intermediate step1}
\end{eqnarray}
From the equality
\begin{equation*}
(1+u) \ln(1+u) = u + \sum_{k=2}^{\infty}
\frac{(-u)^k}{k(k-1)} \, , \quad -1 < u \leq 1
\end{equation*}
then it follows from \eqref{eq: intermediate step1} that for every
$n > \frac{\delta^2}{\gamma^2}$
\begin{eqnarray*}
&& n D\biggl(\frac{\delta'+\gamma}{1+\gamma} \Big|\Big|
\frac{\gamma}{1+\gamma}\biggr) \\
&& = \frac{n \gamma}{1+\gamma} \left[\frac{\delta^2}{2n}
\left(\frac{1}{\gamma^2} + \frac{1}{\gamma} \right) +
\frac{\delta^3}{6 n^{\frac{3}{2}}} \left(\frac{1}{\gamma}
- \frac{1}{\gamma^3} \right) + \ldots \right] \\
&& = \frac{\delta^2}{2 \gamma} - \frac{\delta^3 (1-\gamma)}{6
\gamma^2} \, \frac{1}{\sqrt{n}} + \ldots \\
&& = \frac{\delta^2}{2 \gamma} + O\left(\frac{1}{\sqrt{n}}\right).
\end{eqnarray*}
Substituting this into the exponent on the right-hand side of
\eqref{eq: exponent from first refined version with delta'}
gives \eqref{eq: concentration1}.

\subsection{Analysis Related to Corollary~\ref{corollary: 3rd corollary}}
Let $\{X_k, \mathcal{F}_k\}_{k=0}^{\infty}$ be a discrete-parameter martingale that
satisfies the conditions in Theorem~\ref{theorem: first
refined concentration inequality}. From Corollary~\ref{corollary:
3rd corollary}, it follows that for every $\alpha \geq 0$
\begin{eqnarray*}
&& \pr(|X_n-X_0| \geq \alpha \sqrt{n}) \\
&& \leq 2 \exp\left(-n \gamma \left[\left(1+\frac{\delta'}{\gamma}\right)
\ln \left(1+\frac{\delta'}{\gamma} \right) - \frac{\delta'}{\gamma}
\right] \right)
\end{eqnarray*}
where $\delta'$ is introduced in \eqref{eq: new delta}. By
substituting \eqref{eq: B} and \eqref{eq: new delta} into the last
inequality, it follows that for every $\alpha \geq 0$
\begin{equation}
\pr(|X_n-X_0| \geq \alpha \sqrt{n})
\leq 2 \exp\left(-\frac{\delta^2}{2\gamma} \;
B\left(\frac{\delta}{\gamma \sqrt{n}} \right) \right). \label{eq:
explicit and simple concentration inequality - ver1}
\end{equation}
The power series expansion around zero of the function $B$
in \eqref{eq: B} is given by
\begin{eqnarray*}
&& B(u) = \sum_{k=0}^{\infty} \frac{2 (-1)^k \,
u^k}{(k+1)(k+2)} \; , \quad |u| \leq 1.
\end{eqnarray*}
Therefore, if $n \gg 1$, substituting this equality in \eqref{eq:
explicit and simple concentration inequality - ver1} implies that
for every $\alpha \geq 0$
\begin{eqnarray}
&& \hspace*{-1.4cm} \pr(|X_n - X_0| \geq \alpha \sqrt{n}) \nonumber \\
&& \hspace*{-1.4cm} \leq 2 \exp\Biggl(-\frac{\delta^2}{2\gamma} \;
\left( 1 - \frac{1}{3} \frac{\delta}{\gamma \sqrt{n}} +
\frac{\delta^2}{6 \gamma^2 n} - \ldots \right) \Biggr) \nonumber \\
&& \hspace*{-1.4cm} = 2 \exp\left(-\frac{\delta^2}{2\gamma}\right)
\left(1+ O\Bigl(\frac{1}{\sqrt{n}}\Bigr)\right) \nonumber
\end{eqnarray}
which gives the inequality in \eqref{eq: concentration1}.

\subsection{Analysis Related to Theorem~\ref{theorem: second inequality}}
\label{subsubsection: Analysis related to Theorem 4}
From \eqref{eq: 2nd concentration inequality}, for every $\alpha
\geq 0$,
\begin{eqnarray*}
&& \hspace*{-0.7cm} \pr(|X_n - X_0| \geq \alpha \sqrt{n}) \\
&& \hspace*{-0.7cm} \leq 2 \left\{ \inf_{x \geq 0} \, e^{-\delta'
x} \left[1 + \sum_{l=2}^{m-1} \frac{(\gamma_l - \gamma_m) x^l}{l!}
+ \gamma_m (e^x-1-x) \right] \right\}^n
\end{eqnarray*}
where
\begin{equation*}
\delta' = \frac{\frac{\alpha}{\sqrt{n}}}{d} =
\frac{\delta}{\sqrt{n}}, \quad \gamma_l \triangleq
\frac{\mu_l}{d^l}, \; \; \forall \; l = 2, \ldots, m.
\end{equation*}
The optimization of the free non-negative parameter $x$ in the
above upper bound is obtained by minimizing the exponent of this
bound. This gives the equation

\small \vspace*{-0.3cm}
\begin{equation*}
\frac{\mathrm{d}}{\mathrm{d}x} \left\{-\delta' x + \ln \left[1 +
\sum_{l=2}^{m-1} \frac{(\gamma_l - \gamma_m) x^l}{l!} + \gamma_m
(e^x-1-x) \right] \right\} = 0
\end{equation*}
\normalsize that is equivalent to the equation \small
\begin{equation} \frac{(\gamma_2-\gamma_m)x +
\gamma_m (e^x-1) + \sum_{l=3}^{m-1} \frac{(\gamma_l - \gamma_m)
x^{l-1}}{(l-1)!}}{1 + \sum_{l=2}^{m-1} \frac{(\gamma_l - \gamma_m)
x^l}{l!} + \gamma_m (e^x-1-x)} = \frac{\delta}{\sqrt{n}} \, .
\label{eq: optimization equation of 2nd concentration inequality
for small deviations}
\end{equation}
\normalsize Note that if $n \gg 1$, then $\delta' \approx 0$ but
the values of $\{\gamma_l\}_{l=2}^m$ stay fixed. Hence, in this
case, the solution of \eqref{eq: optimization equation of 2nd
concentration inequality for small deviations} is approximately
zero. As in the previous analysis, we start with an approximate
analysis to get an approximate solution of \eqref{eq: optimization
equation of 2nd concentration inequality for small deviations}.
Since the above upper bound is valid for every $x \geq 0$, we then
perform an exact analysis with the approximated value of $x$ that
forms a solution of the optimization equation in \eqref{eq:
optimization equation of 2nd concentration inequality for small
deviations}.

For $x \approx 0$, we calculate a first order approximation of the
left-hand side of \eqref{eq: optimization equation of 2nd
concentration inequality for small deviations}. Note that
\begin{eqnarray*}
&& (\gamma_2 - \gamma_m)x + \gamma_m (e^x-1) + \sum_{l=3}^{m-1}
\frac{(\gamma_l - \gamma_m) x^{l-1}}{(l-1)!} \\
&& = \gamma_2 x + O(x^2),
\end{eqnarray*}
and
\begin{equation*}
1 + \sum_{l=2}^{m-1} \frac{(\gamma_l - \gamma_m) x^l}{l!} +
\gamma_m (e^x-1-x) = 1 + O(x^2)
\end{equation*}
so the left-hand side of \eqref{eq: optimization equation of 2nd
concentration inequality for small deviations} is equal to
$\gamma_2 x + O(x^2)$. Hence, if $n \gg 1$, then $x \approx
\frac{\delta}{\gamma_2 \sqrt{n}}$ is an approximated solution of
\eqref{eq: optimization equation of 2nd concentration inequality
for small deviations}. Following this approximation, we chose
sub-optimally the value of $x$ to be
\begin{equation}
x = \frac{\delta}{\gamma_2 \sqrt{n}} \label{eq: suboptimal choice
of x}
\end{equation}
and proceed with an exact analysis that relies on the
concentration inequality in Theorem~\ref{theorem: second
inequality}. Substituting $\delta' = \frac{\delta}{\sqrt{n}}$, and
the replacement of the infimum in the upper bound by the value of
the bound at $x$ in \eqref{eq: suboptimal choice of x} gives that,
for every $\alpha \geq 0$,
\begin{eqnarray*}
&& \hspace*{-0.7cm} \pr(|X_n - X_0| \geq \alpha \sqrt{n})
\\[0.1cm]
&& \hspace*{-0.7cm} \leq 2 \left\{ e^{-\delta' x} \left[1 +
\sum_{l=2}^{m-1} \frac{(\gamma_l - \gamma_m) x^l}{l!} + \gamma_m
(e^x-1-x) \right] \right\}^n \\[0.1cm]
&& \hspace*{-0.7cm} = 2
\exp\left(-\frac{\delta^2}{\gamma_2}\right) \left[1 +
\sum_{l=2}^{m-1} \frac{\gamma_l x^l}{l!} + \gamma_m \left(e^x -
\sum_{l=0}^{m-1} \frac{x^l}{l!} \right) \right]^n \\[0.1cm]
&& \hspace*{-0.7cm} = 2
\exp\left(-\frac{\delta^2}{\gamma_2}\right) \left[1 +
\frac{\gamma_2 x^2}{2} + O(x^3) \right]^n \\[0.1cm]
&& \hspace*{-0.7cm} = 2
\exp\left(-\frac{\delta^2}{\gamma_2}\right) \left[1 +
\frac{\delta^2}{2 \gamma_2 n} + O\Bigl(n^{-\frac{3}{2}}\Bigr)
\right]^n \\[0.1cm]
&& \hspace*{-0.7cm} = 2
\exp\left(-\frac{\delta^2}{2\gamma_2}\right) \, \left(1 +
O\bigl(n^{-\frac{1}{2}}\bigr) \right).
\end{eqnarray*}
Note that $\gamma_2 = \gamma$ in
\eqref{eq: notation} and Theorem~\ref{theorem: second inequality}.
This proves \eqref{eq: concentration1} via the concentration
inequality in Theorem~\ref{theorem: second inequality} for every
$m \geq 2$.

\subsection{Analysis Related to Corollary~\ref{corollary: specialization
of the second inequality}} The concentration inequality in
\eqref{eq: concentration1} was proved to be a consequence of
Theorem~\ref{theorem: second inequality} for an arbitrary even $m \geq 2$.
Since Corollary~\ref{corollary: specialization of the second
inequality} is a special case of Theorem~\ref{theorem: second
inequality} for $m=2$, then it follows
that \eqref{eq: concentration1} results in from
Corollary~\ref{corollary: specialization of the second inequality}
without a need for a separate analysis.

\section{Proof of Proposition~\ref{proposition: Theorem 5 implies Corollary 3 and Proposition 4}}
\label{appendix: proposition related to Theorem 5}

Consider the setting of Theorem~\ref{theorem: first refined
concentration inequality} where $\{X_k,
\mathcal{F}_k\}_{k=0}^{\infty}$ is a discrete-parameter
real-valued martingale such that
\begin{eqnarray*}
&& |X_k - X_{k-1}| \leq d, \quad \expectation[ (X_k - X_{k-1})^2 \, | \, \mathcal{F}_{k-1}] \leq
\sigma^2
\end{eqnarray*}
a.s. for every $k \in \naturals$. Let $S_k \triangleq \frac{X_k -
X_0}{d}$, so $\{S_k, \mathcal{F}_k\}_{k=0}^{\infty}$ is a
martingale sequence with $S_0 = 0$. Also, a.s.
\begin{eqnarray*}
&& Y_k \triangleq S_k - S_{k-1} = \frac{X_k - X_{k-1}}{d} \leq 1 \\
&& Q_n \triangleq \sum_{j=1}^n \expectation(Y_j^2 \, | \,
\mathcal{F}_{j-1}) \leq \frac{n \sigma^2}{d^2} = \gamma n
\end{eqnarray*}
where $\gamma \triangleq \frac{\sigma^2}{d^2}$ is introduced in
\eqref{eq: notation}. Hence, $Q_n \leq \gamma n$ a.s., and it
follows from
Theorem~\ref{theorem: from the book of Dembo and Zeitouni} that
for every $\alpha \geq 0$
\begin{eqnarray*}
&& \pr(X_n - X_0 \geq \alpha n) \\[0.1cm]
&& = \pr(S_n \geq \delta n, Q_n \leq \gamma n) \\[0.1cm]
&& \leq \exp\left(-\frac{\delta^2 n}{2\gamma} \;
B\left(\frac{\delta}{\gamma} \right) \right)
\end{eqnarray*}
where the last inequality follows from \eqref{eq: inequality in
exercise of Amir and Ofer's book}. From the definition of the
function $B$ in \eqref{eq: B} then, for every $\alpha \geq 0$,
\begin{eqnarray*}
&& \pr(X_n - X_0 \geq \alpha n) \\
&& \leq \exp \left(-n \gamma \biggl[\biggl(1+
\frac{\delta}{\gamma}\biggr) \ln\biggl(1+\frac{\delta}{\gamma}\biggr)
-\frac{\delta}{\gamma}\biggr] \right).
\end{eqnarray*}
By applying the last inequality to the martingale sequence
$\{-X_k, \mathcal{F}_k\}_{k=0}^{\infty}$, then the
same inequality also holds as an upper bound on
$\pr(X_n - X_0 \leq -\alpha n)$ for an arbitrary $\alpha \geq 0$.
Finally, the use of the union bound gives the two-sided
concentration inequality in \eqref{eq: 3rd corollary}.
This shows that Corollary~\ref{corollary: 3rd corollary}
is a consequence of
Theorem~\ref{theorem: from the book of Dembo and Zeitouni}.

Similarly, it follows from
\eqref{eq: inequality in exercise of Amir and Ofer's book}
that, for every $\alpha \geq 0$,
\begin{equation*}
\pr(X_n - X_0 \geq \alpha \sqrt{n})
\leq \exp\left(-\frac{\delta^2}{2\gamma} \;
B\left(\frac{\delta}{\gamma \sqrt{n}} \right) \right).
\end{equation*}
The latter inequality coincides with \eqref{eq: explicit and
simple concentration inequality - ver1}. As is shown in
Appendix~\ref{appendix: proof of the statement about the similar
scaling of the concentration inequalities}, \eqref{eq:
concentration1} follows from~\eqref{eq: explicit and simple
concentration inequality - ver1}, so a concentration inequality of
the form of~\eqref{eq: concentration1} in
Proposition~\ref{proposition: a similar scaling of the
concentration inequalities} follows as a consequence of
Theorem~\ref{theorem: from the book of Dembo and Zeitouni}. This
completes the proof of Proposition~\ref{proposition: Theorem 5
implies Corollary 3 and Proposition 4}.

\section{Analysis Related to the Moderate Deviations Principle
in Section~\ref{subsection: MDP for real-valued i.i.d. RVs}}
\label{appendix: MDP}

It is demonstrated in the following that, in contrast to Azuma's
inequality, both Theorems~\ref{theorem: first refined
concentration inequality} and~\ref{theorem: second inequality}
provide upper bounds on $$\pr\left( \Big|\sum_{i=1}^n X_i \Big|
\geq \alpha n^{\eta} \right), \quad \forall \, \alpha \geq 0$$
which coincide with the correct asymptotic result in \eqref{eq:
MDP for i.i.d. real-valued RVs}. It is proved under the further
assumption that there exists some constant $d > 0$ such that
$|X_k| \leq d$ a.s. for every $k \in \naturals$.
Let us define the martingale sequence
$\{S_k, \mathcal{F}_k\}_{k=0}^n$ where
\begin{eqnarray*}
S_k \triangleq \sum_{i=1}^k X_i, \quad \mathcal{F}_k \triangleq
\sigma(X_1, \ldots, X_k)
\end{eqnarray*}
for every $k \in \{1, \ldots, n\}$ with $S_0 = 0$ and $\mathcal{F}_0 = \{\emptyset, \mathcal{F}\}$.

\subsubsection{Analysis related to Azuma's inequality}
The martingale sequence $\{S_k, \mathcal{F}_k\}_{k=0}^n$ has
uniformly bounded jumps, where $|S_k - S_{k-1}| = |X_k| \leq d$
a.s. for every $k \in \{1, \ldots, n\}$. Hence it follows from Azuma's
inequality that, for every $\alpha \geq 0$,
\begin{equation*}
\pr\left( |S_n| \geq \alpha n^{\eta} \right)
\leq 2 \exp\left(-\frac{\alpha^2 n^{2\eta-1}}{2d^2}\right)
\end{equation*}
and therefore
\begin{equation}
\lim_{n \rightarrow \infty} n^{1-2 \eta} \; \ln \pr\bigl(
|S_n| \geq \alpha n^{\eta} \bigr) \leq
-\frac{\alpha^2}{2d^2}. \label{eq: MDP scaling from Azuma's
inequality for the sum of i.i.d. real-valued RVs}
\end{equation}
This differs from the limit in \eqref{eq: MDP for i.i.d.
real-valued RVs} where $\sigma^2$ is replaced by $d^2$, so Azuma's
inequality does not provide the correct asymptotic result in
\eqref{eq: MDP for i.i.d. real-valued RVs} (unless $\sigma^2 =
d^2$, i.e., $|X_k|=d$ a.s. for every $k$).

\subsubsection{Analysis related to Theorem~\ref{theorem: first refined concentration inequality}}
The analysis here is a slight modification of the analysis in
Appendix~\ref{appendix: Analysis related to Theorem 2} with the
required adaptation of the calculations for $\eta \in
(\frac{1}{2}, 1)$. It follows from Theorem~\ref{theorem: first
refined concentration inequality} that, for every $\alpha \geq 0$,
\begin{equation*}
\pr(|S_n| \geq \alpha n^{\eta}) \leq 2 \exp\left(-n \,
D\biggl(\frac{\delta'+\gamma}{1+\gamma} \Big|\Big|
\frac{\gamma}{1+\gamma}\biggr) \right)
\end{equation*}
where $\gamma$ is introduced in \eqref{eq: notation}, and
$\delta'$ in \eqref{eq: new delta} is replaced with
\begin{equation}
\delta' \triangleq \frac{\frac{\alpha}{n^{1-\eta}}}{d} =
\delta n^{-(1-\eta)}
\label{eq: new delta'}
\end{equation}
due to the definition of $\delta$ in \eqref{eq: notation}.
Following the same analysis as in Appendix~\ref{appendix: Analysis
related to Theorem 2}, it follows that for every $n \in \naturals$
\begin{eqnarray*}
&& \pr(|S_n| \geq \alpha n^{\eta}) \\
&& \leq 2 \exp\left( -\frac{\delta^2 n^{1-2\eta}}{2\gamma} \left[
1 + \frac{\alpha (1-\gamma)}{3 \gamma d} \cdot n^{-(1-\eta)} +
\ldots \right] \right)
\end{eqnarray*}
and therefore (since, from \eqref{eq: notation},
$\frac{\delta^2}{\gamma} = \frac{\alpha^2}{\sigma^2})$
\begin{equation}
\lim_{n \rightarrow \infty} n^{1-2 \eta} \; \ln \pr\bigl(
|S_n| \geq \alpha n^{\eta} \bigr) \leq
-\frac{\alpha^2}{2 \sigma^2}. \label{eq: MDP scaling from Theorem
2 for the sum of i.i.d. real-valued RVs}
\end{equation}
Hence, this upper bound
coincides with the exact asymptotic result in \eqref{eq: MDP for
i.i.d. real-valued RVs}.

\subsubsection{Analysis related to Theorem~\ref{theorem: second inequality}}
It is shown in the following that Theorem~\ref{theorem: second
inequality} coincides with the exact asymptotic result in
\eqref{eq: MDP for i.i.d. real-valued RVs} for an arbitrary even number $m \geq
2$. To this end, it is sufficient to prove it w.r.t. the looser
version in Corollary~\ref{corollary: first loosened version of
Theorem 4}. Due to Proposition~\ref{proposition: first proposition
on the loosened version of Theorem 4}, the tightness of the bound
in Corollary~\ref{corollary: first loosened version of Theorem 4}
is improved by increasing the even value of $m \geq 2$. It is
therefore enough to show that choosing $m=2$, which provides the
weakest bound in Corollary~\ref{corollary: first loosened version
of Theorem 4}, is already good enough to get the exact asymptotic
result in \eqref{eq: MDP for i.i.d. real-valued RVs}. But, for
$m=2$, Theorem~\ref{theorem: second inequality} and
Corollary~\ref{corollary: first loosened version of Theorem 4}
coincide (and both imply the inequality in
Corollary~\ref{corollary: specialization of the second
inequality}). This suggests to follow the analysis in
Appendix~\ref{subsubsection: Analysis related to Theorem 4} for
$m=2$ with the slight required modification of this analysis for
$\eta \in (\frac{1}{2}, 1)$ (instead of the case where it is
one-half as in Appendix~\ref{subsubsection: Analysis related to
Theorem 4}). In the following, we refer to the martingale sequence
$\{S_k, \mathcal{F}_k\}_{k=1}^n$ as above. Based on the analysis
in Appendix~\ref{subsubsection: Analysis related to Theorem 4},
the sub-optimal value of $x$ in \eqref{eq: suboptimal choice of x}
is modified to $x = \left(\frac{\delta}{\gamma}\right)
n^{-(1-\eta)}$ where, from \eqref{eq: notation}, $\gamma =
\frac{\sigma^2}{d^2}$ and $\delta = \frac{\alpha}{d}$. Hence, $x =
\frac{\delta'}{\gamma}$ with $\delta'$ in \eqref{eq: new delta'}.
Following the analysis in Appendix~\ref{subsubsection: Analysis
related to Theorem 4} for the special case of $m=2$ with the
sub-optimal $x$ as above, then for every
$\alpha \geq 0$

\small \vspace*{-0.2cm}
\begin{eqnarray*}
&& \hspace*{-0.7cm} \pr(|S_n| \geq \alpha n^{\eta}) \\
&& \hspace*{-0.7cm} \leq 2 \left\{ e^{-\delta' x}
\Bigl[1+\gamma (e^x-1-x) \Bigr] \right\}^n \\
&& \hspace*{-0.7cm} = 2 \exp\left(-\frac{n \delta'^2}{\gamma}\right)
\left(1 + \frac{\gamma x^2}{2} + \frac{\gamma x^3}{6} + O(x^4) \right)^n \\
&& \hspace*{-0.7cm} = 2 \exp\left(-\frac{\alpha^2 n^{2\eta-1}}{\gamma d^2}\right)
\left(1 + \frac{\alpha^2 n^{-2(1-\eta)}}{2 \gamma d^2}
+ \frac{\alpha^3 n^{-3(1-\eta)}}{6 \gamma^2 d^3} + \ldots \right)^n \\
&& \hspace*{-0.7cm} \leq 2 \exp\left(-\frac{\alpha^2
n^{2\eta-1}}{\sigma^2}\right)
\exp\left(\frac{\alpha^2 n^{2\eta-1} \bigl(1+O(n^{-(1-\eta)})\bigr)}{2 \sigma^2}\right) \\
&& \hspace*{-0.7cm} \leq 2 \exp\left(-\frac{\alpha^2 n^{2\eta-1}
\bigl(1+O(n^{-(1-\eta)})\bigr)}{2 \sigma^2}\right)
\end{eqnarray*}
\normalsize so the asymptotic result in \eqref{eq: MDP scaling
from Theorem 2 for the sum of i.i.d. real-valued RVs} also follows
in this case, thus coinciding with the exact asymptotic result in
\eqref{eq: MDP for i.i.d. real-valued RVs}.

\section{Proof of Proposition~\ref{proposition: Fisher information}}
\label{appendix: Fisher information} The proof of \eqref{eq:
relation between the Chernoff information and Fisher information}
is based on calculus, and it is similar to the proof of the limit
in \eqref{eq: relation between the divergence and Fisher
information} that relates the divergence and Fisher information.
For the proof of \eqref{eq: relation between the improved lower
bound on the error exponent and Fisher information}, note that

\vspace*{-0.3cm}
\small
\begin{equation}
\hspace*{-0.25cm} C(P_{\theta}, P_{\theta'})
\geq E_{\text{L}}(P_{\theta}, P_{\theta'}) \geq \min_{i=1,2}
\left\{\frac{\delta_i^2}{2 \gamma_i} -
\frac{\delta_i^3}{6 \gamma_i^2 (1+\gamma_i)} \right\}.
\label{eq: bounds on the improved lower bound}
\end{equation}
\normalsize The left-hand side of \eqref{eq: bounds on the
improved lower bound} holds since $E_{\text{L}}$ is a lower bound
on the error exponent, and the exact value of this error exponent
is the Chernoff information. The right-hand side of \eqref{eq:
bounds on the improved lower bound} follows from Lemma~\ref{lemma:
inequality for lower bounding the divergence} (see \eqref{eq:
lower bound on the lower bound of the exponents}) and the
definition of $E_{\text{L}}$ in \eqref{eq: E_L}. By definition
$\gamma_i \triangleq \frac{\sigma_i^2}{d_i^2}$ and $\delta_i
\triangleq \frac{\varepsilon_i}{d_i}$ where, based on \eqref{eq:
delta1 and delta2},
\begin{equation}
\varepsilon_1 \triangleq D(P_\theta || P_{\theta'}),
\quad \varepsilon_2 \triangleq D(P_\theta' || P_\theta).
\label{eq: epsilon1,2}
\end{equation}
The term on the left-hand side of \eqref{eq: bounds on the
improved lower bound} therefore satisfies
\begin{eqnarray*}
&& \frac{\delta_i^2}{2 \gamma_i} -
\frac{\delta_i^3}{6 \gamma_i^2 (1+\gamma_i)} \\
&& = \frac{\varepsilon_i^2}{2 \sigma_i^2} -
\frac{\varepsilon_i^3 d_i^3}{6 \sigma_i^2 (\sigma_i^2 + d_i^2)} \\
&& \geq \frac{\varepsilon_i^2}{2 \sigma_i^2}
\left( 1 - \frac{\varepsilon_i d_i}{3} \right)
\end{eqnarray*}
so it follows from \eqref{eq: bounds on the improved lower bound}
and the last inequality that
\begin{equation}
\hspace*{-0.25cm} C(P_{\theta}, P_{\theta'}) \geq
E_{\text{L}}(P_\theta, P_{\theta'}) \geq \min_{i=1,2} \left\{
\frac{\varepsilon_i^2}{2 \sigma_i^2} \left( 1 -
\frac{\varepsilon_i d_i}{3} \right) \right\}. \label{eq: 2nd ver.
for the bounds on the improved lower bound}
\end{equation}
Based on the continuity assumption of the indexed family
$\{P_{\theta}\}_{\theta \in \Theta}$,
then it follows from \eqref{eq: epsilon1,2} that
$$\lim_{\theta' \rightarrow \theta} \varepsilon_i=0,
\quad \forall \, i \in \{ 1, 2\}$$ and also, from \eqref{eq: d1}
and \eqref{eq: d2} with $P_1$ and $P_2$ replaced by $P_\theta$ and
$P_\theta'$ respectively, then
$$\lim_{\theta' \rightarrow \theta} d_i = 0, \quad \forall \, i \in \{ 1, 2\}.$$
It therefore follows from \eqref{eq: relation between the Chernoff
information and Fisher information} and \eqref{eq: 2nd ver. for
the bounds on the improved lower bound} that
\begin{equation}
\hspace*{-0.4cm} \frac{J(\theta)}{8} \geq \lim_{\theta'
\rightarrow \theta} \frac{E_{\text{L}}(P_{\theta}, P_{\theta'})}{(\theta -
\theta')^2} \geq \lim_{\theta' \rightarrow \theta} \, \min_{i=1,2}
\left\{ \frac{\varepsilon_i^2}{2 \sigma_i^2 (\theta-\theta')^2}
\right\}. \label{eq: bounds on the limit}
\end{equation}
The idea is to show that the limit on the right-hand side of this
inequality is $\frac{J(\theta)}{8}$ (same as the left-hand side),
and hence, the limit of the middle term is also
$\frac{J(\theta)}{8}$.
\begin{eqnarray}
&& \hspace*{-0.7cm} \lim_{\theta' \rightarrow \theta}
\frac{\varepsilon_1^2}{2 \sigma_1^2 (\theta-\theta')^2}
\nonumber \\[0.1cm]
&& \hspace*{-0.7cm} \stackrel{(\text{a})}{=}
\lim_{\theta' \rightarrow \theta} \frac{D(P_\theta ||
P_{\theta'})^2}{2 \sigma_1^2 (\theta-\theta')^2} \nonumber \\[0.1cm]
&& \hspace*{-0.7cm} \stackrel{(\text{b})}{=} \frac{J(\theta)}{4}
\lim_{\theta' \rightarrow \theta} \frac{D(P_\theta ||
P_{\theta'})}{\sigma_1^2} \nonumber \\[0.1cm]
&& \hspace*{-0.7cm} \stackrel{(\text{c})}{=} \frac{J(\theta)}{4}
\lim_{\theta' \rightarrow \theta} \frac{D(P_\theta ||
P_{\theta'})}{\sum_{x \in \mathcal{X}} P_\theta(x) \left(
\ln \frac{P_\theta(x)}{P_{\theta'}(x)} - D(P_\theta ||
P_{\theta'}) \right)^2} \nonumber \\[0.1cm]
&& \hspace*{-0.7cm} \stackrel{(\text{d})}{=} \frac{J(\theta)}{4}
\lim_{\theta' \rightarrow \theta} \frac{D(P_\theta ||
P_{\theta'})}{\sum_{x \in \mathcal{X}} P_\theta(x) \left(
\ln \frac{P_\theta(x)}{P_{\theta'}(x)} \right)^2 \; - \; D(P_\theta ||
P_{\theta'})^2} \nonumber \\[0.1cm]
&& \hspace*{-0.7cm} \stackrel{(\text{e})}{=} \frac{J(\theta)^2}{8}
\lim_{\theta' \rightarrow \theta}
\frac{(\theta-\theta')^2}{\sum_{x \in \mathcal{X}}
P_\theta(x) \left( \ln \frac{P_\theta(x)}{P_{\theta'}(x)}
\right)^2 \; - \; D(P_\theta || P_{\theta'})^2} \nonumber \\[0.1cm]
&& \hspace*{-0.7cm} \stackrel{(\text{f})}{=} \frac{J(\theta)^2}{8}
\lim_{\theta' \rightarrow \theta}
\frac{(\theta-\theta')^2}{\sum_{x \in \mathcal{X}}
P_\theta(x) \left( \ln \frac{P_\theta(x)}{P_{\theta'}(x)}
\right)^2} \nonumber \\[0.1cm]
&& \hspace*{-0.7cm} \stackrel{(\text{g})}{=} \frac{J(\theta)}{8}
\label{eq: first calculated limit}
\end{eqnarray}
where equality~(a) follows from \eqref{eq: epsilon1,2},
equalities~(b), (e) and~(f) follow from \eqref{eq: relation
between the Chernoff information and Fisher information},
equality~(c) follows from \eqref{eq: sigma1 squared for the jumps
of the martingale U} with $P_1 = P_{\theta}$ and $P_2 =
P_{\theta'}$, equality~(d) follows from the definition of the
divergence, and equality~(g) follows by calculus (the required
limit is calculated by using L'H\^{o}pital's rule twice) and from
the definition of Fisher information in \eqref{eq: Fisher
information}. Similarly, also
$$ \lim_{\theta' \rightarrow \theta} \frac{\varepsilon_2^2}{2
\sigma_2^2 (\theta-\theta')^2} = \frac{J(\theta)}{8}$$ so
$$ \lim_{\theta' \rightarrow \theta} \, \min_{i=1,2}
\left\{ \frac{\varepsilon_i^2}{2 \sigma_i^2 (\theta-\theta')^2}
\right\} = \frac{J(\theta)}{8}.$$ Hence, it follows from
\eqref{eq: bounds on the limit} that $ \lim_{\theta' \rightarrow
\theta} \frac{E_{\text{L}}(P_\theta, P_{\theta'})}{(\theta - \theta')^2}
= \frac{J(\theta)}{8}.$ This completes the proof of \eqref{eq:
relation between the improved lower bound on the error exponent
and Fisher information}.

We prove now equation \eqref{eq: relation between the loosened
lower bound on the error exponent and Fisher information}. From
\eqref{eq: d1}, \eqref{eq: d2}, \eqref{eq: delta1 and delta2} and
\eqref{eq: tilde E_L} then
\begin{equation}
\widetilde{E}_{\text{L}}(P_\theta, P_{\theta'}) = \min_{i=1,2}
\frac{\varepsilon_i^2}{2 d_i^2}
\end{equation}
with $\varepsilon_1$ and $\varepsilon_2$ in \eqref{eq:
epsilon1,2}. Hence,
\begin{equation*}
\lim_{\theta' \rightarrow \theta}
\frac{\widetilde{E}_{\text{L}}(P_\theta, P_{\theta'})}{(\theta' -
\theta)^2} \leq \lim_{\theta' \rightarrow \theta}
\frac{\varepsilon_1^2}{2 d_1^2 (\theta' - \theta)^2}
\end{equation*}
and from \eqref{eq: first calculated limit} and last inequality
then it follows that
\begin{eqnarray}
&& \hspace*{-0.5cm} \lim_{\theta' \rightarrow \theta}
\frac{\widetilde{E}_{\text{L}}(P_\theta, P_{\theta'})}{(\theta' - \theta)^2} \nonumber \\
&& \hspace*{-0.5cm} \leq \frac{J(\theta)}{8}
\lim_{\theta' \rightarrow \theta} \frac{\sigma_1^2}{d_1^2} \nonumber \\
&& \hspace*{-0.5cm} \stackrel{(\text{a})}{=}
\frac{J(\theta)}{8} \lim_{\theta' \rightarrow \theta}
\frac{\sum_{x \in \mathcal{X}} P_\theta(x)
\left( \ln \frac{P_\theta(x)}{P_{\theta'}(x)} -
D(P_\theta || P_{\theta'}) \right)^2}{\biggl( \max_{x
\in \mathcal{X}} \left|  \ln \frac{P_\theta(x)}{P_{\theta'}(x)}
- D(P_\theta || P_{\theta'}) \right| \biggr)^2}. \nonumber \\
\label{eq: second calculated limit}
\end{eqnarray}
It is clear that the second term on the right-hand side of
\eqref{eq: second calculated limit} is bounded between zero and
one (if the limit exists). This limit can be made arbitrarily small,
i.e., there exists an indexed family of probability mass functions
$\{P_{\theta}\}_{\theta \in \Theta}$ for which the second term on
the right-hand side of \eqref{eq: second calculated limit} can be
made arbitrarily close to zero. For a concrete example, let
$\alpha \in (0,1)$ be fixed, and $\theta \in \reals^+$ be a
parameter that defines the following indexed family of probability
mass functions over the ternary alphabet $\mathcal{X} = \{0, 1,
2\}$: $$ P_{\theta}(0) = \frac{\theta(1-\alpha)}{1+\theta}, \quad
P_{\theta}(1) = \alpha, \quad P_{\theta}(2) =
\frac{1-\alpha}{1+\theta}. $$ Then, it follows by calculus that
for this indexed family
\begin{equation*}
\lim_{\theta' \rightarrow \theta} \frac{\sum_{x \in \mathcal{X}}
P_\theta(x) \left( \ln \frac{P_\theta(x)}{P_{\theta'}(x)}
- D(P_\theta || P_{\theta'}) \right)^2}{\biggl( \max_{x \in
\mathcal{X}} \left|  \ln \frac{P_\theta(x)}{P_{\theta'}(x)} -
D(P_\theta || P_{\theta'}) \right| \biggr)^2} = (1-\alpha) \theta
\end{equation*}
so, for any $\theta \in \reals^{+}$, the above limit can be made
arbitrarily close to zero by choosing $\alpha$ close enough to~1.
This completes the proof of \eqref{eq: relation between the
loosened lower bound on the error exponent and Fisher
information}, and also the proof of
Proposition~\ref{proposition: Fisher information}.

\section{Proof of the properties in \eqref{eq: properties of Y-sequence for OFDM signals}}
\label{appendix: OFDM} Consider an OFDM signal from
Section~\ref{subsection: Concentration of the Crest-Factor for
OFDM Signals}. The sequence in \eqref{eq: martingale sequence for
OFDM} is a martingale due to Remarks~\ref{remark: construction of
martingales} and~\ref{remark: construction of martingales
(cont.)}. It is claimed that $|Y_i - Y_{i-1}| \leq
\frac{2}{\sqrt{n}} $ (a.s.). To show this, note that from
\eqref{eq: CF}, for every $i \in \{0, \ldots, n-1\}$
\begin{eqnarray*}
&& \hspace*{-0.7cm} Y_i = \expectation_{X_i, \ldots, X_{n-1}}
\Bigl[ \, \max_{0 \leq t \leq T} \bigl|s(t; X_0, \ldots, X_{i-1},
X_i, \ldots, X_{n-1}) \bigr| \\
&& \hspace*{4cm} \Big| \, X_0, \ldots, X_{i-1}\Bigr].
\end{eqnarray*}
The conditional expectation for the RV $Y_{i-1}$
refers to the case where only $X_0, \ldots, X_{i-2}$ are revealed.
Let $X'_{i-1}$ stand for the next RV in the sequence $\{X_j\}_{j=0}^{n-1}$
whose value is not revealed (note that for $i=n-1$, an expectation
is not required). Then, for every $1 \leq i \leq n$,

\vspace*{-0.2cm}
\small
\begin{eqnarray*} && \hspace*{-0.7cm} Y_{i-1} =
\expectation_{X'_{i-1}, X_i, \ldots, X_{n-1}}\Bigl[ \, \max_{0
\leq t \leq T} \bigl|s(t; X_0, \ldots, X'_{i-1}, X_i, \ldots,
X_{n-1})\bigr| \\
&& \hspace*{4cm} \Big| \, X_0, \ldots, X_{i-2} \Bigr].
\end{eqnarray*}
\normalsize Since $|\expectation(Z)| \leq E(|Z|)$, then for $i \in
\{1, \ldots, n\}$

\vspace*{-0.2cm}
\small
\begin{equation}
\hspace*{-0.2cm} |Y_i - Y_{i-1}| \leq \expectation_{X'_{i-1}, X_i,
\ldots, X_{n-1}} \Bigl[|U-V| \; \Big| \; X_0, \ldots, X_{i-1}
\Bigr] \label{eq: bounded differences for the sequence Y}
\end{equation}
\normalsize where
\begin{eqnarray*}
&& U \triangleq \max_{0 \leq t \leq T} \bigl|s(t; X_0, \ldots,
X_{i-1},
X_i, \ldots, X_{n-1})\bigr| \\
&& V \triangleq \max_{0 \leq t \leq T} \bigl|s(t; X_0, \ldots,
X'_{i-1}, X_i, \ldots, X_{n-1})\bigr|.
\end{eqnarray*}
It therefore implies that
\begin{eqnarray}
&& |U-V| \leq \max_{0 \leq t \leq T} \bigl|s(t; X_0, \ldots,
X_{i-1}, X_i, \ldots, X_{n-1}) \nonumber \\
&& \hspace*{2.5cm} - s(t; X_0, \ldots, X'_{i-1}, X_i, \ldots,
X_{n-1})\bigr| \nonumber \\[0.15cm]
&& \hspace*{1.3cm} = \max_{0 \leq t \leq T} \frac{1}{\sqrt{n}} \,
\Bigr|\bigl(X_{i-1} - X'_{i-1}\bigr) \exp\Bigl(\frac{j \, 2\pi i
t}{T}\Bigr)\Bigr| \nonumber \\[0.15cm]
&& \hspace*{1.3cm} = \frac{|X_{i-1} - X'_{i-1}|}{\sqrt{n}} \leq
\frac{2}{\sqrt{n}} \label{eq: bound on |U-V|}
\end{eqnarray}
where the last equality holds since $|X_{i-1}|=|X'_{i-1}|=1$. It
therefore follows from \eqref{eq: bounded differences for the
sequence Y} that, for every $i \in \{1, \ldots, n\}$, the
inequality $ |Y_i - Y_{i-1}| \leq \frac{2}{\sqrt{n}}$ holds a.s.
In the following, an upper bound on the conditional variance
$$\text{Var}(Y_i \, | \, \mathcal{F}_{i-1}) = \expectation \bigl[
(Y_i - Y_{i-1})^2 \, | \, \mathcal{F}_{i-1} \bigr]$$ is derived
for every $i \in \{1, \ldots, n\}$. From \eqref{eq: bounded
differences for the sequence Y}, \eqref{eq: bound on |U-V|}, and
since $\bigl(\expectation(Z)\bigr)^2 \leq \expectation(Z^2)$ for a
real-valued RV $Z$, and the RVs $X_i, \ldots X_{n-1}$ are
independent of $X_{i-1}$ and $X'_{i-1}$ (by assumption), then
\begin{equation*}
\vspace*{-0.1cm} (Y_i - Y_{i-1})^2 \leq \frac{1}{n} \cdot
\expectation_{X'_{i-1}} \left[ |X_{i-1} - X'_{i-1}|^2 \, \big| \,
X_{i-1} \right].
\end{equation*}
This implies that
\begin{equation*}
\vspace*{-0.1cm}
\expectation\bigl[(Y_i - Y_{i-1})^2 \, | \mathcal{F}_{i-1}\bigr]
\leq \frac{1}{n} \cdot \expectation_{X'_{i-1}} \bigl[|X_{i-1} -
X'_{i-1}|^2 \, | \, \mathcal{F}_i \bigr]
\end{equation*}
where $\mathcal{F}_i$ is the $\sigma$-algebra that is generated by
$X_0, \ldots, X_{i-1}$. By assumption, the RVs $X_{i-1}$ and
$X'_{i-1}$ get independently each of the $M$ possible values on
the unit circle $e^{\frac{j(2k+1)\pi}{M}}$ for $k = 0, \ldots,
M-1$ with equal probability $\bigl(\frac{1}{M}\bigr)$. The above
conditioning on the $\sigma$-algebra $\mathcal{F}_i$ is equivalent
to the conditioning on the RVs $X_0, \ldots, X_{i-1}$ (i.e., a
conditioning on the first $i$ elements of the sequence
$\{X_j\}_{j=0}^{n-1}$ that serves to construct the OFDM signal in
\eqref{eq: OFDM signal}). Due to the symmetry of the considered
constellation of $M$ points on the unit-circle, one can assume
without any loss of generality that $X_{i-1}$ is equal to
$\exp\bigl({\frac{j \pi}{M}}\bigr)$ (i.e., it is set to be fixed
to one of these $M$ points), and $X'_{i-1}$ gets with equal
probability each of the $M$ possible points on this unit circle.
This gives that

\vspace*{-0.2cm}
\small
\begin{eqnarray}
&& \hspace*{-2.6cm} \expectation \bigl[ (Y_i - Y_{i-1})^2 \, |
\, \mathcal{F}_{i-1} \bigr] \nonumber \\
&& \hspace*{-2.6cm} \leq \frac{1}{n} \, \expectation_{X'_{i-1}}
\bigl[|X_{i-1} - X'_{i-1}|^2 \, | \, \mathcal{F}_i \bigr] \nonumber \\
&& \hspace*{-2.6cm} = \frac{1}{n} \, \expectation \bigl[
|X_{i-1} - X'_{i-1}|^2 \, | \, X_0, \ldots, X_{i-1} \bigr] \nonumber \\
&& \hspace*{-2.6cm} = \frac{1}{n} \, \expectation \bigl[
|X_{i-1} - X'_{i-1}|^2 \, | \, X_{i-1} \bigr] \nonumber \\
&& \hspace*{-2.6cm} = \frac{1}{n} \, \expectation \Bigl[ |X_{i-1}
- X'_{i-1}|^2 \, | \, X_{i-1} = e^{\frac{j \pi}{M}} \Bigr] \nonumber \\
&& \hspace*{-2.6cm} = \frac{1}{nM} \sum_{k=0}^{M-1}
|e^{\frac{j \pi}{M}}-e^{\frac{j (2k+1)\pi}{M}}|^2 \nonumber \\
&& \hspace*{-2.6cm} = \frac{4}{nM} \sum_{k=0}^{M-1} \sin^2 \Bigl(
\frac{\pi k}{M} \Bigr) = \frac{2}{n}. \label{eq: trigonometric identity}
\end{eqnarray}
\normalsize To clarify the last equality, note that if $x \in \reals$ and $m
\in \naturals \cup \{0\}$

\vspace*{-0.3cm}
\small
\begin{eqnarray*}
&& \hspace*{-1.6cm} \sum_{k=0}^m \sin^2(kx)
= \frac{1}{2} \sum_{k=0}^m \bigl(1-\cos(2kx)\bigr) \\
&& \hspace*{-1.6cm} = \frac{m+1}{2} - \frac{1}{2} \, \text{Re}
\biggl\{ \sum_{k=0}^m e^{j 2kx} \biggr\} \\
&& \hspace*{-1.6cm} = \frac{m+1}{2} - \frac{1}{2} \, \text{Re}
\biggl\{\frac{1-e^{2j(m+1)x}}{1-e^{2jx}} \biggr\} \\
&& \hspace*{-1.6cm} = \frac{m+1}{2} - \frac{1}{2} \, \text{Re}
\Biggl\{\frac{\bigl(e^{j(m+1)x}-e^{-j(m+1)x}\bigr)
e^{jmx}}{e^{jx}-e^{-jx}} \Biggr\} \\
&& \hspace*{-1.6cm} = \frac{m+1}{2} - \frac{1}{2} \, \text{Re}
\biggl\{\frac{\sin((m+1)x) \, e^{jmx}}{\sin(x)} \biggr\} \\
&& \hspace*{-1.6cm}
= \frac{m+1}{2} - \frac{\sin\bigl((m+1)x\bigr)
\, \cos(mx)}{2 \sin(x)}
\end{eqnarray*}
\normalsize which then implies the equality in \eqref{eq: trigonometric identity}.

\end{document}